\documentclass[preprint,12pt]{elsarticle}

\usepackage{graphicx}
\usepackage{bm}
\usepackage{natbib}
\usepackage{xspace} 
\usepackage{hyperref}
\usepackage{todonotes}

\usepackage[mathlines]{lineno} 

\usepackage{ifpdf} 

\usepackage{latexsym,amssymb,amsmath,color}

\newcommand{\be}{\begin{equation}}
\newcommand{\ee}{\end{equation}}
\newcommand{\xxi}{\vec{\xi}}

\newcommand{\ave}[1]{\left\langle {#1} \right\rangle}

\newcommand{\pard}[2]{\frac{\partial{#1}}{\partial{#2}}}

\renewcommand{\vec}[1]{{\mathchoice
                     {\mbox{\boldmath$\displaystyle{#1}$}}
                     {\mbox{\boldmath$\textstyle{#1}$}}
                     {\mbox{\boldmath$\scriptstyle{#1}$}}
                     {\mbox{\boldmath$\scriptscriptstyle{#1}$}}}}

\newcommand{\pd}{\pard}
\newcommand{\geoclaw}{{\sc GeoClaw}\xspace}
\newcommand{\clawpack}{{\sc Clawpack}\xspace}
\newcommand{\tohoku}{T\={o}hoku\xspace}
        \newcommand{\argmin}{\arg\!\min}
        
\journal{Ocean Modelling}

\begin{document}

\ifpdf
\DeclareGraphicsExtensions{.pdf, .png, .jpg, .tif}
\else
\DeclareGraphicsExtensions{.png, .jpg, .tif, .eps}
\fi

\begin{frontmatter}

\title{ Quantifying Uncertainties in Fault Slip Distribution during the \tohoku Tsunami using Polynomial Chaos}
\date{\today}

\author[duke,kaust1]{Ihab Sraj\corref{cor1}}
\ead{ihab.sraj@duke.edu}
\author[cu]{Kyle T. Mandli}
\author[duke,kaust2]{Omar M. Knio}
\author[ut]{Clint N. Dawson}
\author[kaust1,kaust2]{Ibrahim Hoteit}
\address[duke]{Department of Mechanical Engineering and Materials Science, Duke University, 144
Hudson Hall, Durham, North Carolina 27708, USA}
\address[kaust1]{Division of Physical Sciences and Engineering, King Abdullah University of Science and Technology, Thuwal, Saudi Arabia}
\address[cu]{Department of Applied Physics and Applied Mathematics, Columbia
University, 500 W. 120th St., New York, NY 10027, USA}
\address[kaust2]{Division of Computer, Electrical and Mathematical Sciences and Engineering, King Abdullah University of Science and Technology, Thuwal, Saudi Arabia}
\address[ut]{Institute for Computational Engineering and Science, University of Texas at Austin, 201 E 24th ST. Stop C0200, Austin, TX 78712-1229, USA}

\cortext[cor1]{Corresponding author}
\date{\today}

\begin{abstract}
An efficient method for inferring Manning's $n$ coefficients using water surface elevation data was
presented in Sraj \emph{et al.} \cite{sraj:2014} focusing on a test case based on data collected during the \tohoku earthquake
and tsunami. Polynomial chaos expansions were used to build an inexpensive
surrogate for the numerical model \geoclaw, which were then used to perform a
sensitivity analysis in addition to the inversion. In this paper, a new analysis
is performed with the goal of inferring the fault slip distribution of the 
\tohoku earthquake using a similar problem setup. The same approach to
constructing the PC surrogate did not lead to a converging expansion, 
however an alternative approach based on Basis-Pursuit DeNoising 
was found to be suitable. Our result shows that the fault slip distribution
can be inferred using water surface elevation data whereas the inferred values
minimizes the error between observations and the numerical model. The numerical approach and the resulting 
inversion are presented in this work.
\end{abstract}


\end{frontmatter}

\section{Introduction} \label{sec:intro}
Natural disasters impacting coastlines have long been some of the most dangerous
and unpredictable of all natural hazards, and among these tsunamis are particularly devastating.  
Due to their rarity, it is extremely challenging to accurately understand and predict these events, owing to the
uncertainties in their generation, primarily subduction zone earthquakes, and
evolution, primarily the characterization of physical processes and bathymetric
measurements. Quantifying such uncertainties using
available past event data is
critical to help guide decision making during and after an event and also assists in building more accurate models.

One of the greatest sources of uncertainty in tsunami modeling lies with the
earthquake that generates the tsunami.  This is mitigated in the far field by
the nature of the shallow water equations but in the near-field this uncertainty
can lead to significant discrepancies between predicted and actual flooding.
This is particularly troublesome when attempting to forecast tsunami run-up as
the immediately available fault movement is coarse in resolution and highly
uncertain itself. To mitigate this and improve the understanding of these
predictions, we propose an avenue for reconstructing the slip motion based on tsunami
observations immediately available via the DART buoy system.

A number of efforts towards quantification of uncertainty in the context of
tsunamis have been undertaken.  Some studies have looked at fitting multiple
earthquake models, attempting to ascertain the best fit to available data while
allowing for simple variation in their initiation \cite{MacInnes:2013cr}, while
others have looked at other types of generation mechanisms such as land-slide
generated tsunamis \cite{Sarri2012}.  Similar approaches to other problems
within the context of the ocean have also been presented.  Examples of these
include studies examining tidal simulations that employed an adjoint or Kalman
filtering based approach \cite{Das:1992uo, Lardner:1995kn,Verlaan:1997te,
Heemink:2002vt, Mayo:2014}. Recently, the authors also presented an efficient
method for the inversion of Manning's $n$ coefficients that used water surface
elevation data collected during the \tohoku earthquake and
tsunami~\cite{sraj:2014}. The efficiency of the method stems from using a
Polynomial Chaos (PC) surrogate model that approximated the forward model
\geoclaw simulating the tsunami. The surrogate was constructed using a 
non-intrusive spectral projection (NISP) method and was used within a Bayesian
inference formalism to avoid multiple runs of the forward model. We note that
Bayesian inversion of the distribution of fault slip has also been studied using synthetic data of surface displacement~\cite{Fukuda2008}.

The PC method uses polynomials to approximate a forward model (or a function)
and has been employed in the literature in various  applications 
including large-scale models~\citep{Iskandarani2015,Winokur2013,sraj:2013a,sraj:2013b,MatternFennelDowd2012}.
In those applications, traditional spectral projection methods~\citep{sraj:2013a,Reagan:2003,Alexanderian2012}
to construct the PC model were successfully implemented. In recent
studies, however, the spectral projection technique failed to construct
faithfully a PC expansion that represents the forward model~\citep{wang2015,Sraj2016}. 
This was due to the non-linearity of the forward model and 
to the internal noise that was present, leading 
to PC expansion convergence issues. Instead, a compressed sensing technique called Basis-Pursuit-DeNoising (BPDN) was implemented to determine the PC expansion coefficients~\citep{Doostan:2014}. This technique first estimates the noise in the model (if any) and then solves an optimization problem to determine the PC expansion coefficients by
assuming sparsity in the coefficients and fitting the PC surrogate to a set of random model runs subject to the estimated noise. 
BPDN was also recently implemented to build a proxy model for an ocean model with initial and wind forcing uncertainties~\citep{li2015}. 
In that application, there was no noise in the model outputs, however, the BPDN method was used
as it does not require simulations at pre-specified sets of parameters which is a requirement by the NISP method.

In this work, we seek to quantify the uncertainties of the generating earthquake by 
parameterizing the slip field in space. The basic approach is the same as the one employed in~\cite{sraj:2014} 
where a Polynomial Chaos (PC) surrogate is constructed~\citep{MarzoukNajmRahn:2007,MarzoukNajm2009} and used for the inversion process using Bayesian inference~\citep{Malinverno2002}. In the case of the parameterized slip field, however, the specific method of constructing a PC surrogate using the NISP method was not successful.  Instead the BDPN method proved more effective and capable of overcoming the convergence issues of the NISP approach and is the primary contribution presented in this paper. We also present results that show the ability of inferring the fault slip distribution using the DART buoys.

The remainder of the article is laid out as follows. In Section~\ref{sec:setup}, the essential setup of the
forward model is briefly described as well as the earthquake parameterization
considered. In Section~\ref{sec:formu}, the formulation of the inverse problem
including the approaches explored for the construction of the polynomial chaos
surrogate are detailed.  Section~\ref{sec:results} presents results of the PC construction
using both NISP and BPDN methods in addition to results of the forward and inverse problems. Finally a discussion of the results and some conclusions are outlined in Section~\ref{sec:conc}.

\section{Problem Setup} \label{sec:setup}

\begin{figure}[ht]
\centering
\begin{tabular}{clc}        
    \includegraphics[width=0.5\textwidth]{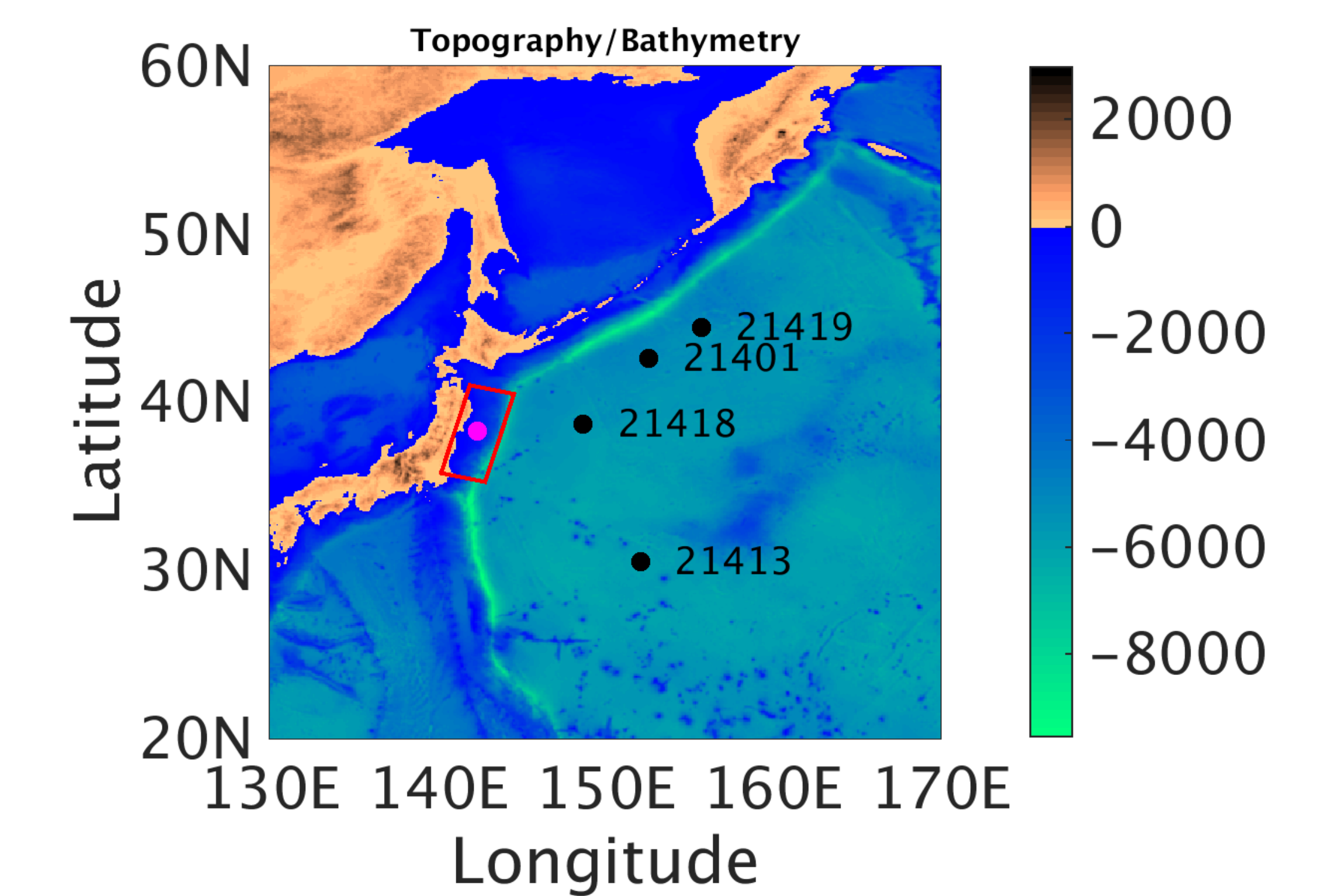} &
\includegraphics[width=0.3\textwidth]{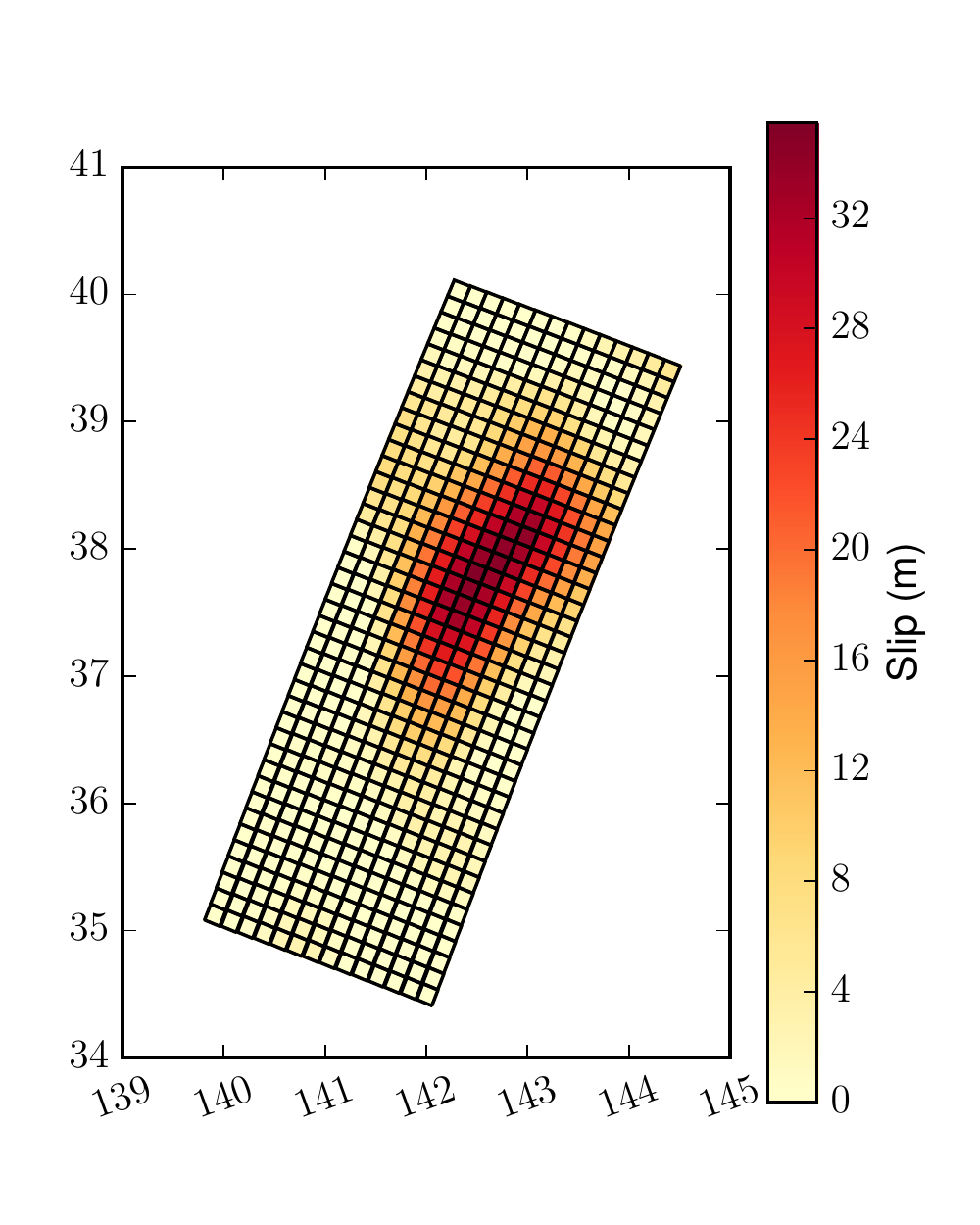} 
\end{tabular}
\caption{(Left) The topography, bathymetry and gauge locations used in the
 simulation, fault is highlighted. (Right) Fault Slip distribution from Ammon
 \emph{et al.} \cite{Ammon:2011dm} with the specified subfault boundaries
 superimposed.}
\label{fig:setup}
\end{figure}

The \tohoku tsunami of 2011 was the most observed tsunami in
history providing us with a wealth of observational data.  The earthquake had an
estimated magnitude of  9.0 ($\text{M}_\text{w}$) causing massive damage across
Japan due to the earthquake alone.  The epicenter of the earthquake was located
approximately 72 km east of the \tohoku region as indicated in
Figure~\ref{fig:setup}(Left).  This section is devoted to the
description of the forward model used to simulate the tsunami and the
parameterization of the earthquake slip field.

\subsection{Forward Model}

The forward numerical model employed in this study is \geoclaw, a package that has been
used to model a number of geophysical phenomena, mostly notably tsunamis for which it has been validated and approved for hazard mapping projects \cite{GonzalezLeVequeEtAl2011}.  It
solves the non-linear, two-dimensional shallow water equations

\begin{equation} \label{eq:swe}
    \begin{aligned}
    &\pd{}{t} h + \pd{}{x} (hu) + \pd{}{y} (hv) = 0, \\
    &\pd{}{t}(hu) + \pd{}{x} \left(hu^2 + \frac{1}{2} g h^2 \right ) + \pd{}{y} (huv) = ~~ fhv - gh \pd{}{x} b - C_f |\vec{u}| hu, \\
    &\pd{}{t} (hv) + \pd{}{x} (huv) + \pd{}{y} \left (hv^2 + \frac{1}{2} gh^2 \right) = -fhu - gh \pd{}{y} b - C_f |\vec{u}| hv,
    \end{aligned}
\end{equation}
where $h$ is the depth of the water column, $u$ and $v$ the velocities in the
longitudinal and latitudinal directions respectively, $g$ the acceleration due
to gravity, $b$ the bathymetry, $f$ the Coriolis parameter, and $C_f$ the bottom
friction coefficient.  The sea-surface anomaly $\eta$, the difference between a
specified datum, such as mean tide level, and the modeled sea-surface, is $\eta
= h + b$.

\geoclaw is an off-shoot of \clawpack that solves systems of hyperbolic
equations in conservative and non-conservative form.  The primary computational
kernel is the Riemann solver which determines fluctuations, wave speeds and
strengths.  The Riemann solver in \geoclaw contains a number of features
relevant to tsunami modeling including: (1) inundation (flooding at the shore);
(2) well-balanced formulation, providing the ability to handle topographical features
while maintaining steady-states (most notably an ocean at rest); and (3) 
inclusion of entropy correction that handle rarefaction of the flow
\cite{George:2008aa}.  One of the key components that makes \geoclaw effective
at modeling trans-oceanic tsunamis is its use of adaptive mesh refinement (AMR).
AMR allows resolution of the model to follow features of the solution of
interest, such as the wave height difference from sea-level.  \geoclaw
implements these schemes via block-structured AMR as detailed in
\cite{Berger:1984ui, Berger:1998aa}.

Much of the setup for the \tohoku simulations presented was adapted from the 
\geoclaw simulations presented in \cite{MacInnes:2013cr} including the 
refinement strategy.  This includes resolutions ranging from 1 degree in both
longitude and latitude to 75" resolution.  The bathymetry used here is a 
combination of ETOPO 1' and 4' resolution data~\cite{Amante:2009ud}; the finer bathymetry used in
\cite{MacInnes:2013cr} to model inundation appropriately was excluded as this 
study does not include inundation data in the inversion.

\subsection{Parametric Representation of the Slip Distribution} \label{sec:parameters}

\begin{figure}[h]
\centering
\begin{tabular}{clc}        
    \includegraphics[width=0.5\textwidth]{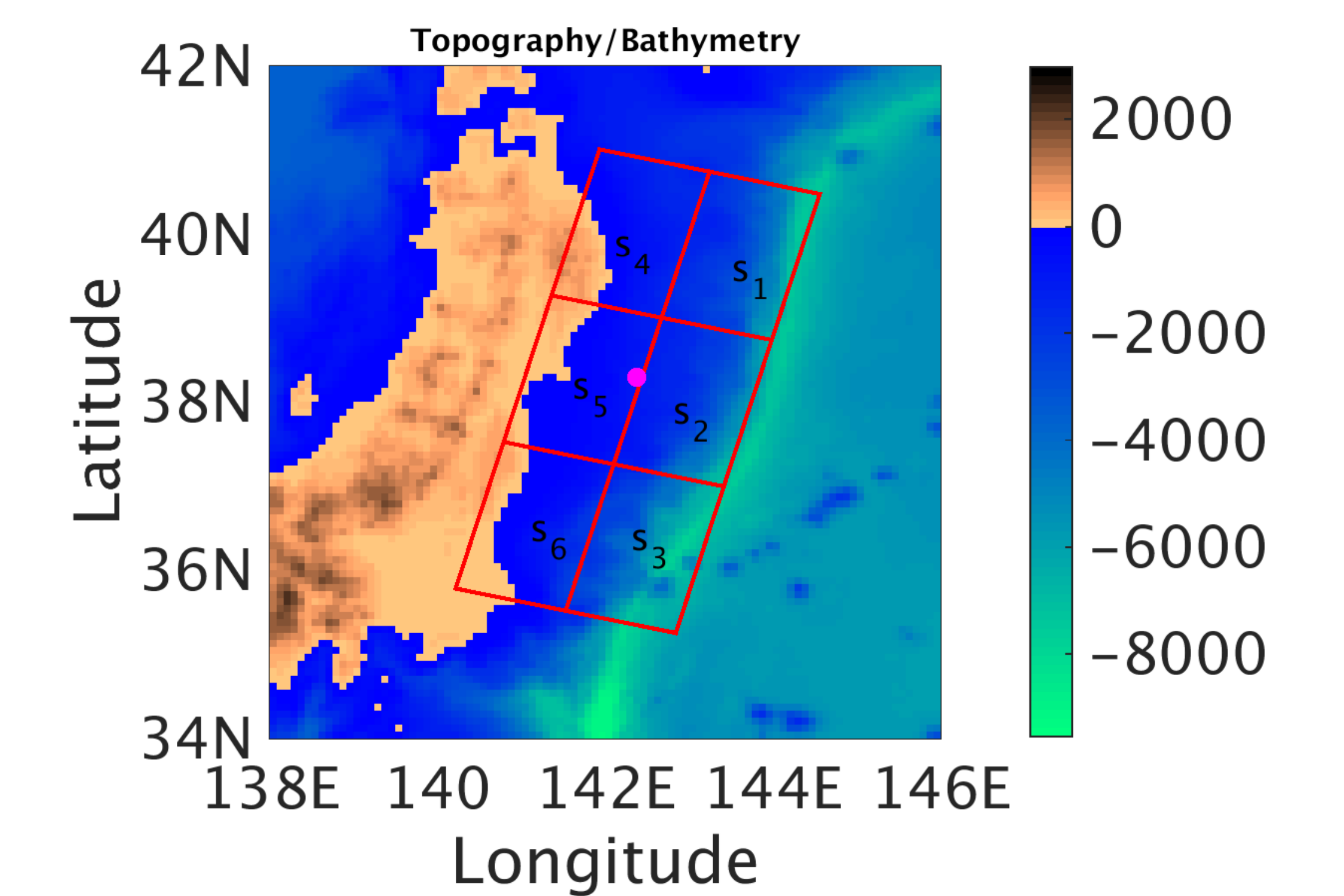}
    \end{tabular}
\caption{ Parameterized slip values in space. The
hypocenter of the earthquake is shown by a marker.}
\label{fig:slips}
\end{figure}

The overall goal of this article is to invert for the source earthquake using
observational data available immediately after the earthquake.  In order to
simplify the investigation of the problem formulation, the base geometry of the fault
was assumed fixed while the slip on the fault is assumed to be uncertain.
Based on previous inversions, the slip magnitude was constrained between the
similar slips proposed in \cite{Ammon:2011dm} of $s_{max} = 30~m$ and no-slip
$s_{min} = 0~m$ (see Figure~\ref{fig:setup} right). Additionally the fault was
broken up into 6 sub-regions of which each can have a unique slip value in the
inversion (see Figure~\ref{fig:slips}) and cover the largest area of slip.  The
initial uncertainty was represented as a non-informative, uniform distribution
with the limits mentioned above.  The uncertainty is then quantified through PC
expansions as in \cite{sraj:2013a,sraj:2013b}.

\section{Formulation} \label{sec:formu}
In this section, we describe the different steps of our method to numerically
solve the inverse problem stated above. In Section~\ref{sec:obs}, we analyze the
available observations used in the Bayesian inference step, outlined in Section~\ref{sec:inference}.
Finally, in Section~\ref{sec:uq}, we provide some details on a key ingredient of
our methodology i.e.\ constructing a surrogate of the forward model for
the sake of accelerating the Bayesian inference.

\subsection{Observations}
\label{sec:obs}
\begin{figure}[ht]
\centering
\begin{tabular}{clc}  
\includegraphics[width=0.5\textwidth]{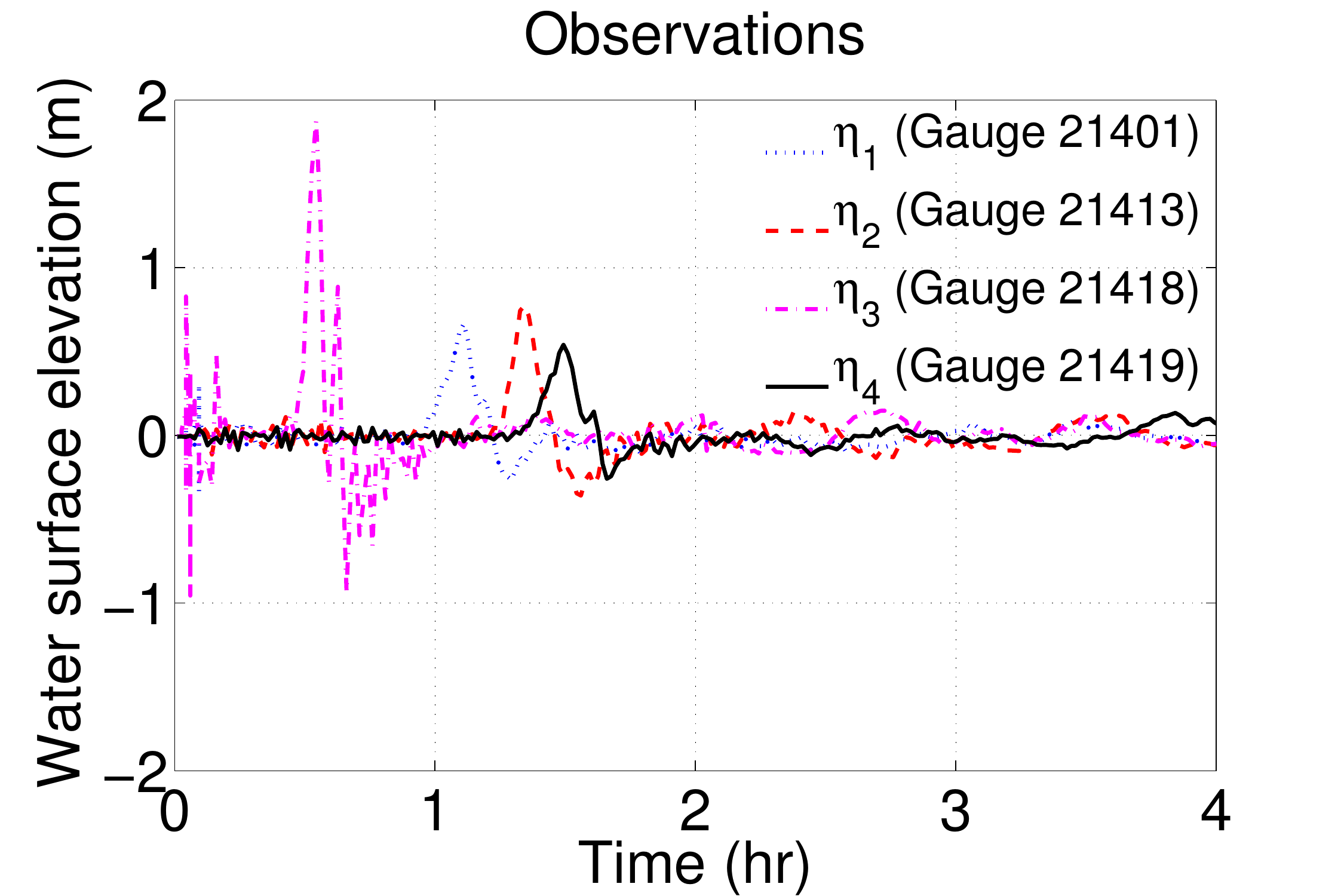} &
\includegraphics[width=0.5\textwidth]{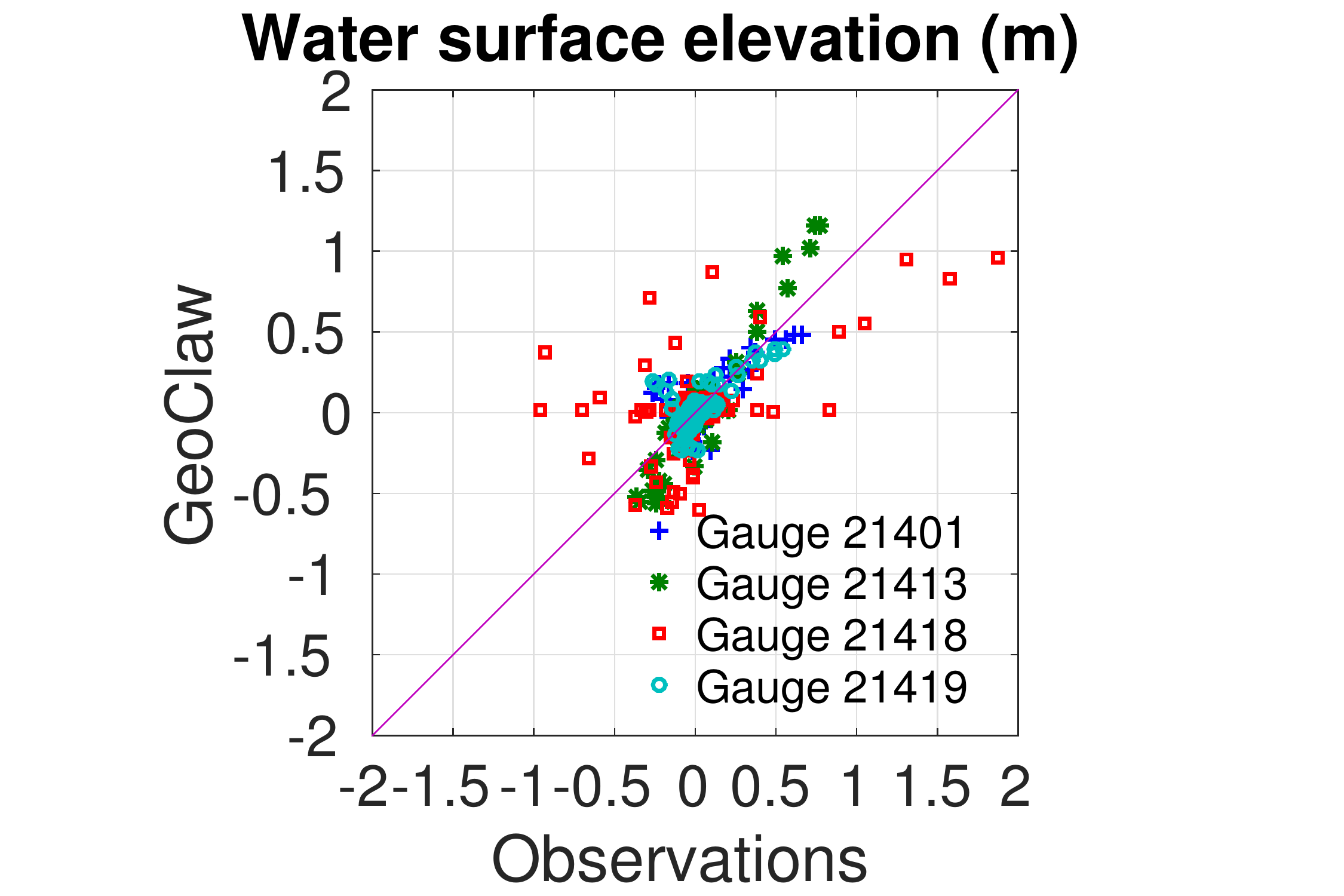} 
\end{tabular}
\caption{(Left) Observed de-tided water surface height at all the DART buoys used.(Right) Scatter plot of the measured water surface elevation against their \geoclaw model counterparts at the four gauges colored differently. The variance of the difference between the two sets of values is:  $7.99\times 10^{-3}~m^2, 9.65\times 10^{-3}~m^2, 4.62\times 10^{-2}~m^2, 5.86\times 10^{-3}~m^2$
at each gauge.}
\label{fig:observations} 
\end{figure}

We use observations consisting of water surface elevation
measurements collected for a period of around $4$ hours during the event at four
different gauge locations. These gauges are part of the Deep-ocean
Assessment and Reporting of Tsunamis (DART) buoy system developed and maintained
by the National Oceanic and Atmospheric Administration (NOAA) with the purpose
of providing early-warning detection and forecasting of tsunami propagation in
the Pacific Ocean \cite{Milburn:1996wm}. The four selected gauges are the
closest to the earthquake source of the \tohoku tsunami denoted by Gauge 21401,
21413, 21418, and 21419. The locations of these buoys are shown in
Figure~\ref{fig:setup} (Left) where the bathymetry and topography of the
numerical domain is also shown. The de-tided water surface elevation data for
the event at the four gauges are shown in Figure~\ref{fig:observations} (Left). The
readers are referred to~\cite{Mungov2013} for details on the data processing
methodologies used for the DART buoy data.

Prior to using these observations for the inference of the fault slip
distribution, we verify the ability of \geoclaw to realistically simulate water
surface elevation during the \tohoku tsunami. To this end, we ran a single
simulation of \geoclaw with default parameters and fault slip
distribution from Ammon \emph{et al.}
\cite{Ammon:2011dm} to predict the water surface elevation at the four gauges.
We compare these with their DART counterparts and plot them in
Figure~\ref{fig:observations} (Right) as a scatter plot for the gauges 21401, 21413, 21418
and 21419. The data points are colored differently for the different gauges and
the variance of the difference between observations and simulations was
calculated to be $7.99\times 10^{-3}~m^2, 9.65\times 10^{-3}~m^2, 4.62 \times 10^{-2}~m^2$
and $5.86\times 10^{-3}~m^2$,  respectively. These variances are consistent with the
distance from the gauges to the epicenter of the earthquake located
approximately 72 kilometers east of Japan. The smallest variance was at gauge
21419 (the farthest gauge from the epicenter) while the higher
variance was at gauge 21418 that can be attributed to its proximity to the
epicenter of the earthquake as well as to the shore region. The scatter plot
along with the calculated variances indicate a reasonable agreement between the
simulations and the observations at the different gauges.  The overall
differences between the simulations and observations can likely be attributed to
uncertainties in the input data such as the Manning's $n$
coefficients~\cite{sraj:2014}, fault slip distribution, errors in the earthquake rupture model,
insufficiently accurate bathymetry in the near-shore region, and to model errors,
such as unresolved effects and approximations inherent in the shallow water
model.

\subsection{Inverse problem}
 \label{sec:inference}
 
Bayesian inference is a well-established probabilistic approach to inverse problems in
which all forms of uncertainty are expressed in terms of random variables. This method
provides complete posterior statistics and not just a single value for the
quantity of interest ($QoI$)~\cite{Tarantola:2005}. Consider a set of $N$ water surface
elevation observations $\vec{\eta}_j =\{{\eta}_j^k\}_{k=1}^N$ measured at the
different DART buoy gauges $j=1,2,3$ and $4$, corresponding to gauges 21401,
21413, 21418 and 21419, respectively. Let $\vec{s}=\{s_i\}_{i=1}^{m=6}$ be a
vector of uncertain parameters representing the six fault slip values. We
consider the forward model  $\vec{G}_j(\vec s) =\{{G}_j^k(\vec s)\}_{k=1}^N$
represented by \geoclaw that predicts the $N$ data at the $j^{th}$ gauge as a
function of the vector of parameters $\vec{s}$ given observations $\vec{\eta}_j$. Bayes's theorem can be applied
that yields:

\begin{equation} \label{eq:bayes}
   \pi(\vec{s}| {\vec \eta}_j) \propto 
   \pi({\vec \eta}_j | \vec{s}) \ \pi(\vec{s}), 
\end{equation}
where $\pi(\vec{s})$ is the prior of $\vec{s}$, $\pi({\vec \eta}_j| \vec s)$ is
the likelihood function and $\pi(\vec{s}| {\vec \eta}_j)$ is the
posterior of $\vec{s}$. The likelihood function $L(\vec{s} | {\vec \eta}_j)  =
\pi({\vec \eta}_j| \vec s)$ can be formulated assuming that independent additive
errors account for the discrepancy between the predicted, $\vec{G}_j(\vec s)
=\{{G}_j^k(\vec{s})\}_{k=1}^N$, and observed $\vec{\eta}_j
=\{{\eta}_j^k\}_{k=1}^N$, values of water surface elevation such that:

\[
    {\epsilon}_j^k = {G}_j^k(\vec s) - {\eta}_j^k, \quad j=1 \ldots 4 , \quad k=1 \ldots N,
\]
where ${\vec \epsilon}_j  = \{{\epsilon}_j^k\}_{k=1}^N$ are assumed to be i.i.d.
random variables with density $p_{\epsilon_j}$.  The likelihood function can
then be written as

\begin{equation} \label{eq:likelihood}
    L(\vec{s} | {\vec \eta}_j) = \prod_{j=1}^4 \prod_{k=1}^N  p_{\epsilon_j} (G_j^k(\vec s) - {\eta}_j^k ).     
\end{equation}

In our application, the measurements may vary significantly from 
one gauge to another and the observations collected may be exposed to different 
measurement errors; therefore, it is reasonable to assume that the errors are normally distributed with
zero mean and a variance that depends on location, i.e. \ $\epsilon_j^k \sim N(0,\sigma_j^2)$
where $\sigma_j^2$ ($j=1 \ldots 4$) is the variance at the different gauges.
Thus the likelihood function can be expressed as

\begin{equation} \label{eq:likelihood2}
    L(\vec{s} | {\vec \eta}_j) = \prod_{j=1}^4\prod_{k=1}^N   \frac{1}{\sqrt{2
    \pi \sigma_j^2}} \exp \left\lbrace \frac{-(G_j^k(\vec s)-\eta^k_j )^2}{2 \sigma_j^2} \right\rbrace, 	
\end{equation}
and the joint posterior in Equation~\ref{eq:bayes} becomes

\[
    \pi(\vec{s}| \eta^k_j) \propto \prod_{j=1}^4 \prod_{k=1}^N  \frac{1}{\sqrt{2 \pi \sigma_j^2}}   \exp \left\lbrace \frac{-(G_j^k(\vec s) - \eta^k_j )^2}{2 \sigma_j^2} \right\rbrace \prod_{i=1}^6   \pi(s_i).
\]
The variance $\sigma_j^2$ is not well known \emph{a priori}, thus it is treated as a hyper-parameter
that becomes an additional parameter for Bayesian inference endowed with a prior which is updated based on available observations. In this 
case the joint posterior is finally expressed as

\begin{equation}  \label{eq:post_coef}
\pi(s_i,\sigma_j^2 | \eta_j^k) 
\propto 
  \prod_{j=1}^4  
\prod_{k=1}^N  
\frac{1}{\sqrt{2 \pi \sigma_j^2}} \exp \left\lbrace \frac{-( G_j^k(\vec s)-\eta^k_j)^2}{2 \sigma_k^2} \right\rbrace
\   \prod_{i=1}^6  
\pi(s_i)  \prod_{j=1}^4  \pi(\sigma_j^2).
\end{equation}

Finally, proper priors are chosen for the uncertain parameters based on some
\emph{a priori} knowledge about them. In our case, we chose a non-informative
uniform prior for all six fault slip values, with $s_i$ in the range  $[s_{min}
- s_{max}]$ so that $\pi(s_i) = \frac{1}{s_{max}-s_{min}}$. Regarding the noise
variance, the only information known is that $\sigma_j^2$ is always positive.
We thus assume a Jeffreys prior \citep{sivia} for $\sigma_j^2$, expressed as:

\begin{equation} 
\pi(\sigma_j^2) =  \begin{cases}
		\displaystyle \frac{1}{\sigma_j^2} &\text{for~} \sigma_j^2 > 0,  \\
		0 &\text{otherwise}. 
		\end{cases}
\label{eq:var_pr}
\end{equation}

The described Bayesian formulation requires sampling the resulting posterior
(Equation~\ref{eq:post_coef}) to estimate the joint posterior of the parameters. Markov
Chain Monte Carlo (MCMC) methods are convenient and popular sampling strategies
that require a large number of posterior evaluations. We rely on an
adaptive Metropolis MCMC algorithm ~\cite{Haario2001,Gareth2009} to efficiently
sample the posterior distribution. In addition, we build a surrogate model of
the model response for further reduction in computational time as explained below.

\subsection{Surrogate model}
\label{sec:uq}

To accelerate the process of sampling the
posterior~(Equation~\ref{eq:post_coef}) using MCMC, we build a surrogate
model of the $QoIs$, namely the  $\vec{\eta}_j$'s using a small ensemble of \geoclaw model runs. For this
purpose, we apply a probabilistic method to express the $QoI$ as a function of
the uncertain model inputs, namely the Polynomial Chaos (PC)
method~\citep{LeMaitreKnio2010,Xiu2004}. As the name indicates, the function
would be in the form of a polynomial expansion~\citep{GhanemSpanos1991,Xiu2004}
that is truncated at a specific order. This approach was adopted
in~\cite{sraj:2014} to build a surrogate model for the water surface elevation
and then used to determine statistical properties (mean and variance) as well as
sensitivities~\citep{Crestaux}. Additionally the surrogate model was used for
efficient sampling of the posteriors. We briefly show here the process of
constructing a PC surrogate for the $QoI$; for more details on the PC method the
reader is referred to~\cite{LeMaitreKnio2010}.

\subsubsection{Polynomial Chaos}
\label{sec:pc}
We denote by $G=G(\xxi)$ our $QoI$ which is the water surface elevation produced by \geoclaw;
$\xxi=[\xi_1,...,\xi_m]$ denotes the canonical vector of $m$ random variables that parameterize the uncertain fault slip values
as follows: 

\[
    \xi_{i} = \frac{2s_i-(s_{min}+s_{min})}{(s_{min}-s_{max})}. 
\]

The PC method seeks to represent $G$ as a function of the uncertain input
variables $\xxi$ as

\begin{equation}
 G(\xxi) \approx \sum_{k = 0}^R g_k \psi_k(\xxi),
\label{eq:stochseries}
\end{equation}
 where $g_k$ are the polynomial coefficients to be determined, and
$\psi_k(\xxi)$ are tensor products of the scaled Legendre polynomials~\citep{LeMaitreKnio2010} forming an orthogonal basis
of the space of square integrable functions of the underlying uniform probability
distributions $\rho(\xxi)$ with

\begin{equation}
 \left<\psi_i,\psi_j\right> = \int \psi_i(\xxi) \;\psi_j(\xxi) \; \rho(\xxi) \; \mbox{d}\xxi=\delta_{ij}\ave{\psi_i^2},
\label{eq:inner}
\end{equation}

The PC coefficients $g_k$ can be determined using a number of methods.
In this work, we rely on non-intrusive approaches~\cite
{Berveiller:2006,Reagan:2003}
that use a set of deterministic model runs $G(\xxi)$ evaluated
at particular realizations of $\xxi$. In particular, we relied on two non-intrusive methods 
described in Section~\ref{sec:nisp} and Section~\ref{sec:bpdn} below. The reasoning behind using
these two methods is explained in the results section. 

\subsubsection{Non-Intrusive Spectral Projection}
\label{sec:nisp}
The Non-Intrusive Spectral Projection (NISP) method makes use
of the orthogonality of the polynomial basis and applies a Galerkin projection
to find the PC expansion coefficients~\citep{Constantine:2012,Conrad:2013} as

\[
    g_k = \frac{\left< G, \psi_k \right>}{\left< \psi_k, \psi_k \right>} = 
 \frac{1}{\left< \psi_k, \psi_k \right>} 
 \int G \psi_k(\xxi) \rho(\xxi) \mbox{ d}\xxi.
\]
A numerical quadrature is used to approximate the integrals with

\[
      \int G \psi_k(\xxi) \rho(\xxi) \mbox{ d}\xxi
\approx \sum_{q=1}^{Q} G(\xxi_q) \psi_k(\xxi_q) \omega_q,
\]
where $\xxi_q$ and $\omega_q$ are the multi-dimensional quadrature points and weights, 
respectively, and $Q$ is the total number of nodes in the multi-dimensional quadrature.
$G(\xxi_q)$ is the model prediction evaluated at the quadrature values $\xxi_q$. 
We note that the order of quadrature should be commensurated with the PC truncation order,
and should be high enough to avoid aliasing artifacts.

\subsubsection{Basis-Pursuit DeNoising}
\label{sec:bpdn}
Basis-Pursuit DeNoising (BPDN) is a non-intrusive method for finding the PC
coefficients using a number of random model evaluations. BPDN is based on
the compressed sensing methodology that assumes sparsity  in a signal, in our case
the PC coefficients, and  seeks to determine the non-zero coefficients. using
optimization techniques~\citep{Berveiller:2006,Blatman:2011,Doostan:2014}. Let
${\vec{g}}=[g_0,...,g_R]$ be the vector of PC coefficients to be determined and
$\vec G = [G(\xxi_1),...,G(\xxi_S)]$ be the vector of random model evaluations
at the sampled $\xxi_s$. We also let $\Psi$ be a matrix whose rows are evaluations
of the PC basis functions $\psi_k(\xxi)$ at the sampled $\xxi_s$. We therefore
transform Equation~\ref{eq:stochseries} into the following system in matrix
form to solve for:

\[
      \vec{G} = \Psi \vec{g}.
\]

The sparsity in the system is exploited by constraining the system and minimizing its "energy", which is its $\ell_1$-norm, 
and thus solving the optimization problem

\begin{equation} \label{eq:optim}
    {\cal{O}}_{1,\delta} \approx \left\lbrace \argmin_{{\vec g}} || {\vec{g}} ||_1  : || \vec G - \Psi \vec{g} ||_2 \le \delta \right\rbrace.
\end{equation}   
In this specific method, we assumed the presence of noise $\delta$ in the signal
that is estimated \emph{a priori} in contrast to the Basis-Pursuit (BP)
technique where no noise is assumed~\cite{Donoho:2006}. The noise $\delta$ is
determined using a cross-validation method that assures the computed PC
coefficients not only fit the random model evaluations but also accurately
approximate the model~\cite{Doostan:2014}. The system ${\cal{O}}_{1,\delta}$ is
then solved using standard $\ell_1$-minimization solvers such as the MATLAB package
SPGL1~\citep{spgl1:2007}, that is based on the spectral projected gradient
algorithm~\citep{BergFriedlander:2008}.

\section{Results} \label{sec:results}

\subsection{PC Expansion Construction and Validation}
\begin{figure}[ht]
\centering
\begin{tabular}{clclclc}        
\includegraphics[width=0.5\textwidth]{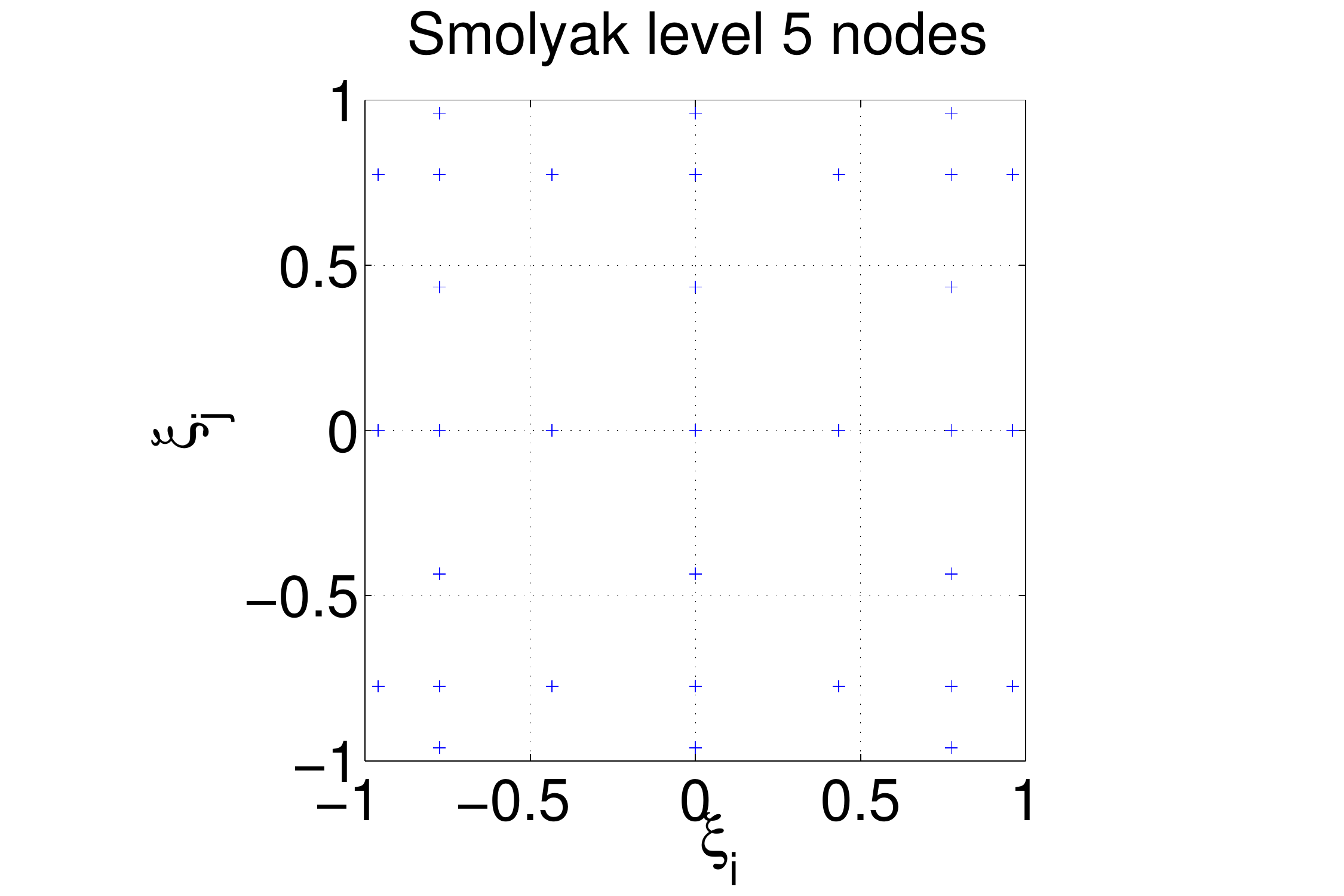} &
\includegraphics[width=0.5\textwidth]{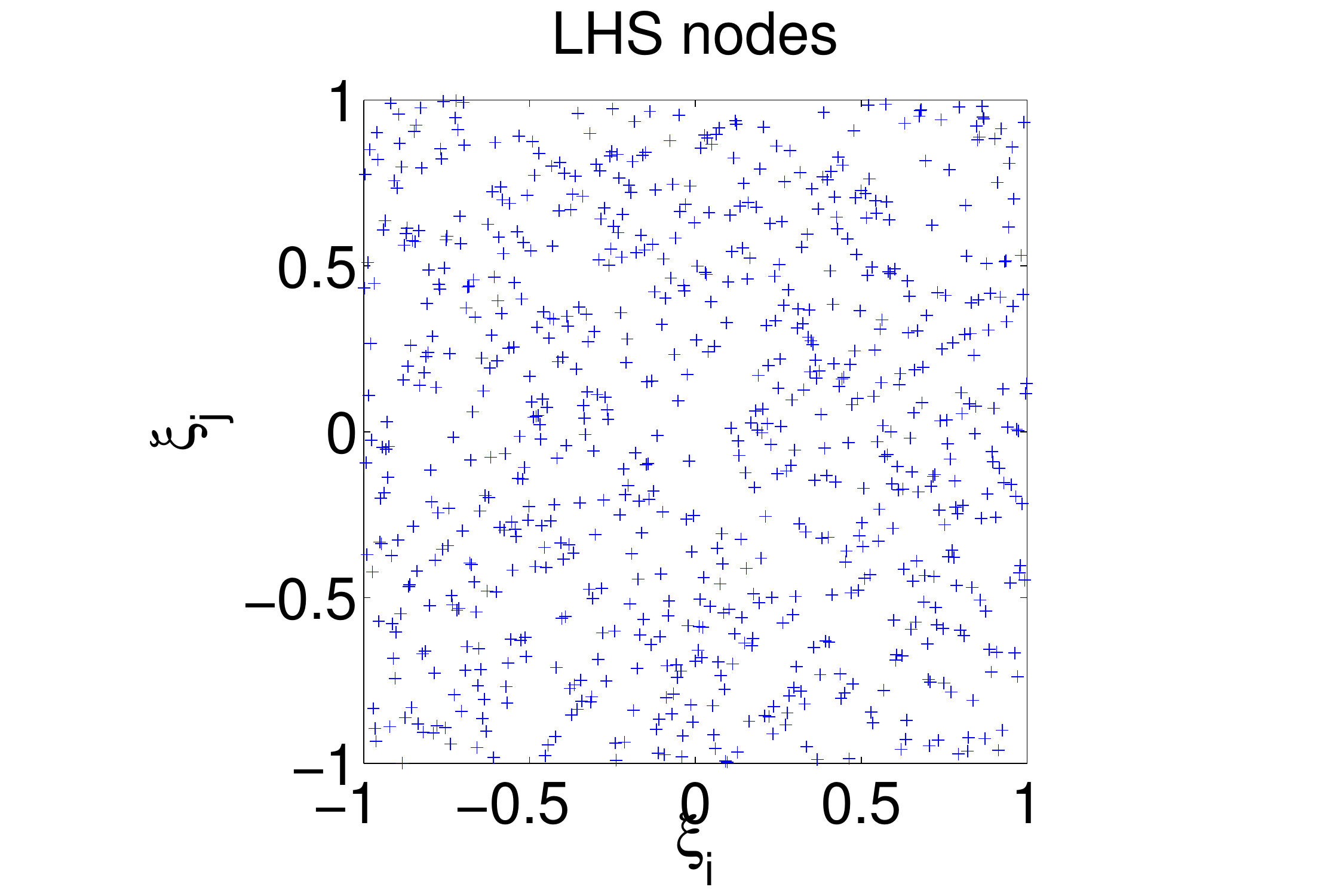} 
\end{tabular}
\caption{2-D projection of the nodes of (Left) Smolyak quadrature and (Right) Latin-Hypercube sample.} 
\label{fig:sample}
\end{figure}

\begin{figure}[h]
\centering
\begin{tabular}{clc}        
\includegraphics[width=0.475\textwidth]{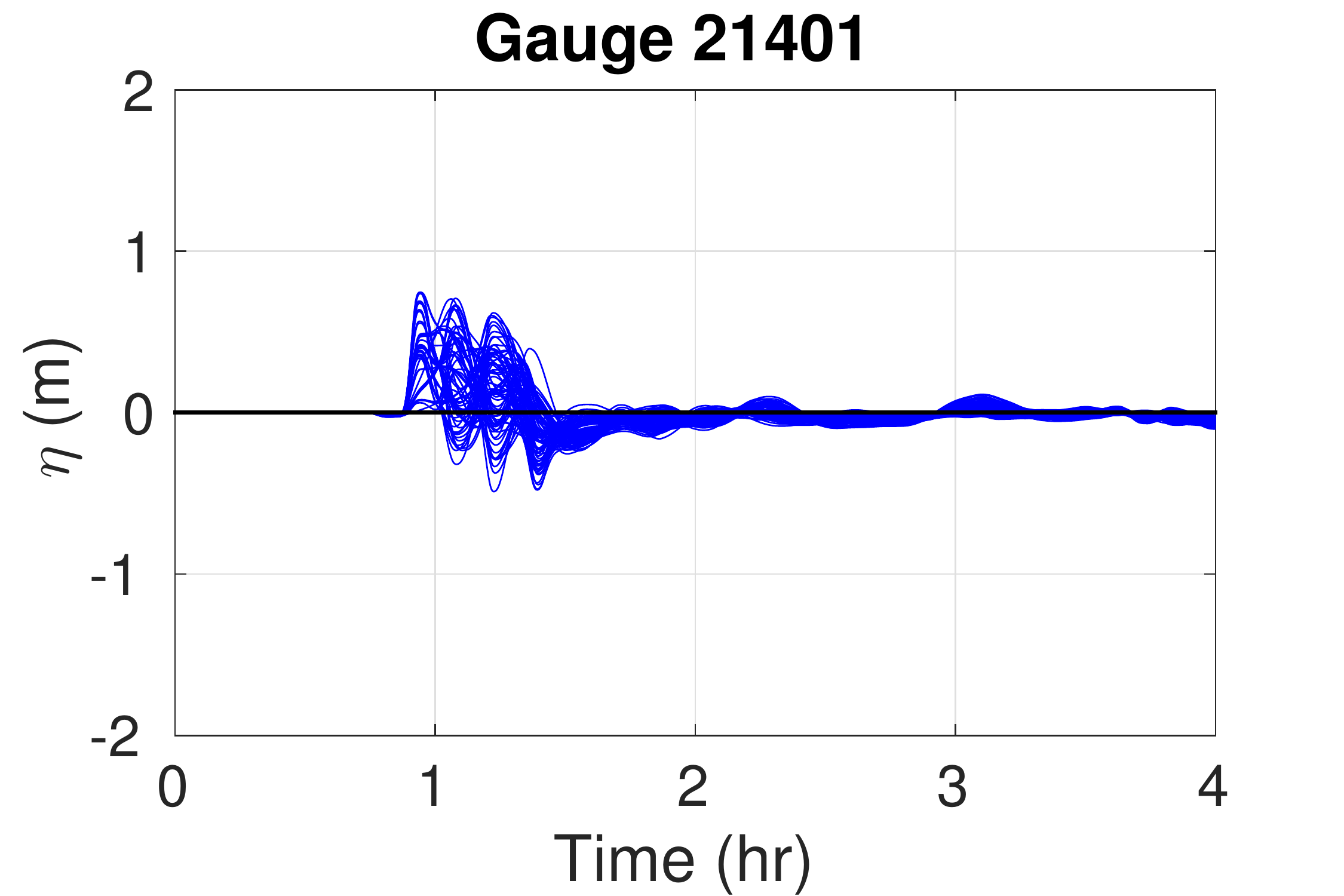} &
\includegraphics[width=0.475\textwidth]{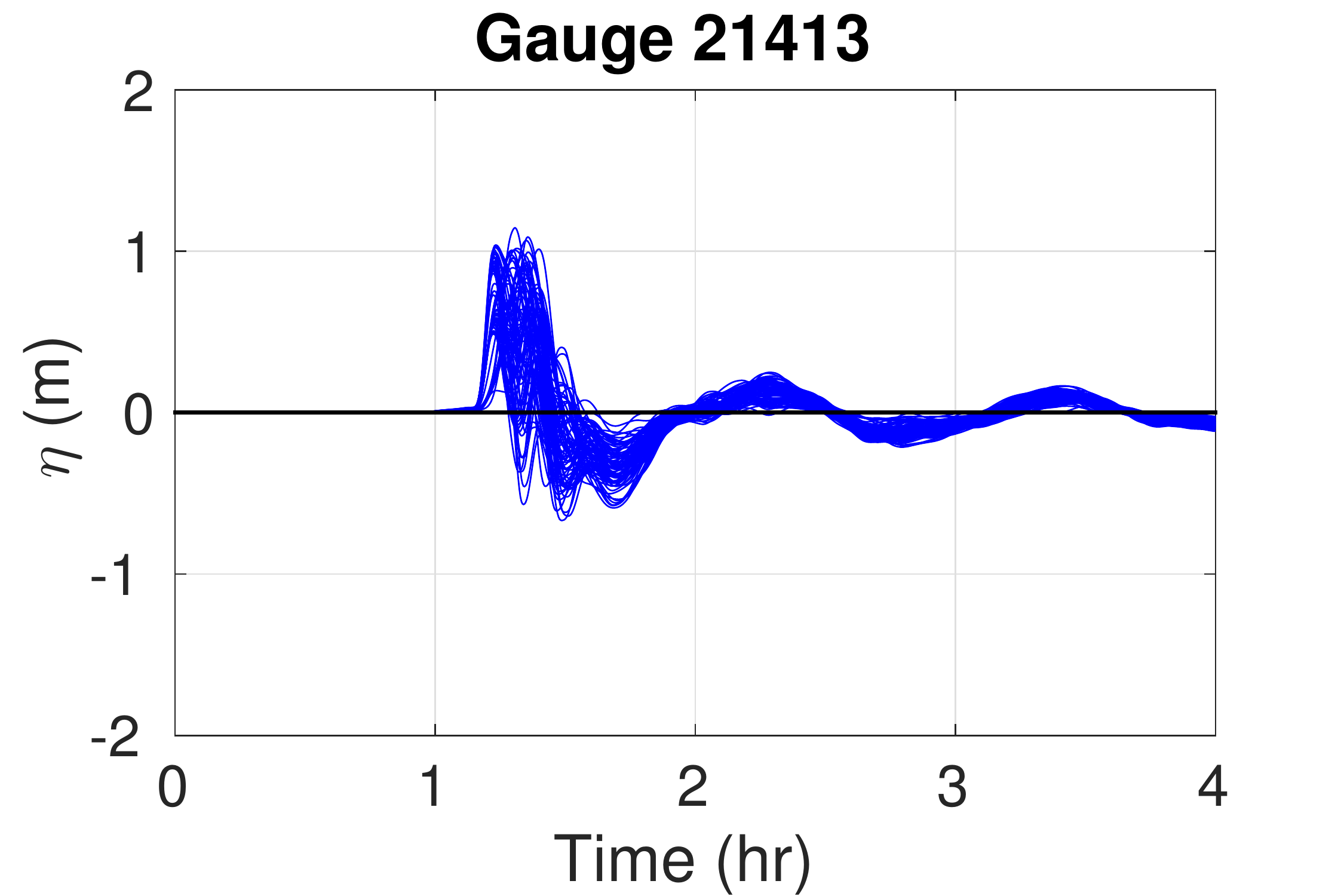} \\
\includegraphics[width=0.475\textwidth]{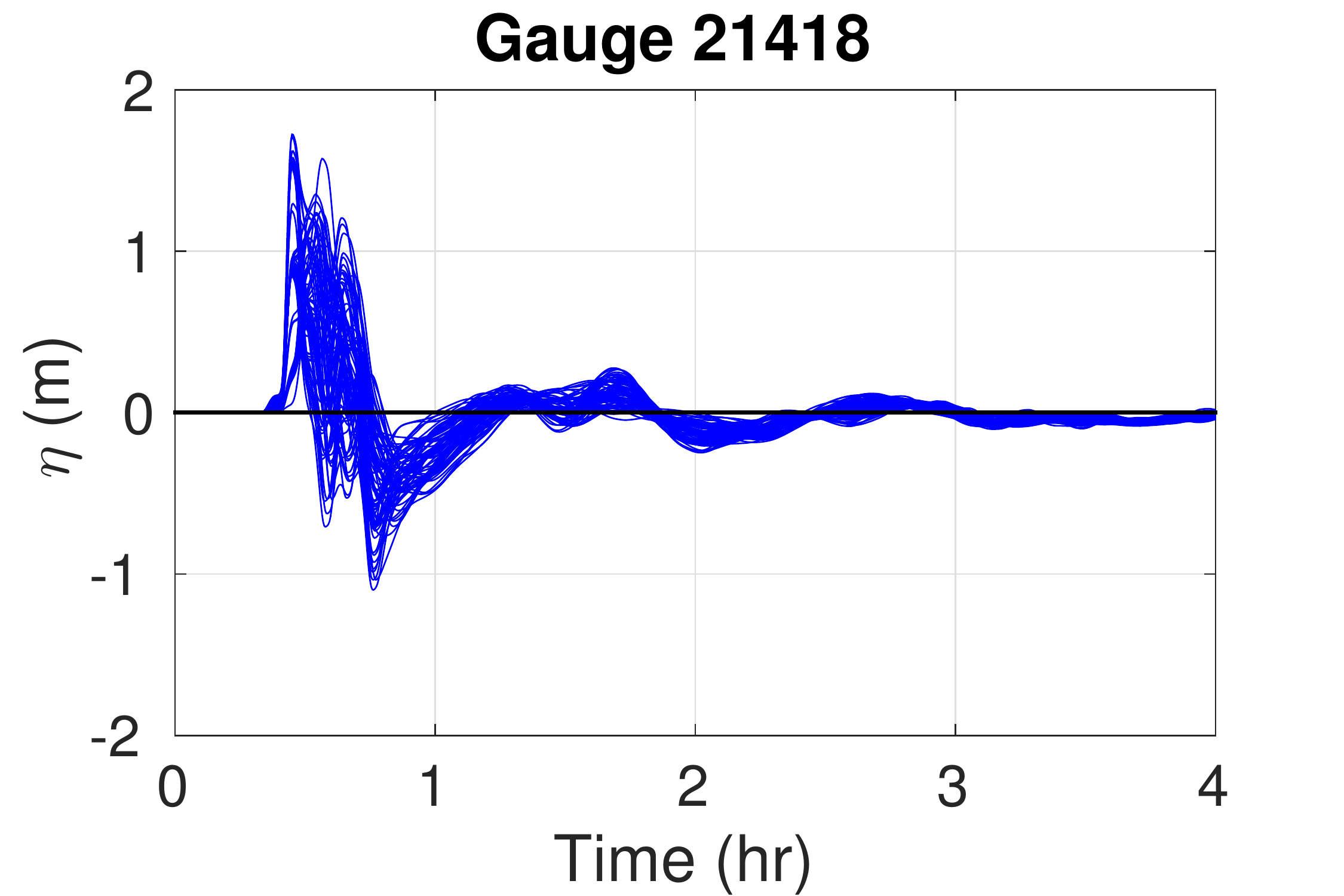} &
\includegraphics[width=0.475\textwidth]{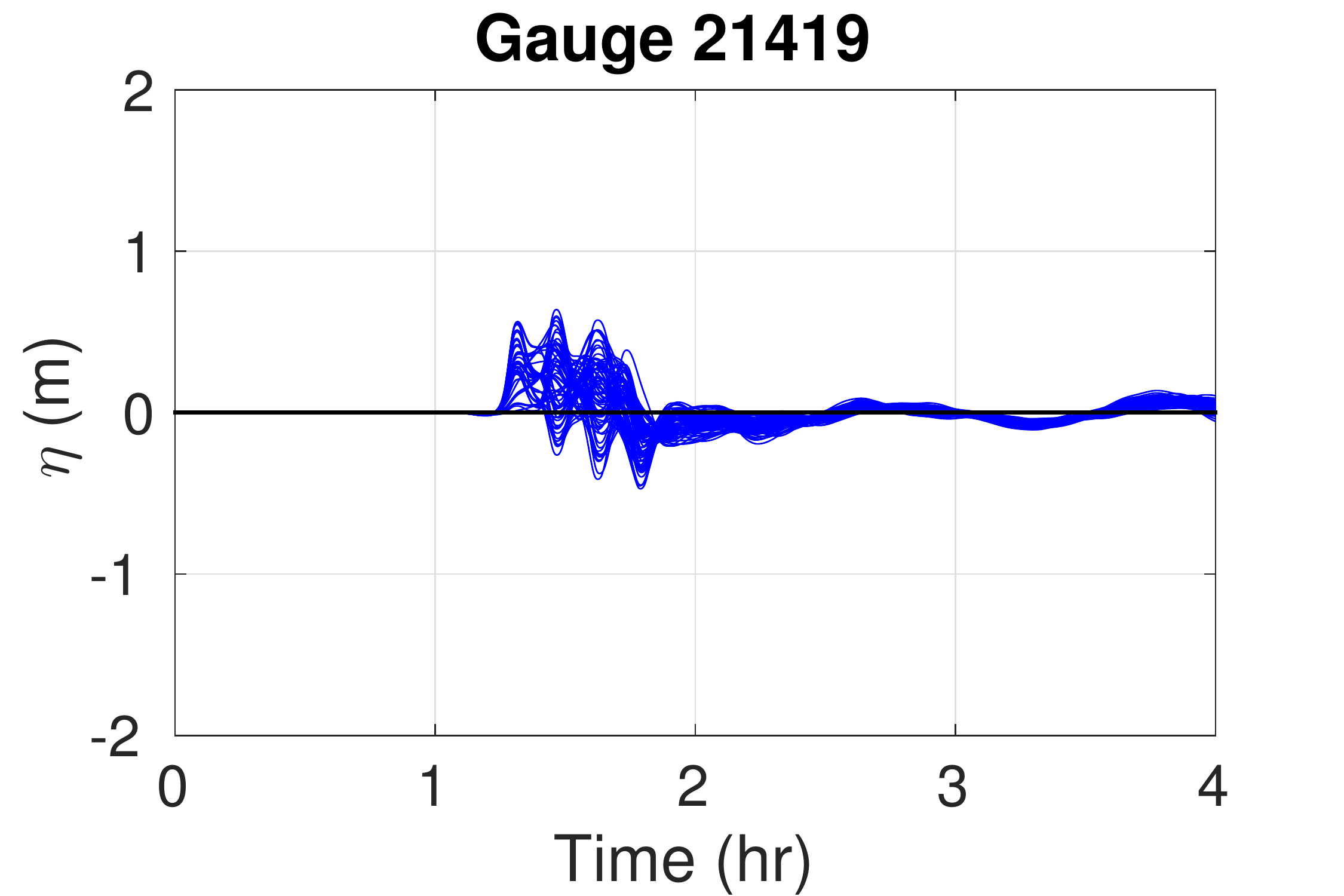} 
\end{tabular}
\caption{1889 realizations of the evolution of the \geoclaw simulated water surface elevation 
corresponding to the Smolyak quadrature nodes of PC order 5 at the different gauges.}
\label{fig:quadrature}
\end{figure}

\begin{figure}[h]
\centering
\begin{tabular}{clc}        
\includegraphics[width=0.475\textwidth]{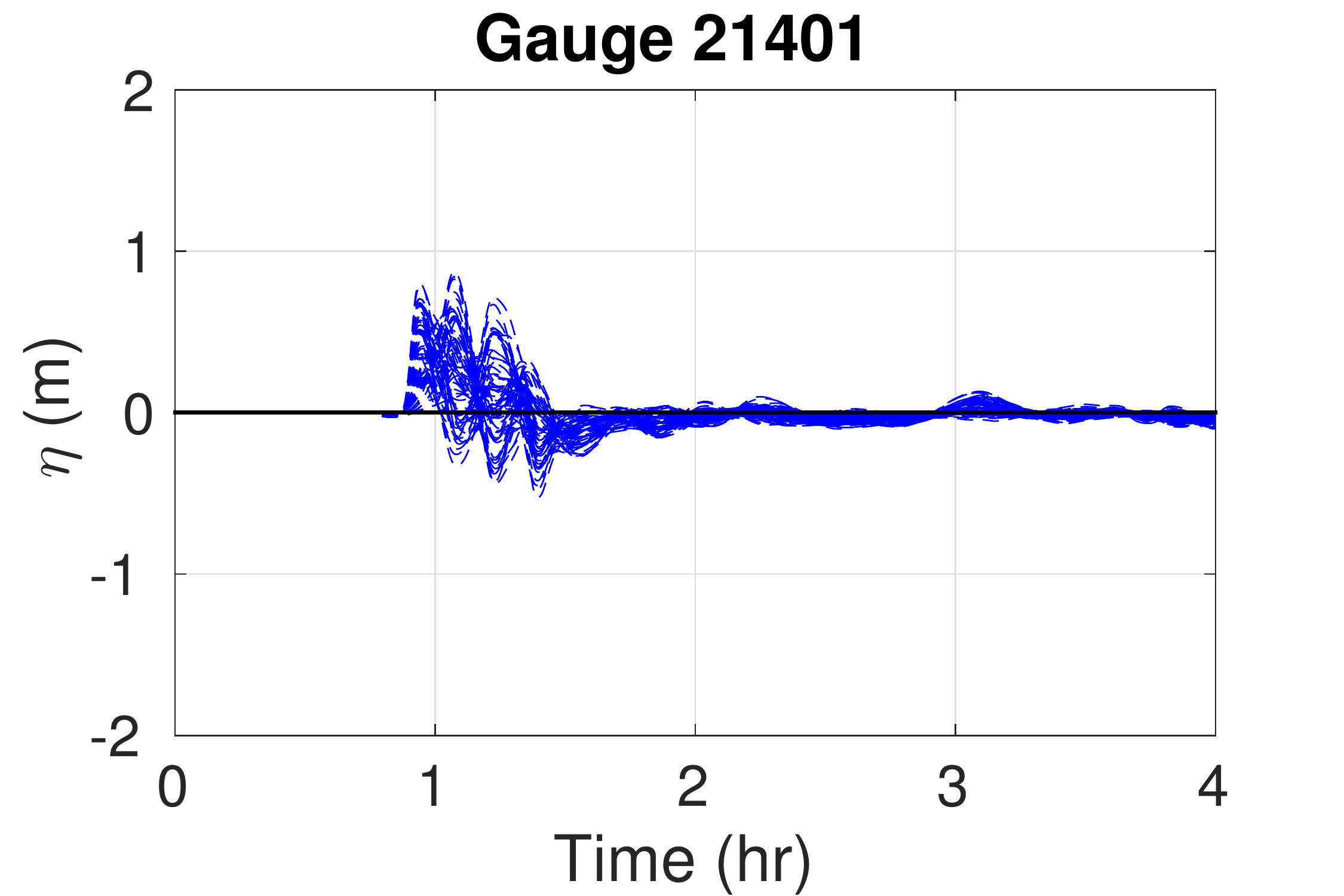} &
\includegraphics[width=0.475\textwidth]{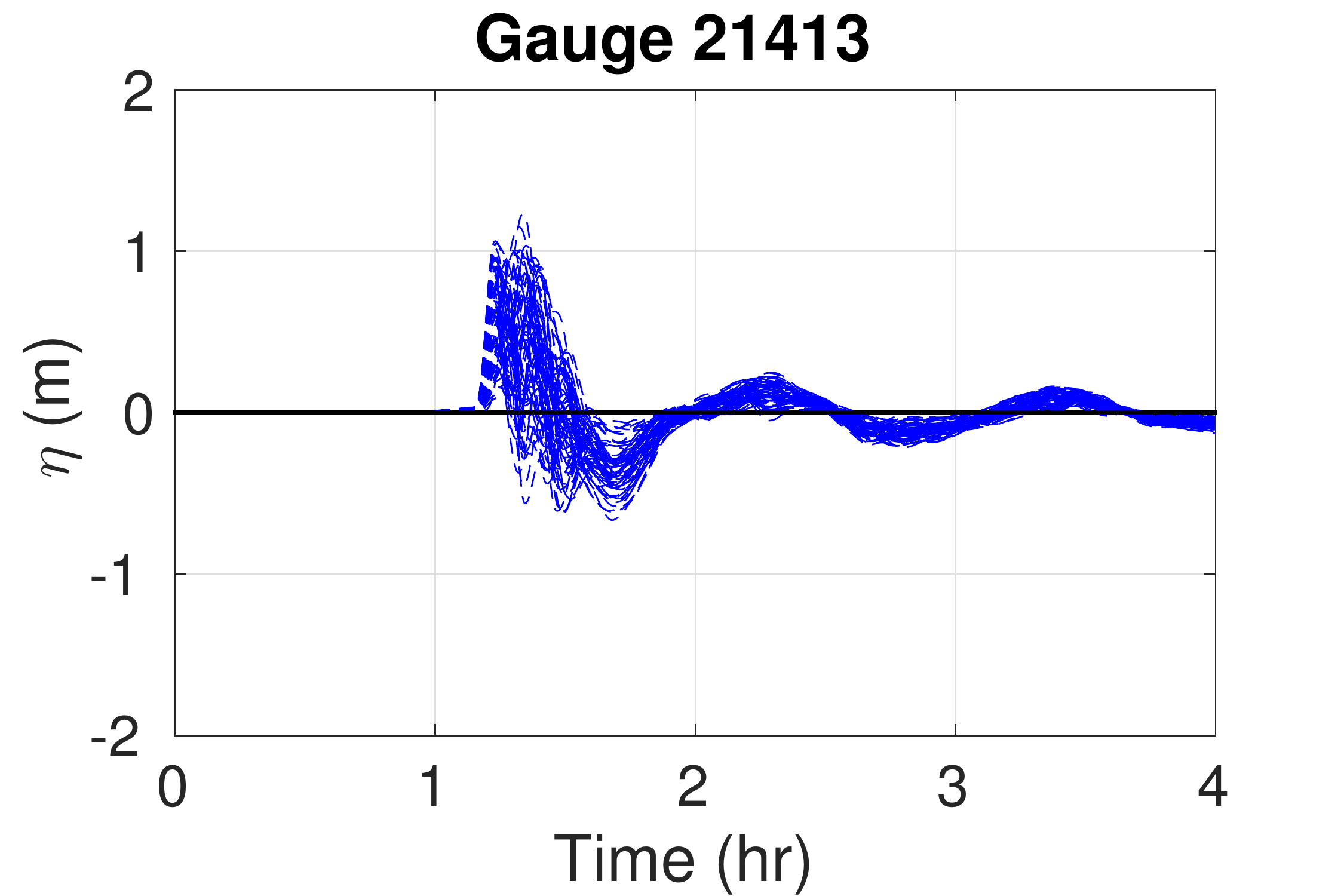} \\
\includegraphics[width=0.475\textwidth]{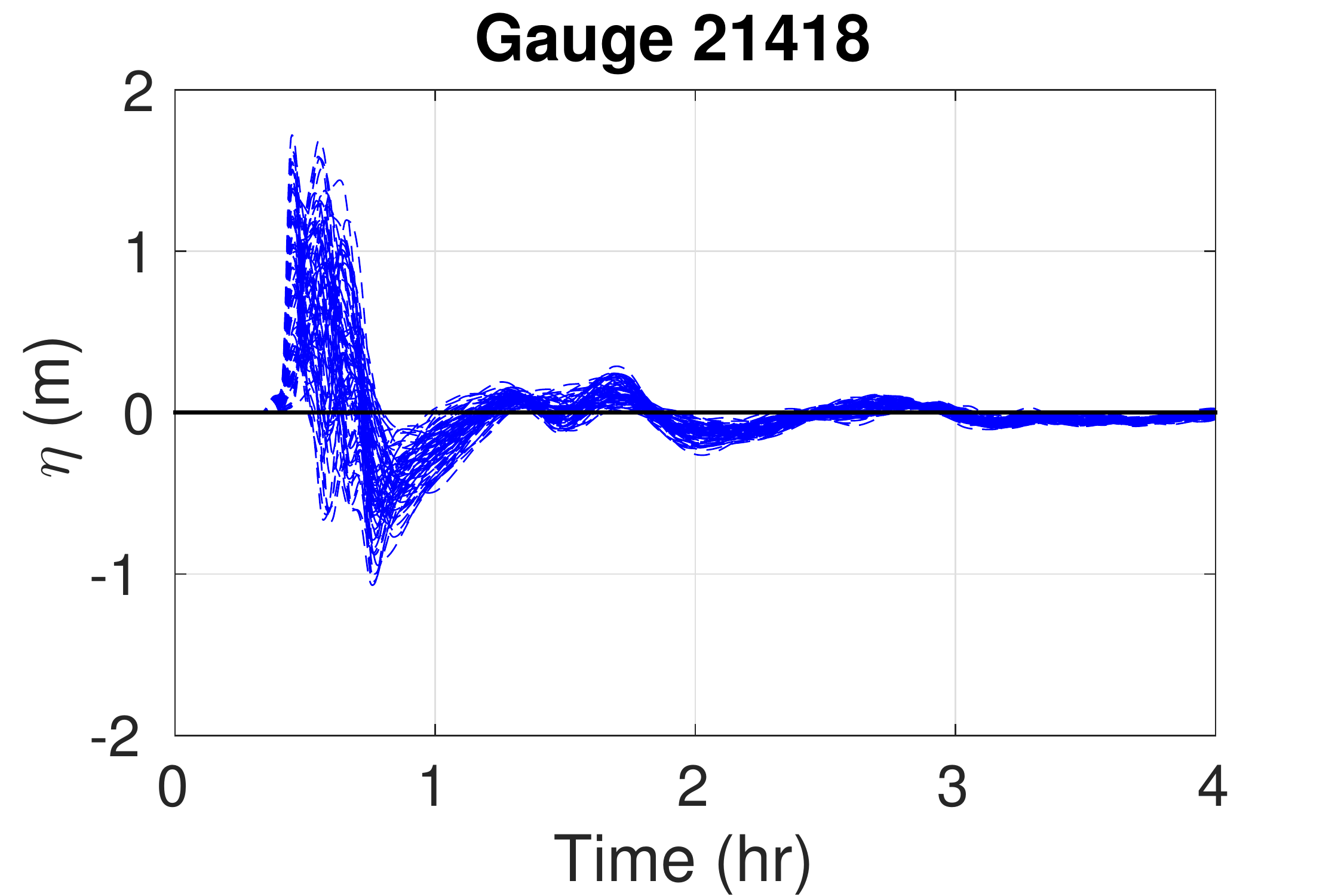} &
\includegraphics[width=0.475\textwidth]{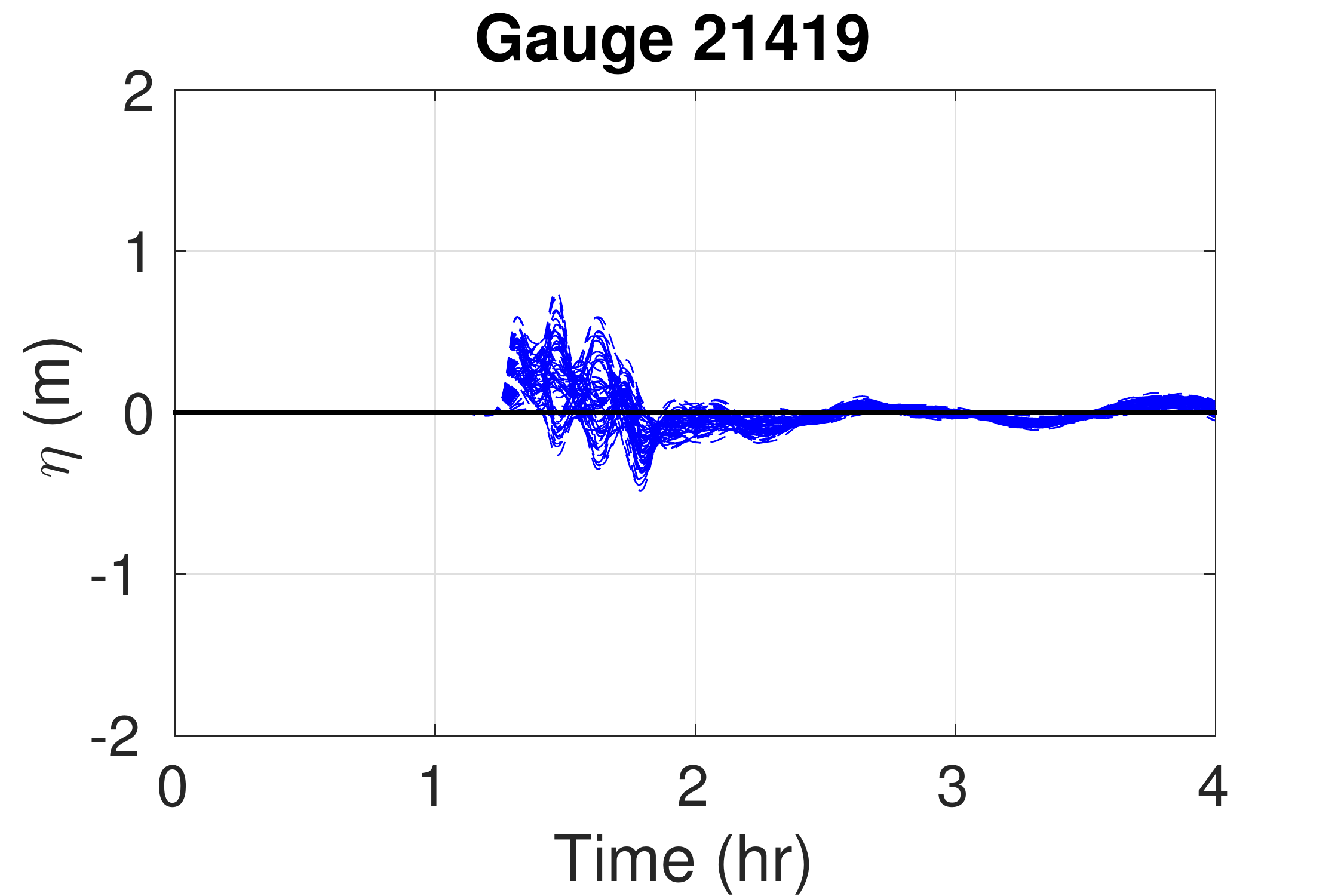} 
\end{tabular}
\caption{729 realizations of the evolution of the \geoclaw simulated water surface elevation 
corresponding to the LHS sample at the different gauges.}
\label{fig:lhs_sample}
\end{figure}

The construction of the PC surrogate for the water surface elevation using non-intrusive methods
requires an ensemble of forward model runs. The shape (distribution) of the ensemble and number of 
members (model runs) is dictated by the particular method employed. In this work, we employ two different methods that require two different ensembles as follows: 
\begin{enumerate}
\item NISP requires a quadrature to compute the PC coefficients~\cite{sraj:2014}. Here, we adopted a sparse nested
Smolyak quadrature~\citep{Petras:2000,Gerstner:2003,Smolyak:1963}. In particular,
Smolyak level 5 grid rule was used requiring a total number of $Q = 1889$
quadrature nodes for the case of $m=6$ uncertain parameters to accurately
approximate PC expansion of order $p = 5$. A two-dimensional projection of the quadrature grid
is shown in Figure~\ref{fig:sample}~(Left) on the $\xxi_1 - \xxi_2$ plane.
The evolution of the water surface elevation predicted by \geoclaw at these nodes
is shown in Figure~\ref{fig:quadrature} at the four different gauges.
\vspace{6mm} 
\item BPDN accommodates both regular and random sampling to determine the PC coefficients. 
Here, we used a Latin-Hyper-Cube (LHS) sample consisting
of 729 \geoclaw realizations whose nodes are shown in
Figure~\ref{fig:sample}~(Right) when projected on the $\xxi_1 - \xxi_2$ plane.
The evolution of the water surface elevation predicted by \geoclaw at these nodes
is shown in Figure~\ref{fig:lhs_sample} at the four different gauges.
\end{enumerate}

\begin{figure}[ht]
\centering
\begin{tabular}{clclclc}        
\includegraphics[width=0.475\textwidth]{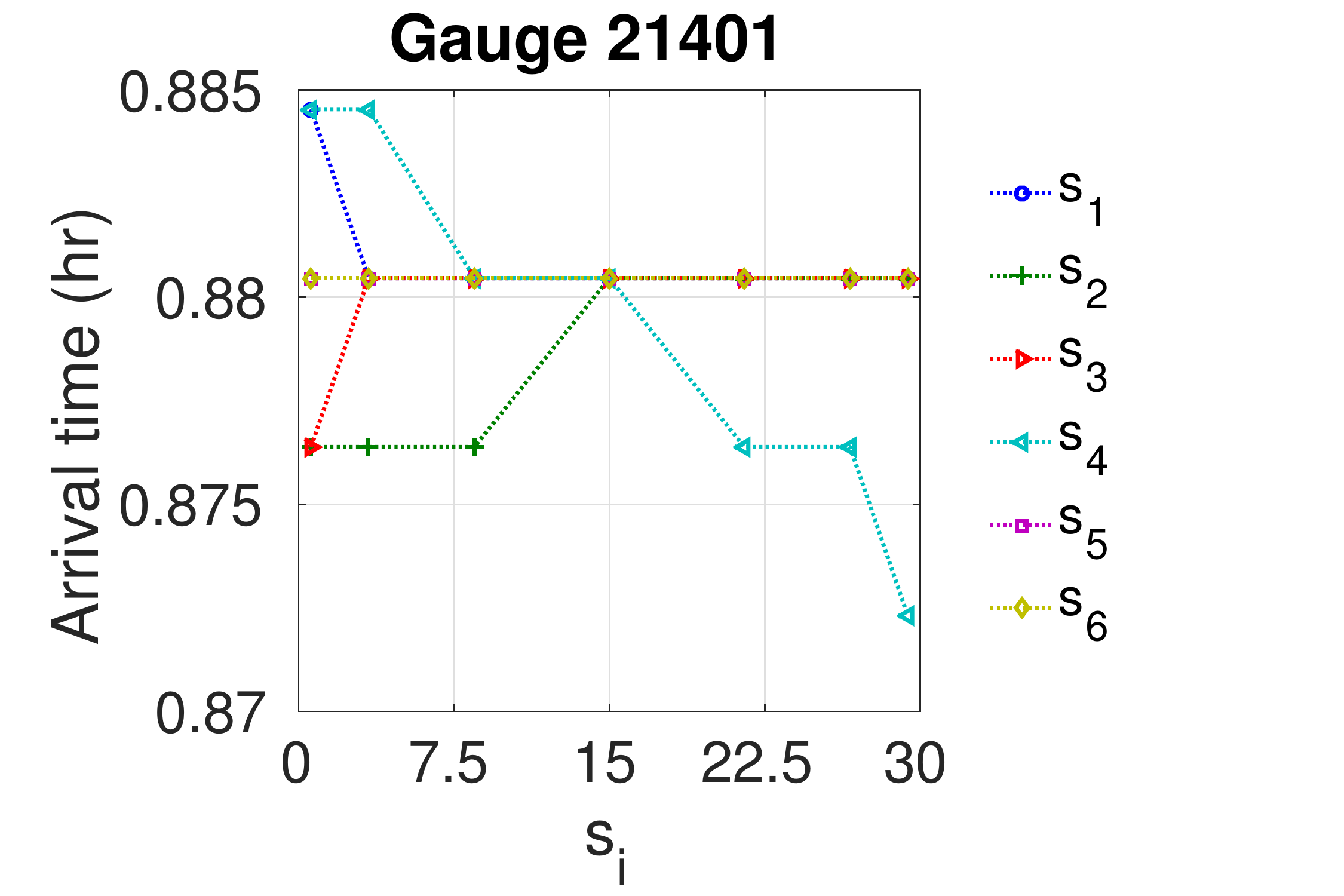} &
\includegraphics[width=0.475\textwidth]{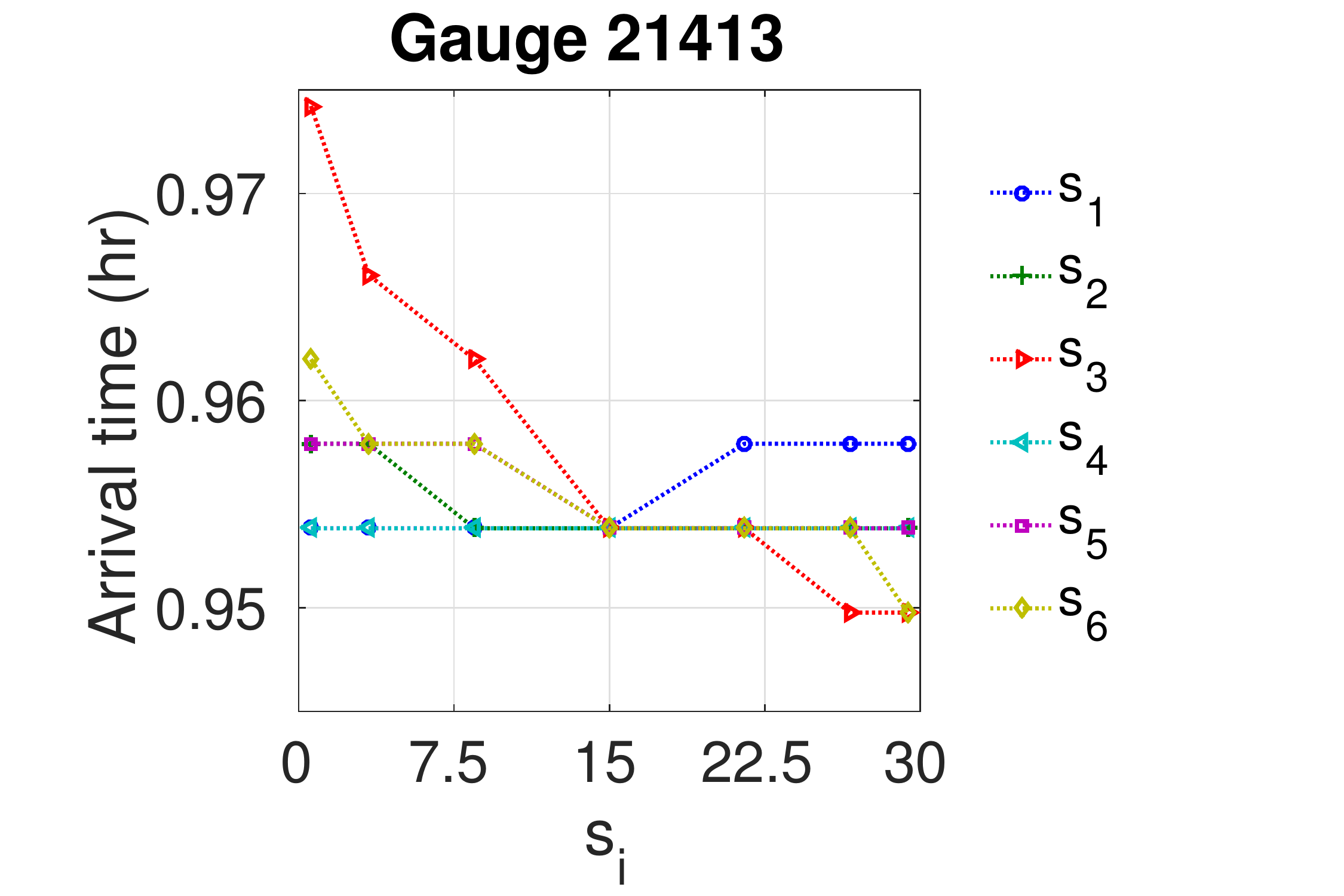} \\
\includegraphics[width=0.475\textwidth]{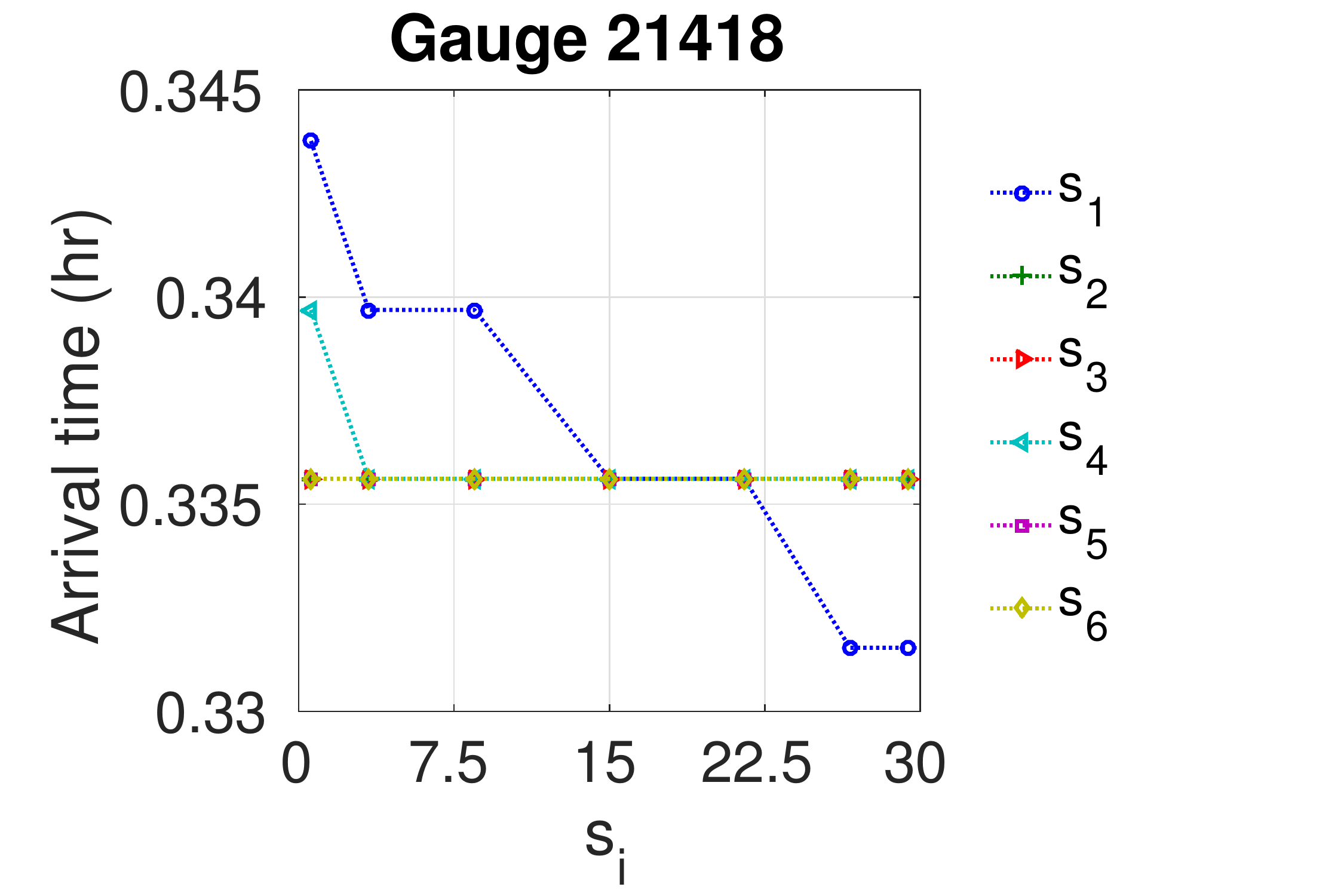} &
\includegraphics[width=0.475\textwidth]{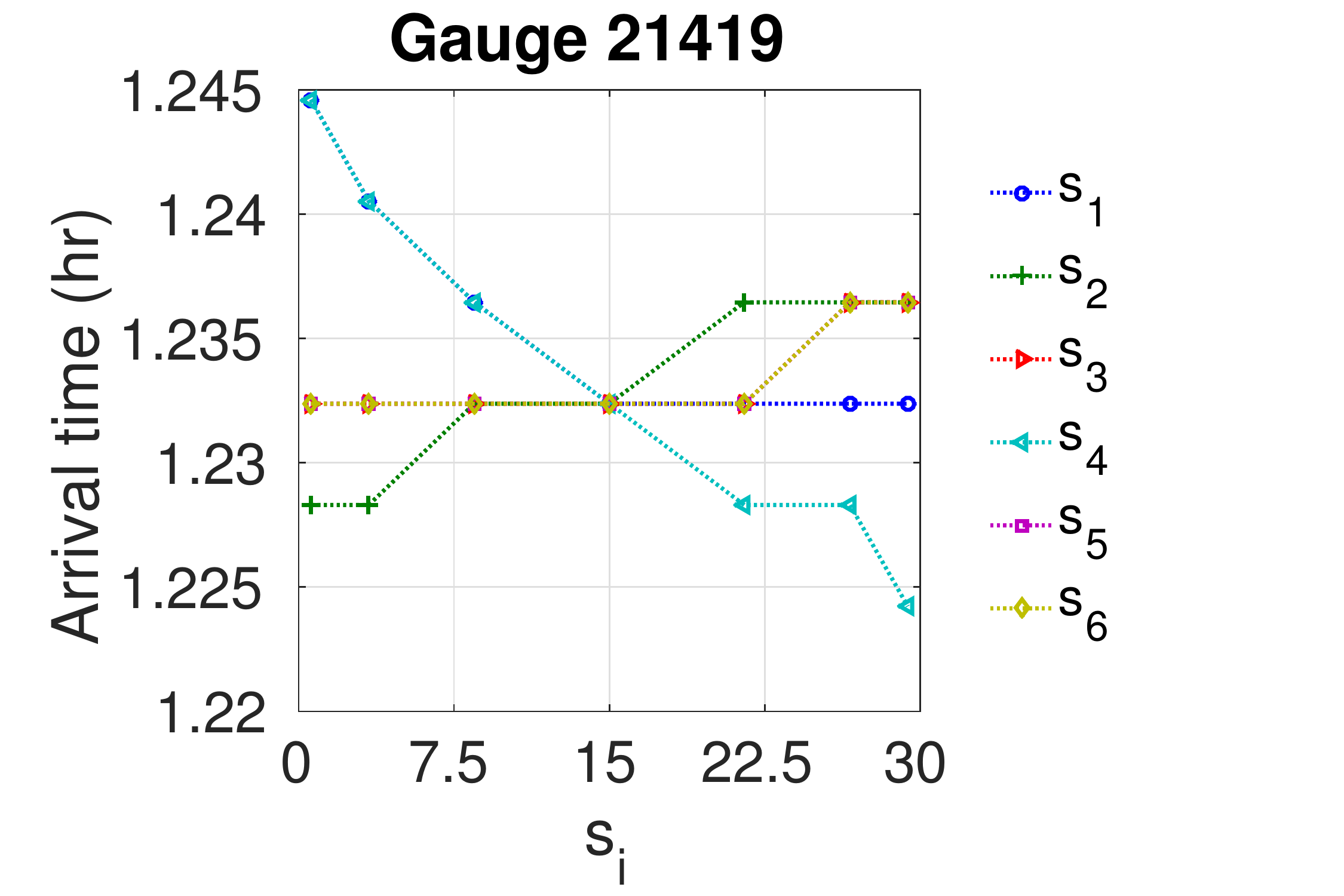} \\
\end{tabular}
\caption{Arrival time function of slip values for the quadrature sample at different gauges.} 
\label{fig:quad_pt}
\end{figure}

\begin{figure}[ht]
\centering
\begin{tabular}{clclclc}        
\includegraphics[width=0.475\textwidth]{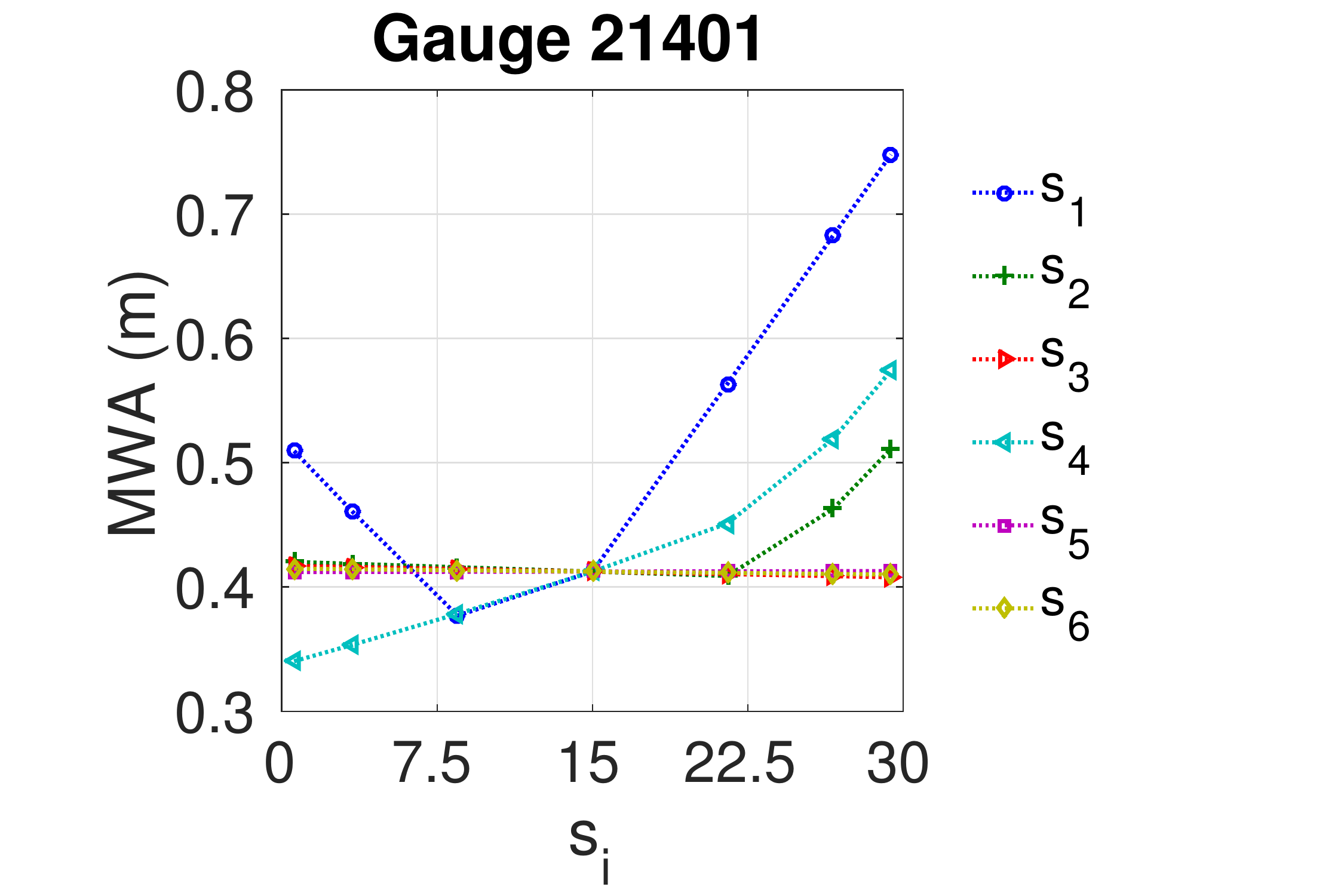} &
\includegraphics[width=0.475\textwidth]{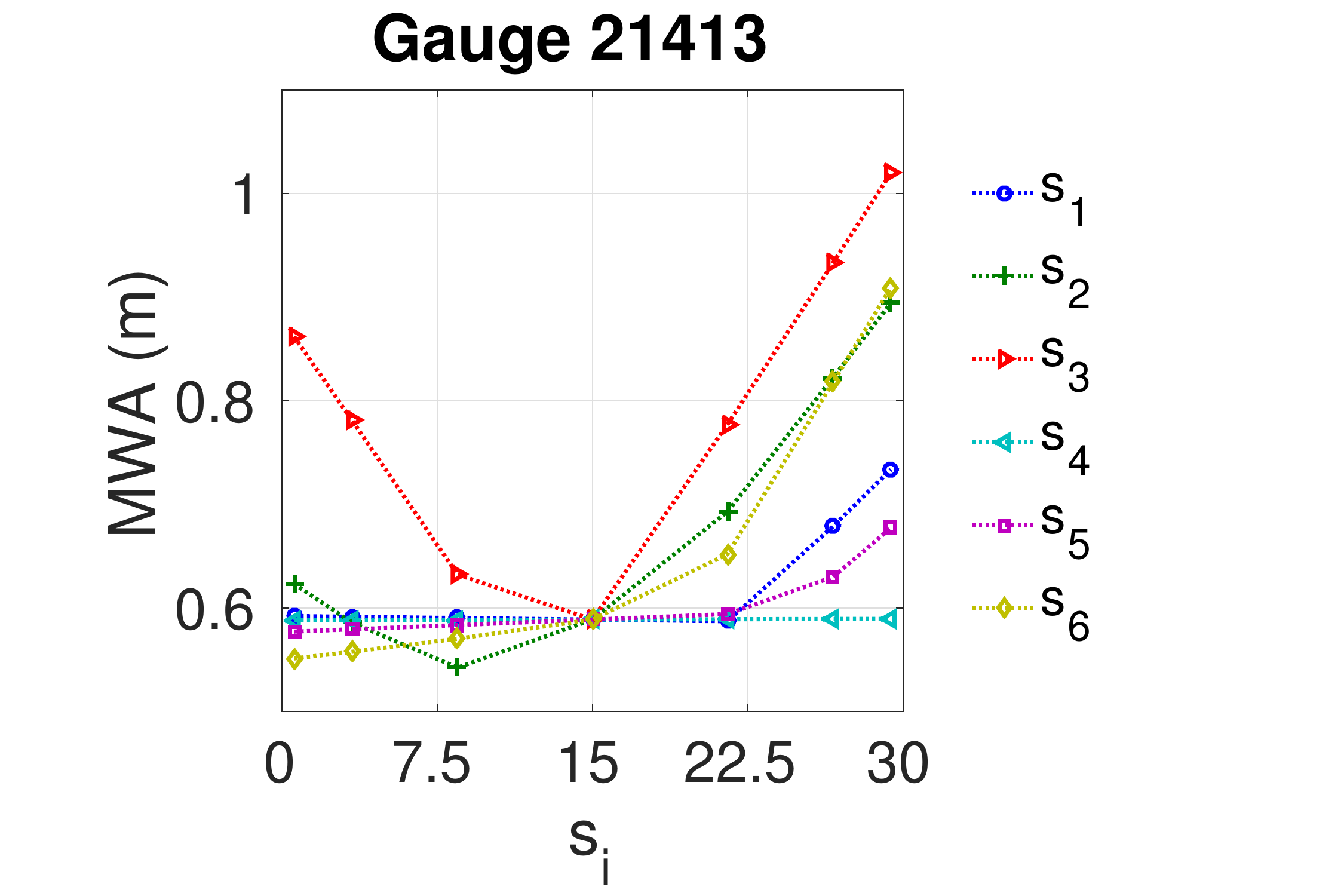} \\
\includegraphics[width=0.475\textwidth]{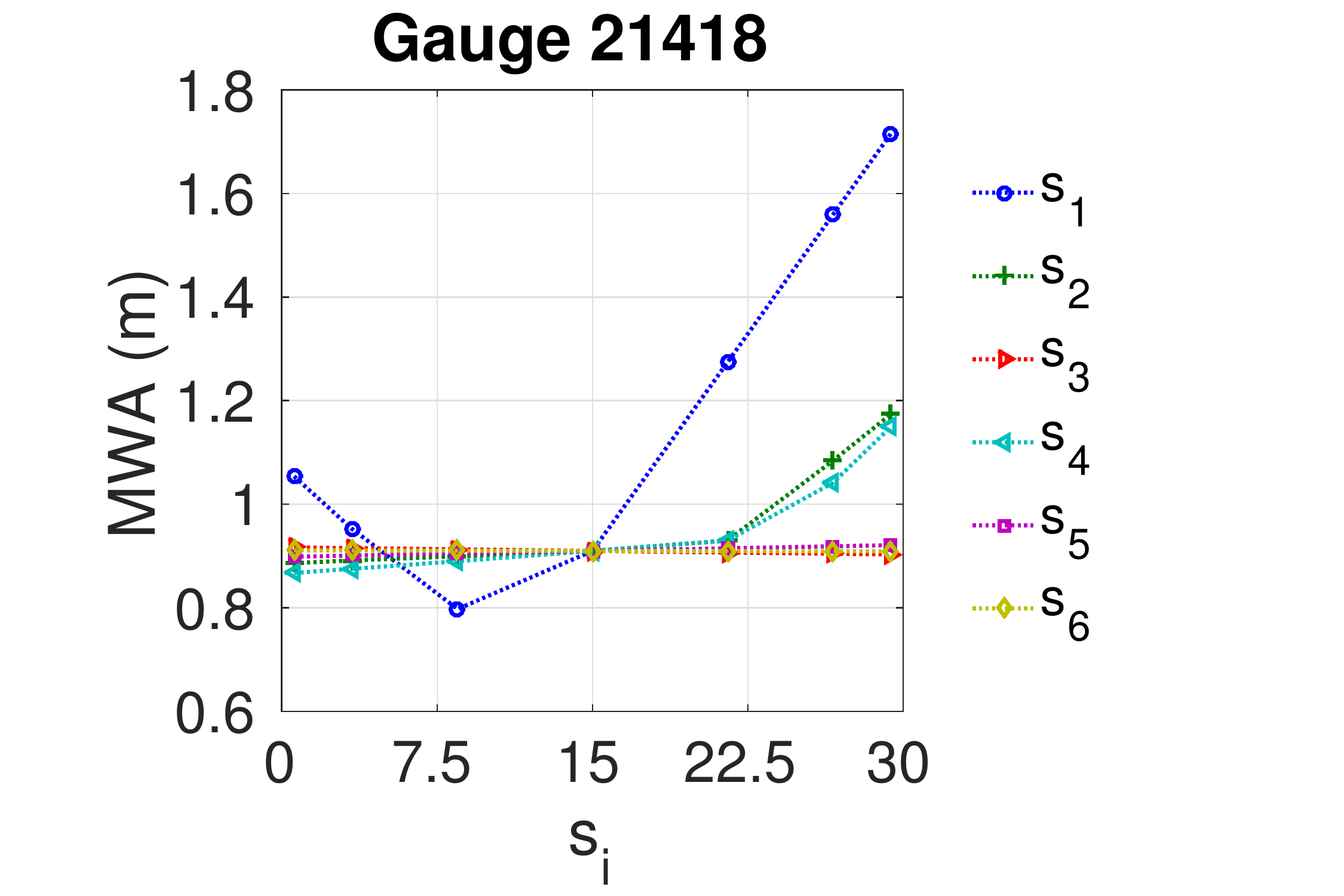} &
\includegraphics[width=0.475\textwidth]{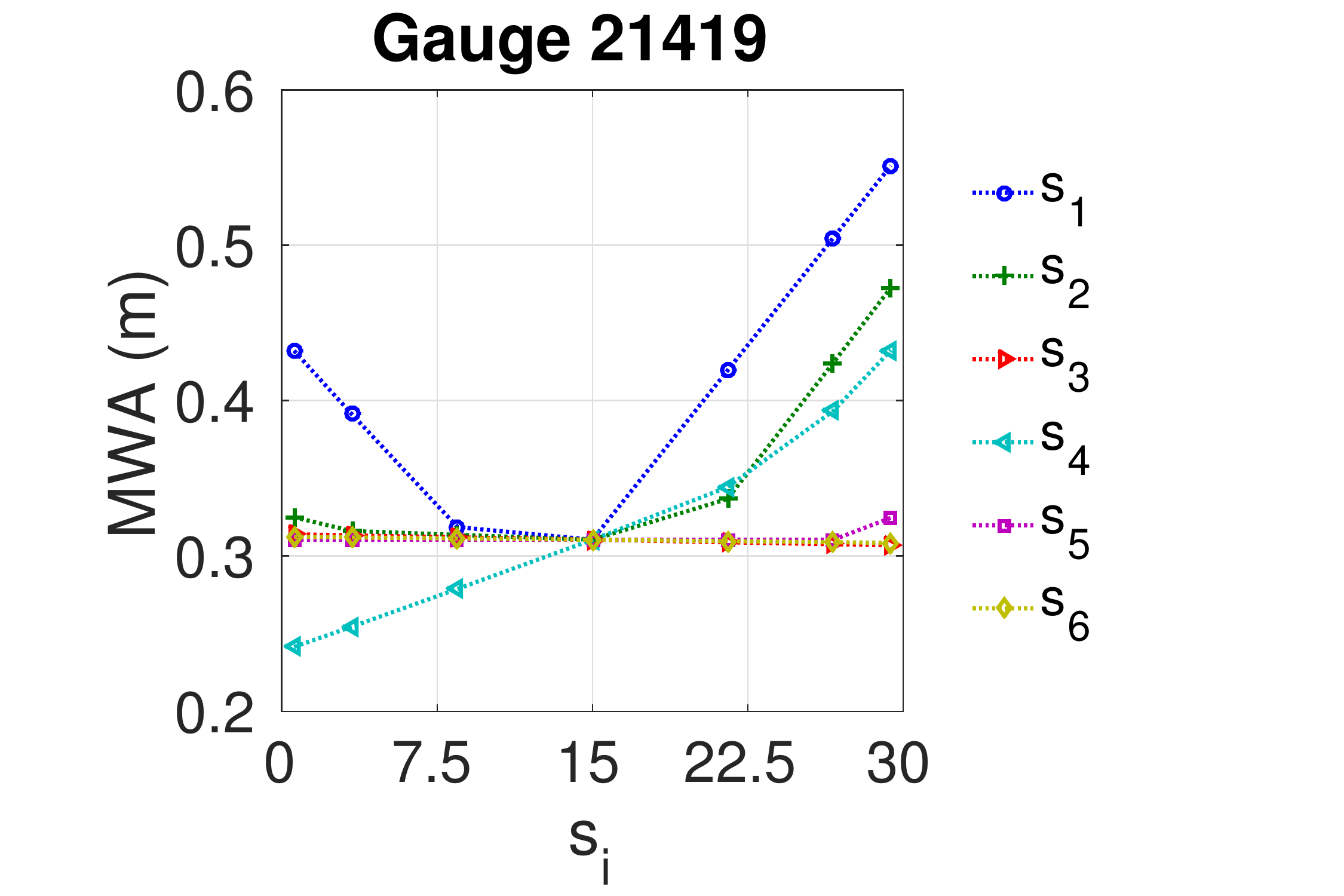} \\
\end{tabular}
\caption{Maximum Wave Amplitude function of slip values for the quadrature sample at different gauges.} 
\label{fig:quad_mwa}
\end{figure}

In both sets of realizations, we notice that the variability in water surface elevation
is significant at all gauges. This variability in the prediction of water surface
elevation persists till the end of the simulations for all gauges as well.
It is noticed that this variability is present in the arrival time in addition to the Maximum Wave Amplitude (MWA).
To confirm, we estimate the arrival time and MWA of the 1889 realizations corresponding to the quadrature sample
and plot them as functions of the different slip values $s_i$ on the subfaults (other values are set to $s_j = 15$) in Figure~\ref{fig:quad_pt} and Figure~\ref{fig:quad_mwa}, respectively. We clearly notice the significant variation of arrival time with the slip values and similarly  for the MWA. These variations are expected to be challenging when computing the PC coefficients as it might require
a high order PC expansion~\cite{Alen:2011}.

Finally, we note that the  average arrival time and average MWA are both consistent with the distance from the gauge to the epicenter of the earthquake (located approximately 72 kilometer east of Japan). For instance,
gauge 21418 is the closest to the source as shown in Figure~\ref{fig:setup} with the shortest arrival time and largest MWA, while 
on the other hand, gauge 21419 is the farthest from  the source with the longest arrival time and smallest MWA.

\label{sec:pccs}
\subsubsection{Non-Intrusive Spectral Projection}
\label{sec:pcnisp}

\begin{figure}[ht]
\centering
\begin{tabular}{clc}        
\includegraphics[width=0.475\textwidth]{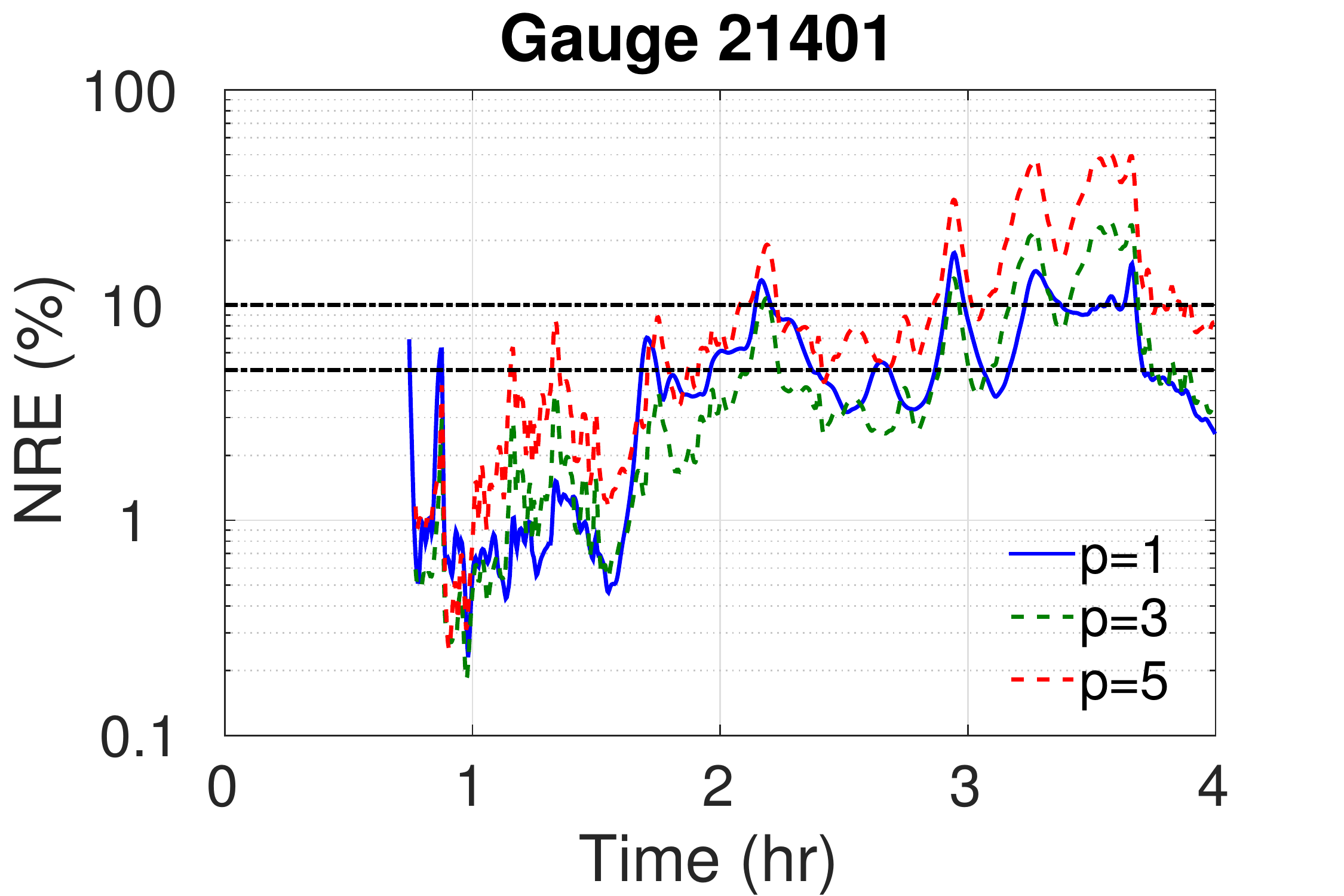} &
\includegraphics[width=0.475\textwidth]{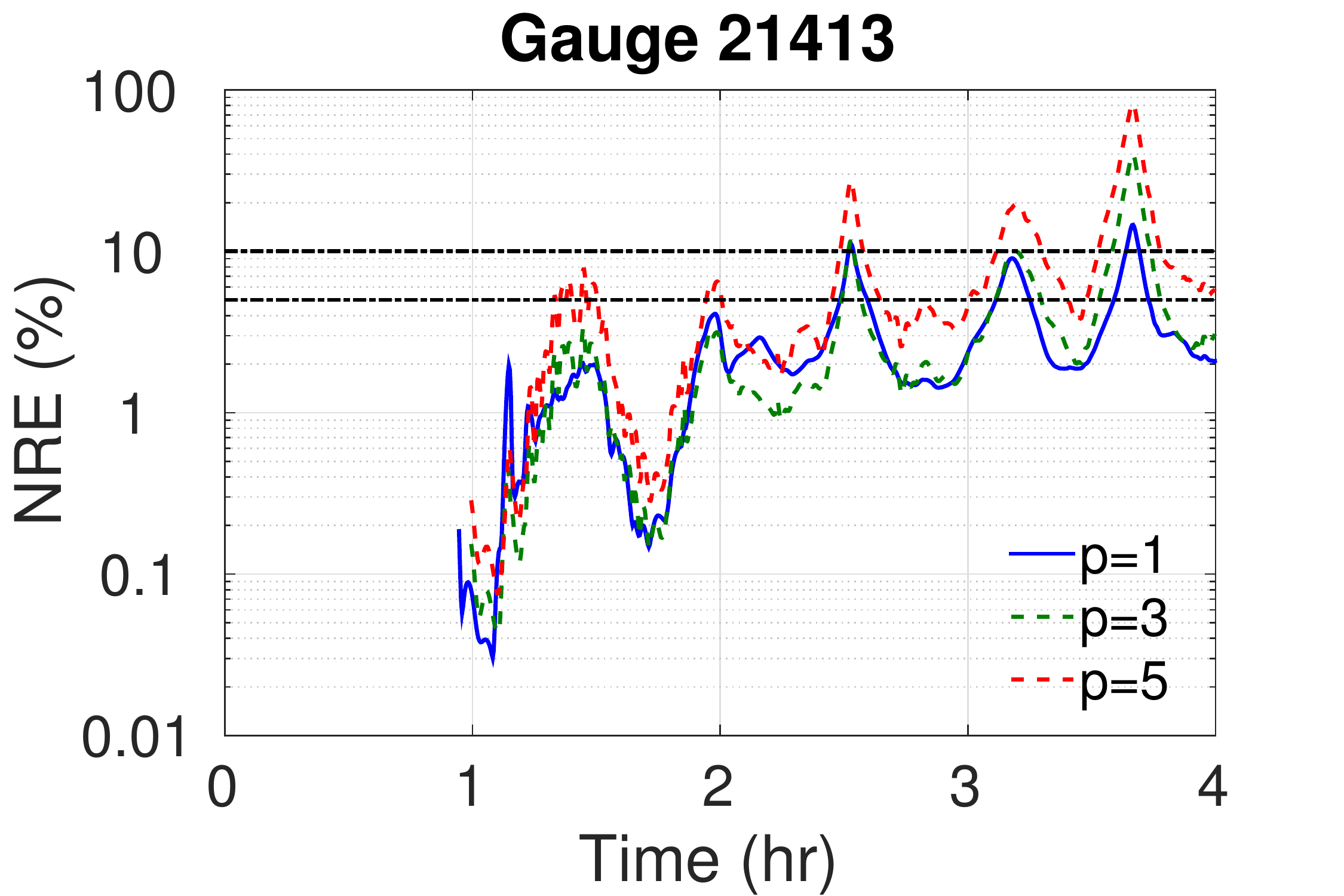} \\
\includegraphics[width=0.475\textwidth]{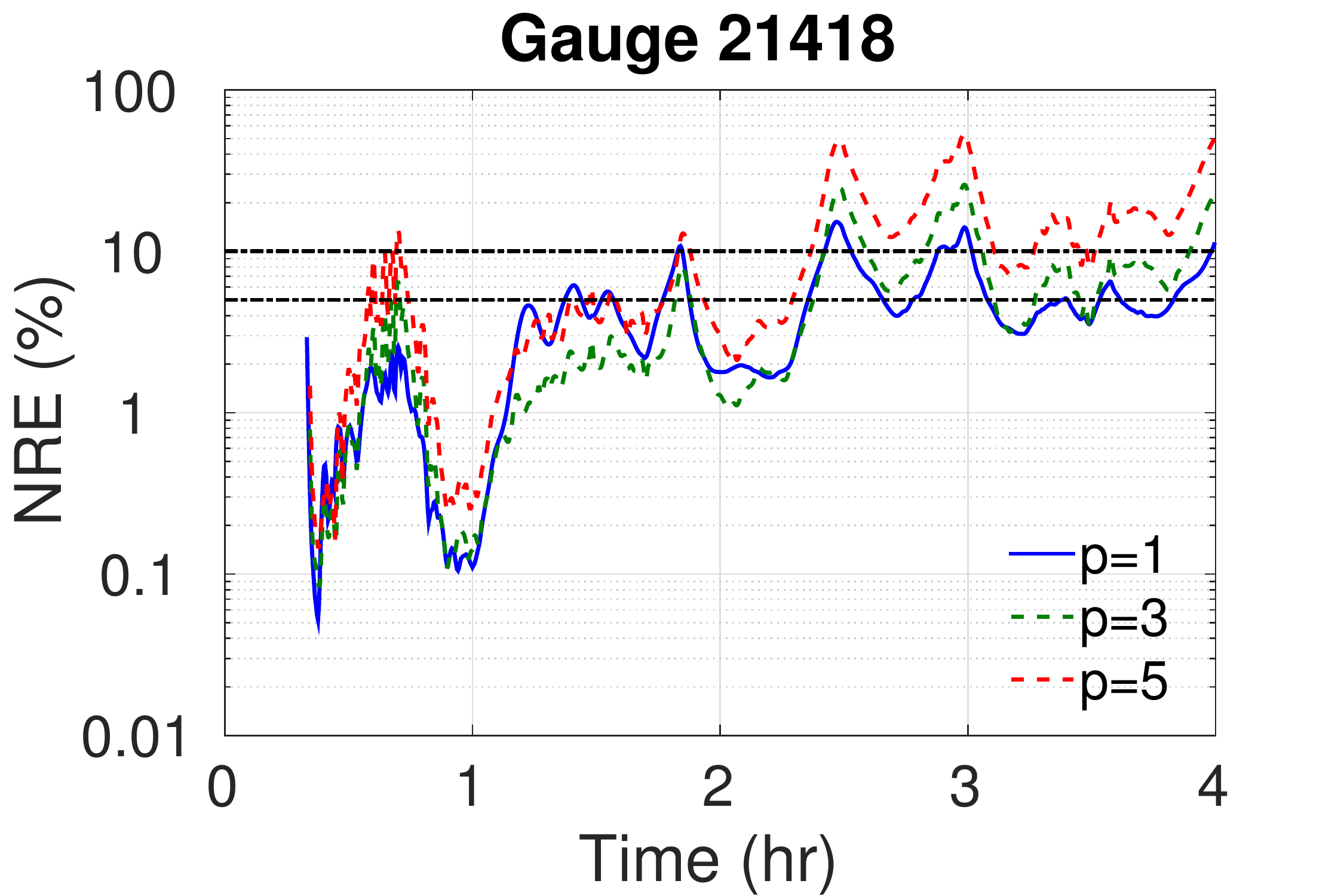} &
\includegraphics[width=0.475\textwidth]{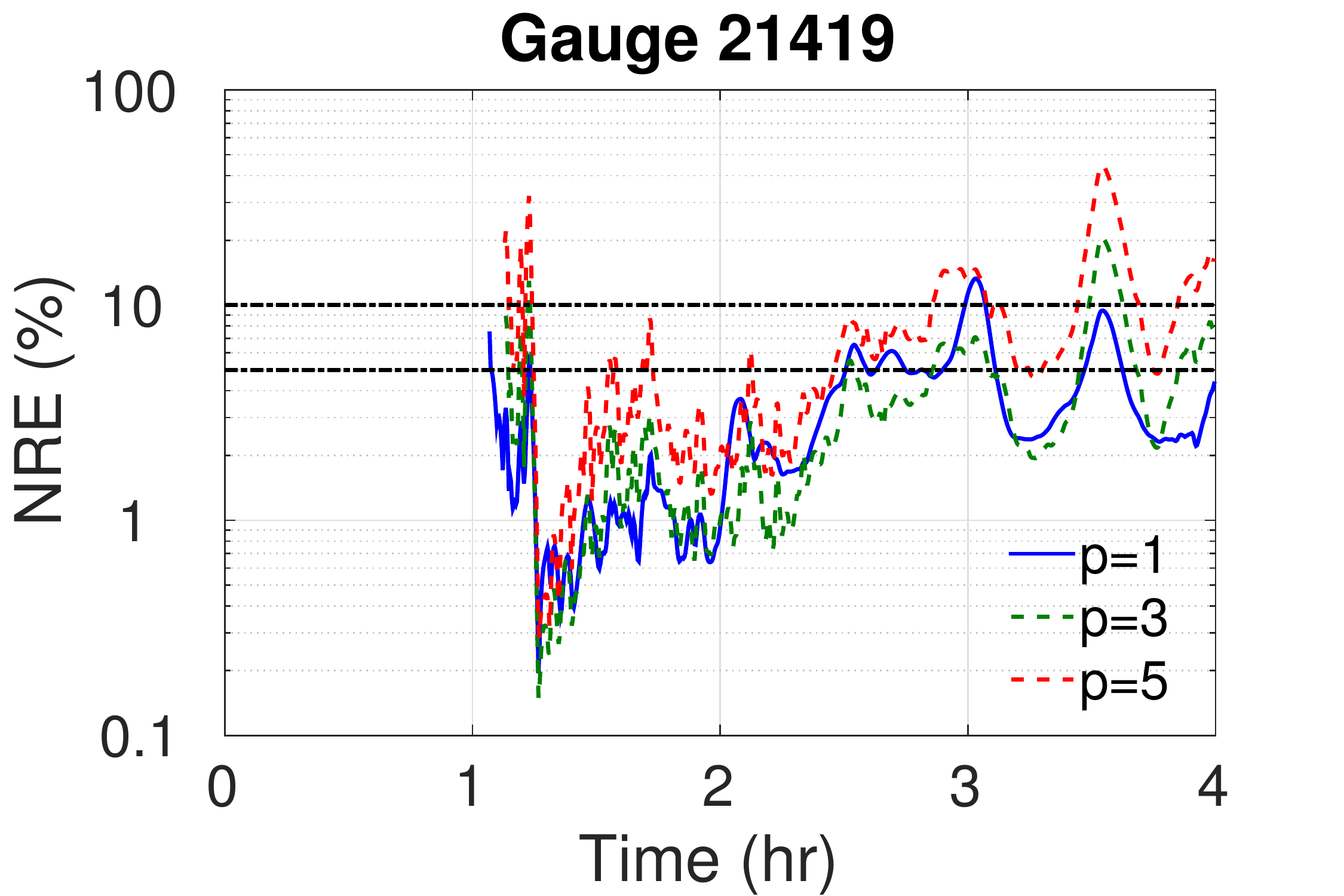} 
\end{tabular}
\caption{Evolution of NRE between the LHS sample and 
the corresponding NISP-estimated PC surrogate at the different gauges.}

\label{fig:error_lhs_nisp}
\end{figure}

The PC expansion coefficients are first computed using the output of the 1889
quadrature ensemble.  The constructed PC surrogate is validated using the normalized
relative error (NRE) that measures the accuracy of predicted values by the PC surrogate
using an independent set of \geoclaw simulations as follows:

\begin{equation} 
   NRE = \frac{\displaystyle
         \left(\sum_{q=1}^S \left|G(\xxi_s) - \sum_{k = 0}^{R}
g_k\psi_k(\xxi_q)\right|^2
         \right)^{1/2}}
        {\displaystyle
          \left(\sum_{q=1}^S \left|G(\xxi_s)\right|^2\right)^{1/2} 
          },
\label{eq:error}
\end{equation}
where $G(\xxi_s)$ is the $QoI$ corresponding to the LHS sample
that were not used in the PC construction process.
The evolution of NRE is shown in Figure~\ref{fig:error_lhs_nisp} for different PC orders 
as indicated. The horizontal dotted lines are guides to the eye indicating the $5\%$ and $10\%$ errors.
The calculated NRE appears to be larger than $10\%$ for PC order $p=1$ at certain times
that amplifies with increasing PC order. This indicates convergence issues in the PC that leads to inaccuracies in
the representation of the $QoI$. This is noticed for all the gauges.

We conclude that the construction of a converging PC expansion using the NISP method was not successful which promoted us to use an to alternative method. The large errors  can be attributed to the large variation in the arrival times
and the MWA that is not tolerated by the NISP method.
One option to overcome this issue is preconditioning the $QoI$.
This idea was proposed in~\cite{Alexanderian2011a,Alexanderian2012}, where appropriate transformations of 
the original time-dependent $QoI$ into a new one having a tight sparse PC expansion, thus
requiring less effort to be projected. Instead we resort here to a recent compressed technique as explained above.

\subsubsection{Basis-Pursuit DeNoising}
\label{sec:pcbpdn}
\begin{figure}[ht]
\centering
\begin{tabular}{clc}        
\includegraphics[width=0.475\textwidth]{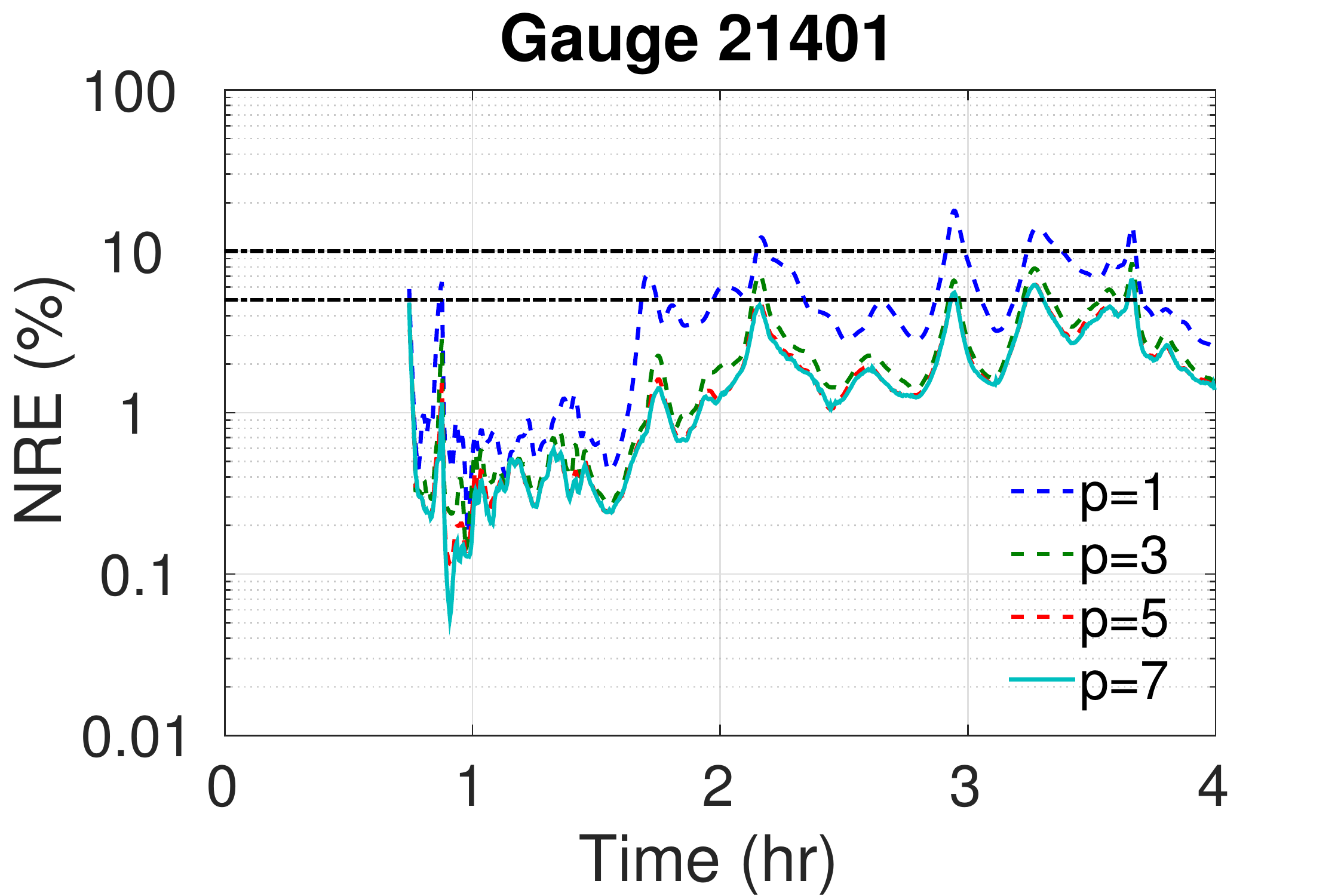} &
\includegraphics[width=0.475\textwidth]{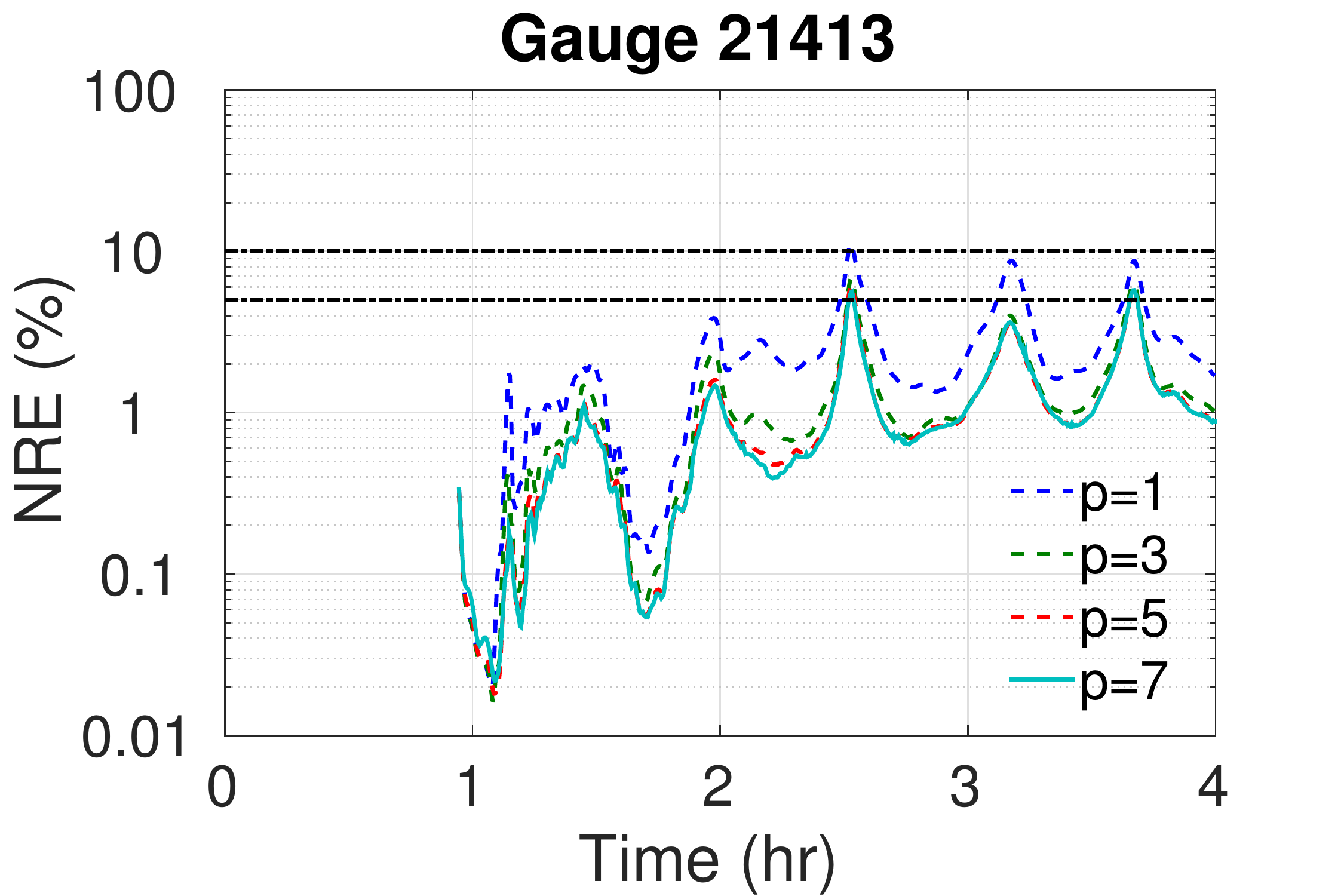} \\
\includegraphics[width=0.475\textwidth]{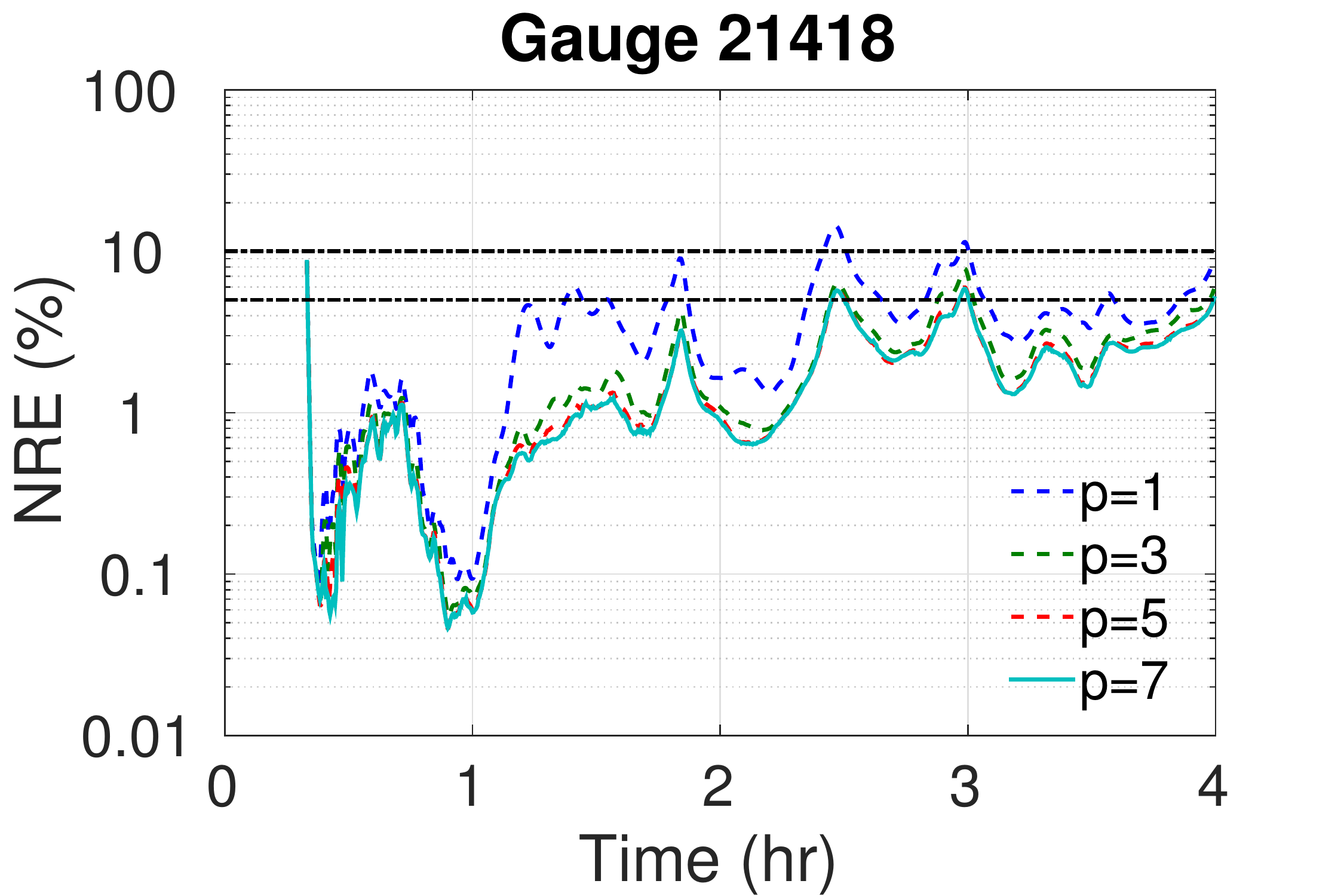} &
\includegraphics[width=0.475\textwidth]{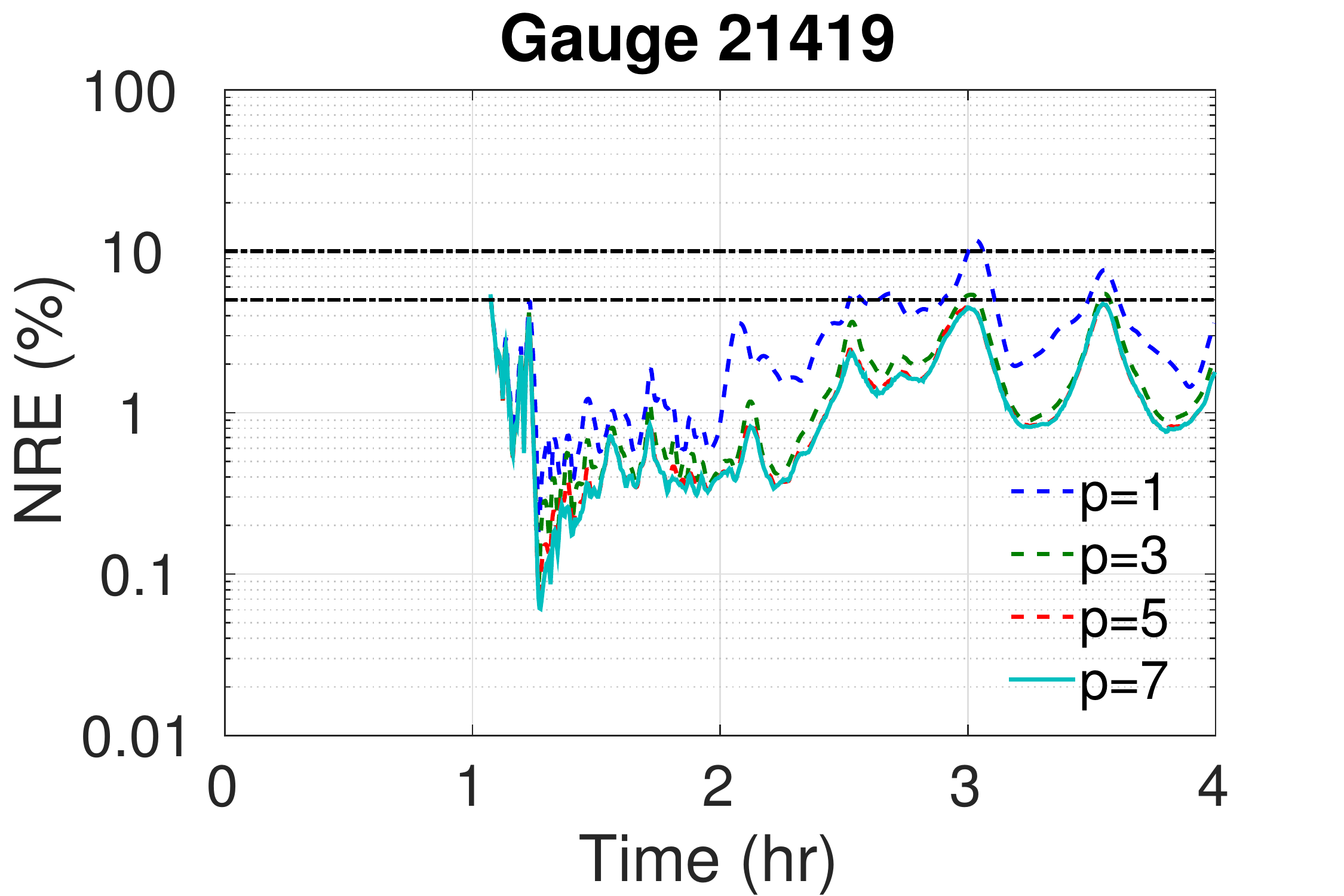} 
\end{tabular}
\caption{Evolution of NRE between the quadrature sample and 
the corresponding BPDN-estimated PC surrogate at the different gauges.}
\label{fig:error_quad_bpdn}
\end{figure}

We next applied the BPDN method to estimate the PC
coefficients~\cite{Doostan:2014} using the LHS sample consisting of 729 \geoclaw
realizations. We again quantified the agreement between the PC surrogate and the \geoclaw
realizations where we now calculate the NRE using the Smolyak quadrature sample 
(not used in the PC coefficients estimation). 
The evolution of error shown in Figure~\ref{fig:error_quad_bpdn} indicate a better agreement (compared to NISP)
whereas the maximum error was found to decrease as the PC order is increased. The average error is less than $5 \%$, indicating that BPDN is successful in constructing a surrogate that yields accurate $QoI$ predictions.

\begin{figure}[ht]
\begin{tabular}{clc}
        
\includegraphics[width=0.475\textwidth]{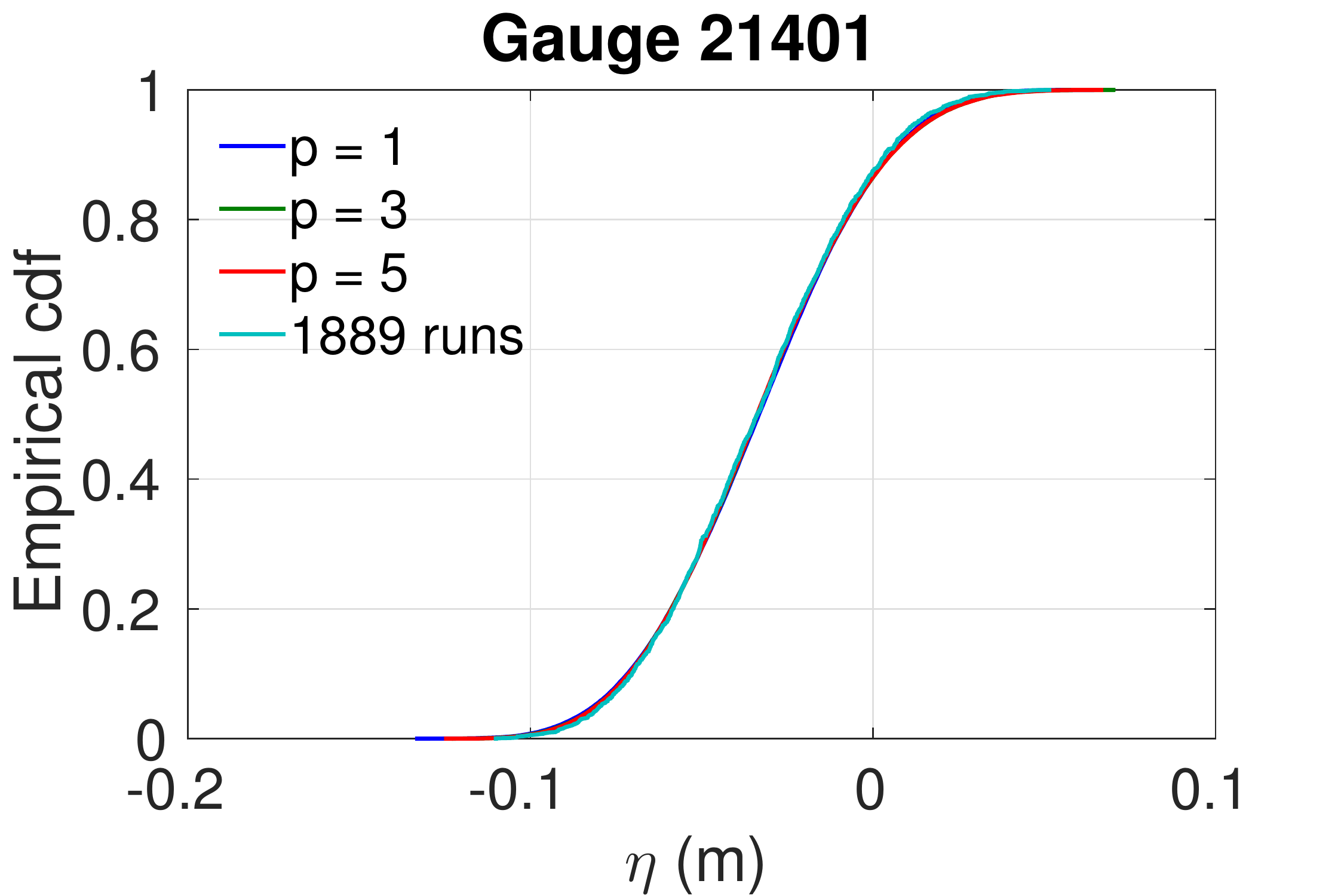} &
\includegraphics[width=0.475\textwidth]{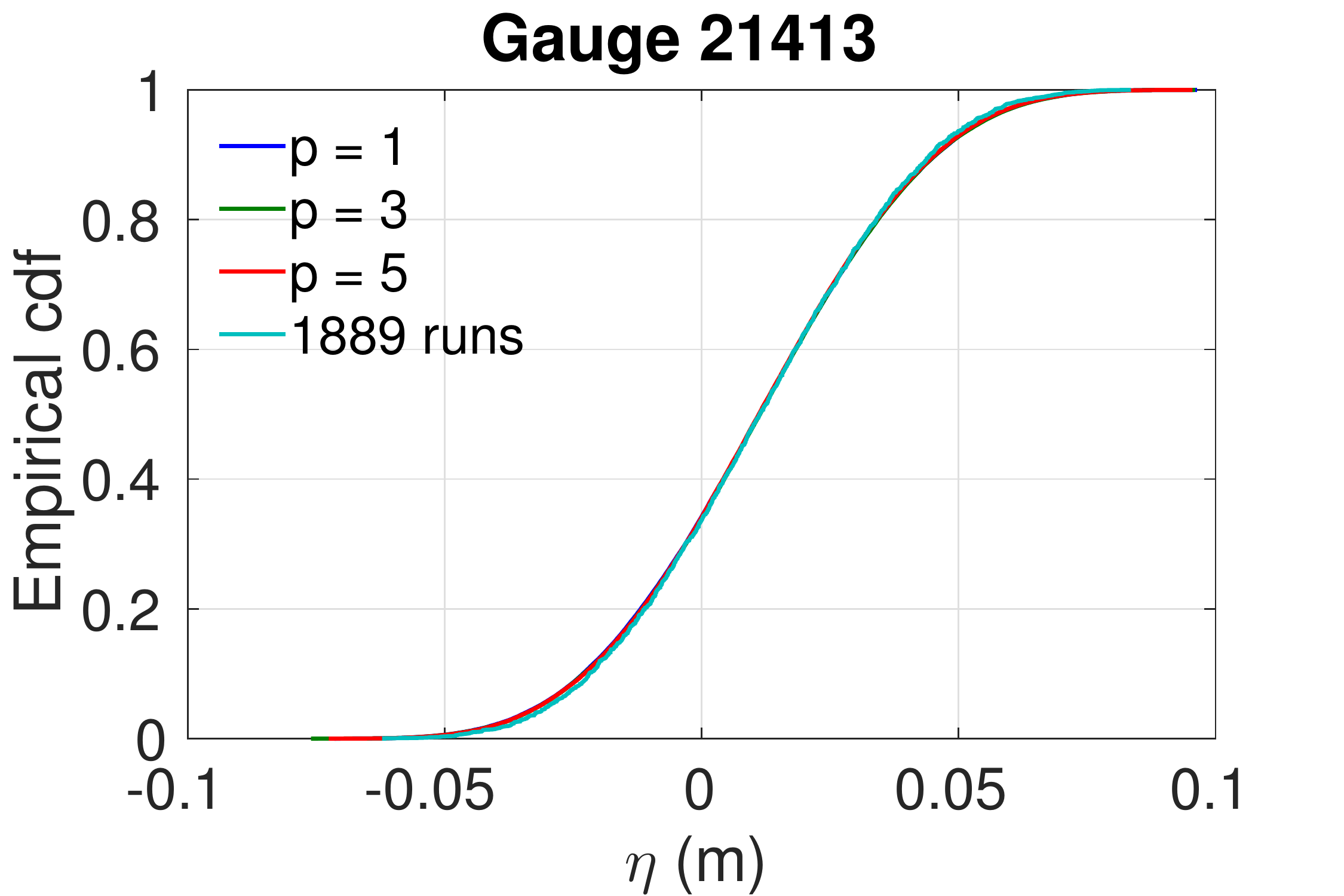} \\
\includegraphics[width=0.475\textwidth]{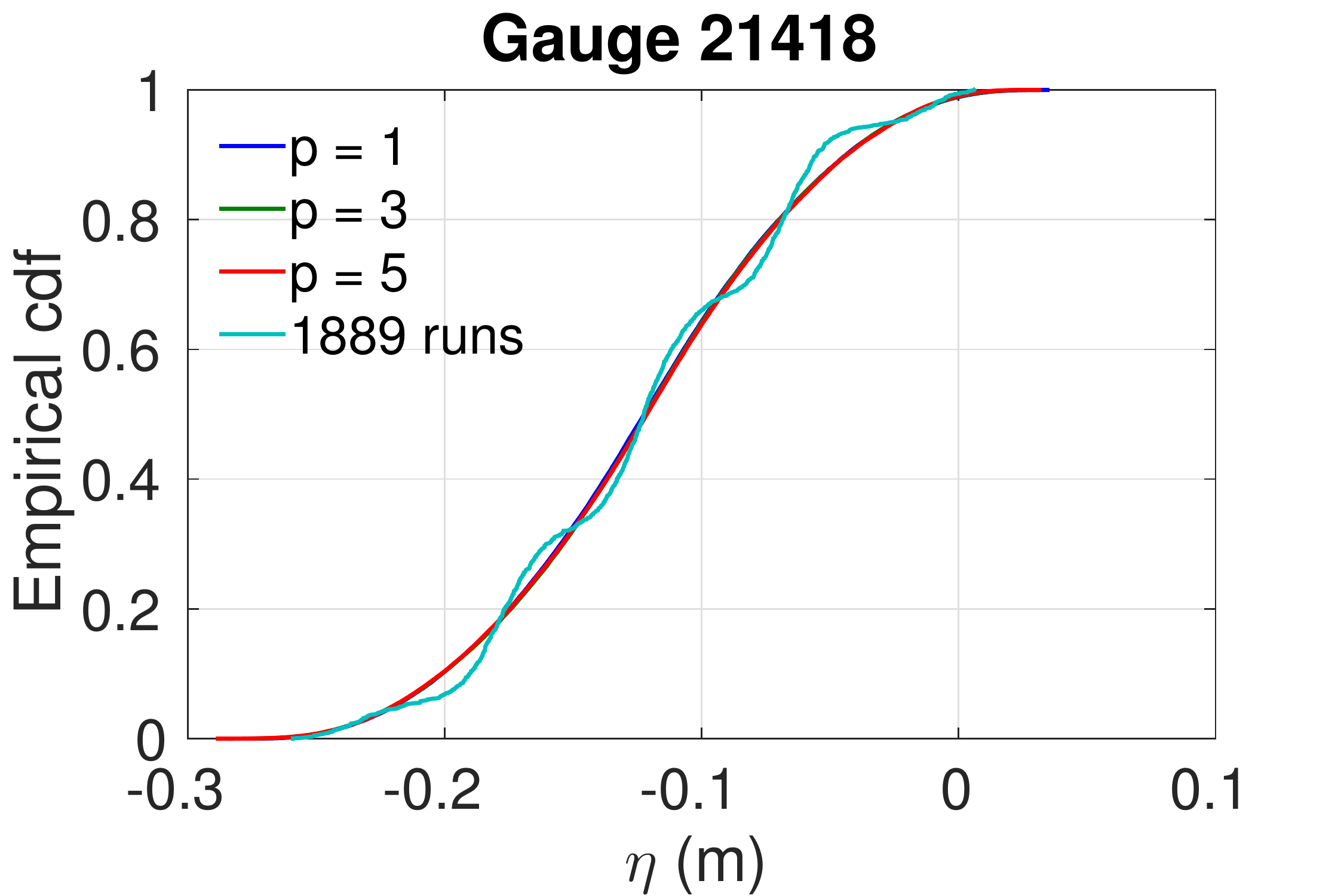} &
\includegraphics[width=0.475\textwidth]{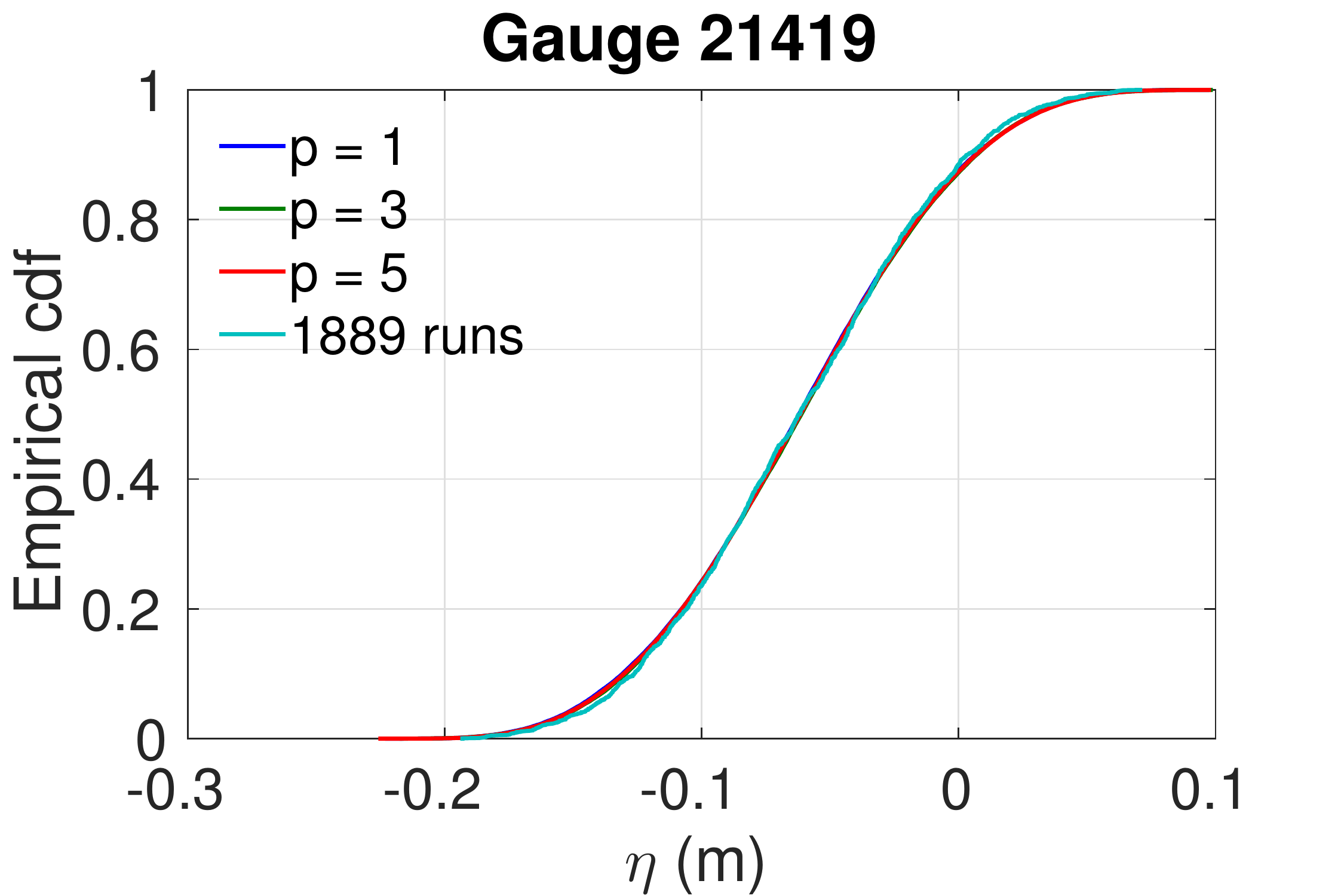}
\end{tabular}
\caption{Empirical CDF of PC-estimated water surface elevation at the different gauges
compared to that of the 1889 runs.}
\label{fig:cdfs}
\end{figure}

We also computed the empirical CDF of water surface elevation 
at the different gauges using samples from the PC surrogate for different orders.
We also computed the CDF using the 1889 \geoclaw model runs and compare them
to the PC-estimated ones and plot them in Figure~\ref{fig:cdfs}. 
These different panel show that the CDFs obtained using higher order bases agree with each other
and with the CDF obtained from the full model runs directly.

In conclusion, these tests provide confidence that the PC expansion is a faithful 
model surrogate that can be used in both the forward and inverse problems. 

\subsection{Statistical Analysis}
\label{sec:sens}

The PC expansion created using an ensemble of \geoclaw simulations
simplifies the calculations of the statistical moments of model output $G$ as
the expectation and variance can be computed from the PC coefficients as follows:

\begin{equation}
 \mu_G = \int G \, \rho(\xxi) \, \mbox{d}\xxi \approx \left< G,\psi_0\right>_{\cal Q} = G_0,  
 \label{eq:mean}
\end{equation}

\begin{equation}
 \sigma^2_G = \int (G- \mu_G)^2 \, \rho(\xxi) \, \mbox{d}\xxi \approx \sum_{k=1}^R G_k^2
\left<\psi_k,\psi_k\right>.
 \label{eq:sigma}
\end{equation}

\begin{figure}[ht]
\begin{tabular}{clc}
        
\includegraphics[width=0.475\textwidth]{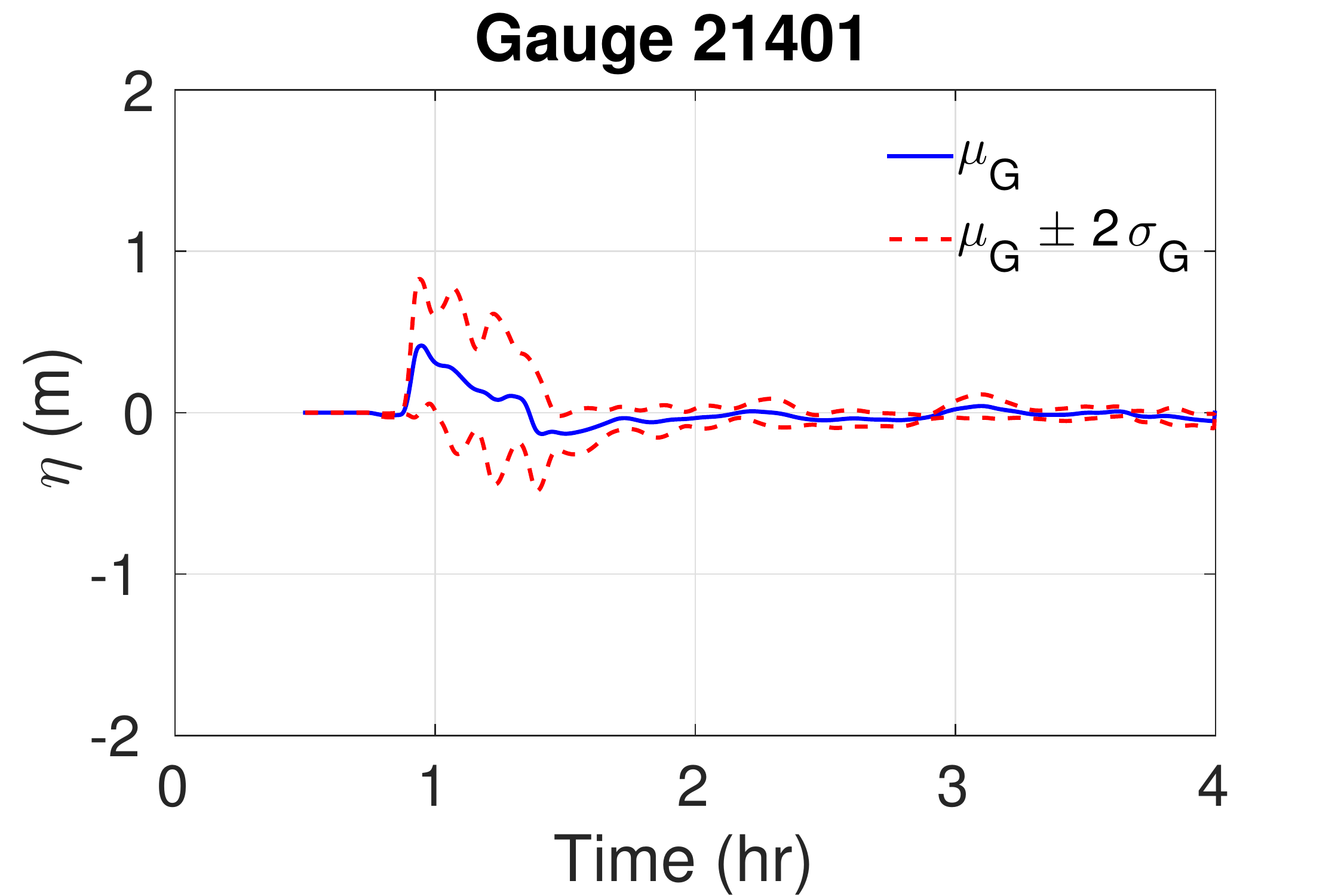} &
\includegraphics[width=0.475\textwidth]{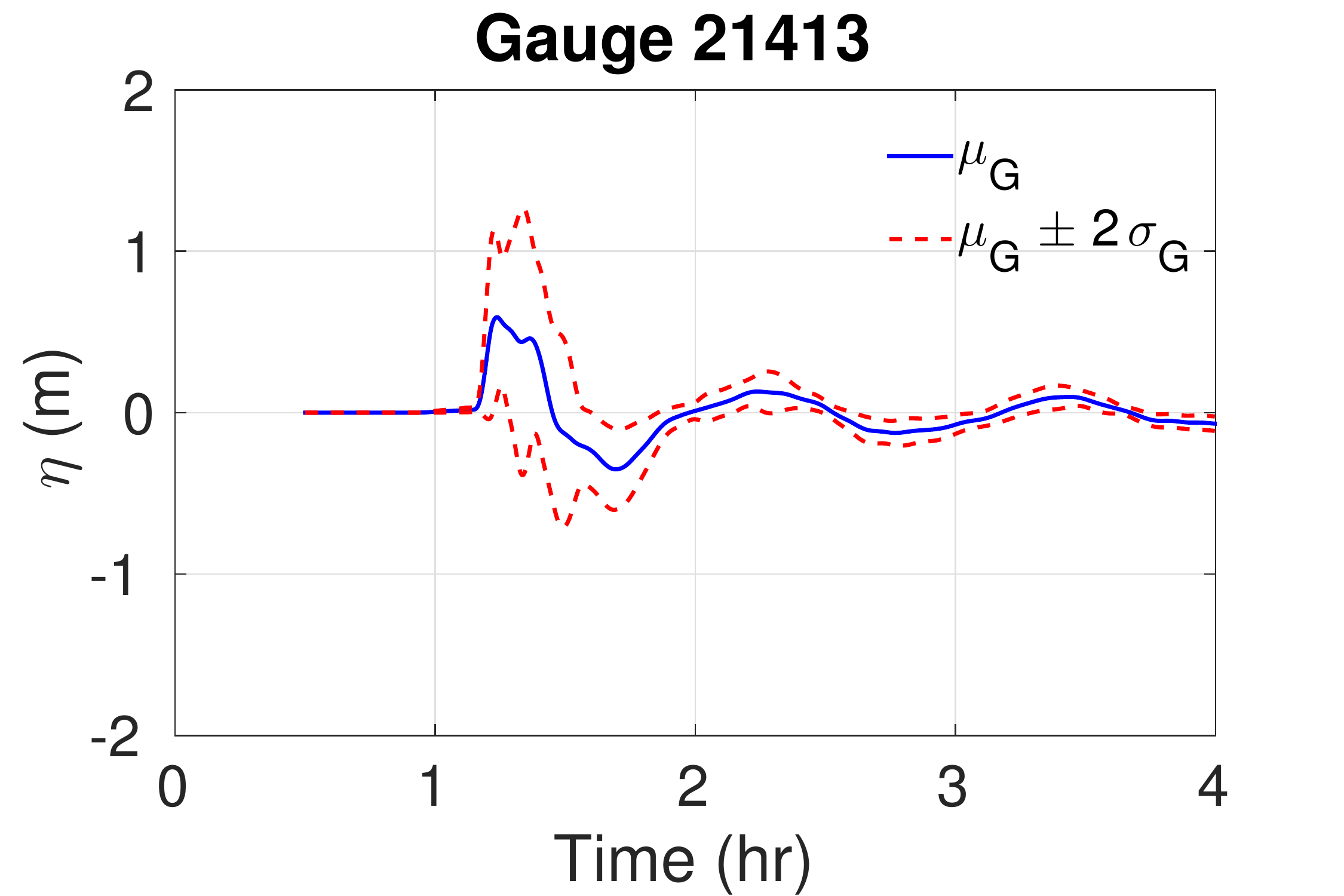} \\
\includegraphics[width=0.475\textwidth]{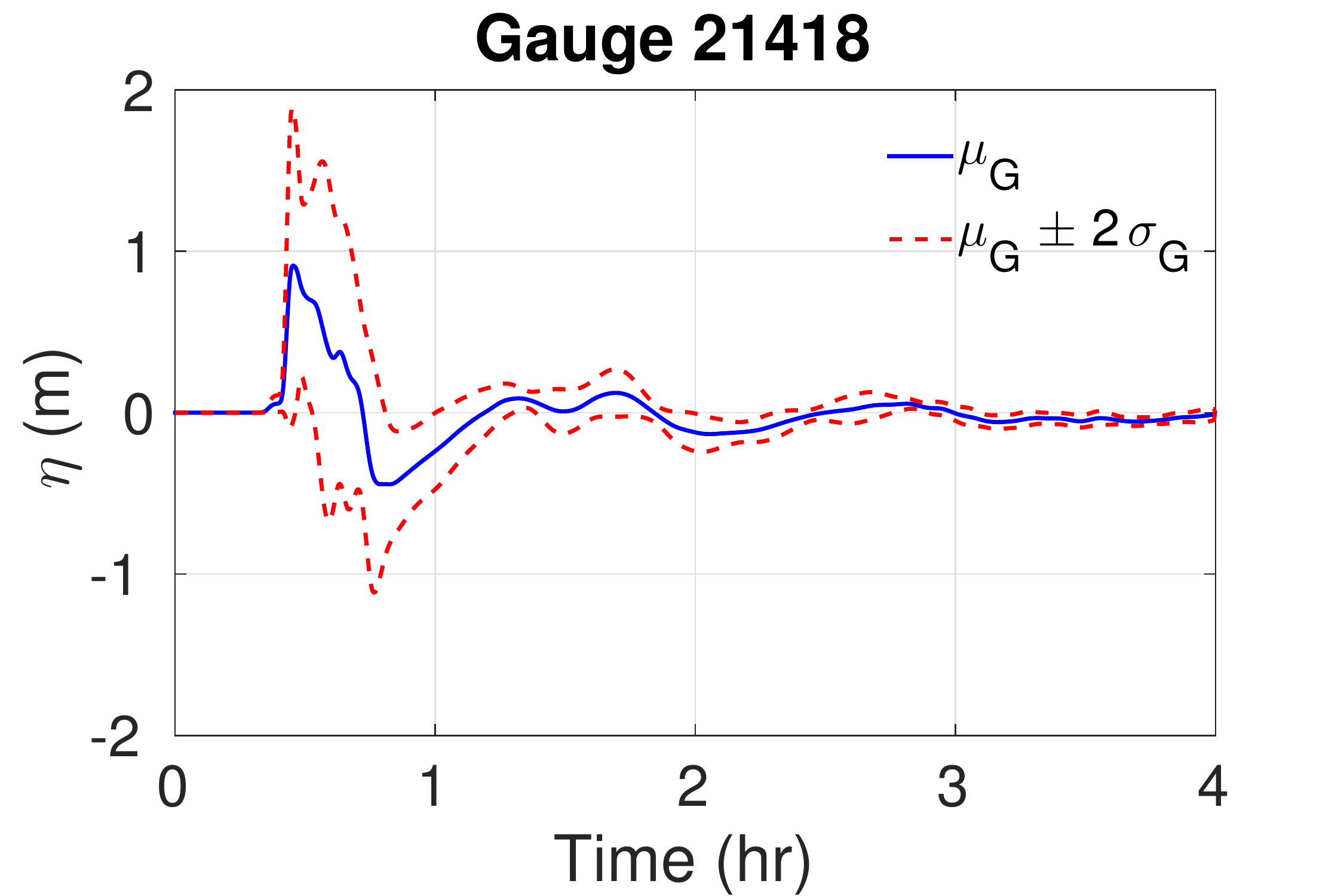} &
\includegraphics[width=0.475\textwidth]{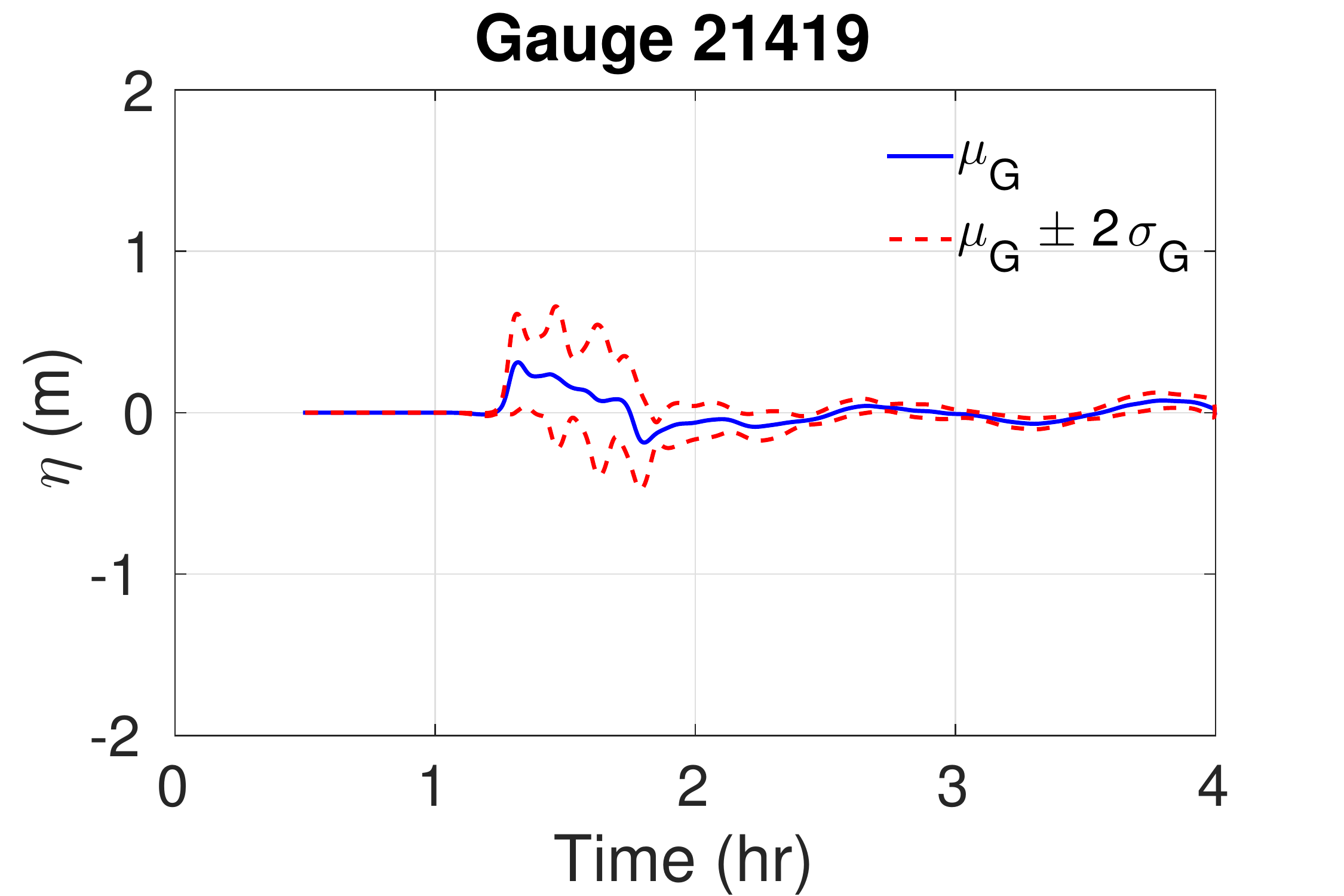}
\end{tabular}
\caption{Evolution of the PC-mean water surface elevation at the different gauges.}
\label{fig:ave}
\end{figure}

The evolution of the mean of the sea-surface elevation $\mu_G$ along with two
standard deviation bounds ($\pm 2\sigma_G$) are thus computed from the PC
coefficients and plotted in Figure~\ref{fig:ave} at the four gauges.
Note that the evolution shown starts at $t=2 hrs$ when the
uncertainty becomes significant. An interesting observation is that the standard
deviation in water surface elevations waxes and wanes as the tsunami evolves.
The narrowing of the variance at these instances is possibly associated with the
waves that arrive due to reflections from a single source and then move away
from the gauge location imposing no variance in the water surface
elevation.

\subsection{Fault Slip Inference} 
\label{sec:infer}

Finally with our PC surrogate in hand we can solve the inverse problem, estimating the fault slip values as well as the variance of the noise in the measured data
using Bayesian inference. For this purpose, we implement an adaptive MCMC method ~\citep{Gareth2009,Haario2001} to sample 
the posterior distributions in Equation~\ref{eq:post_coef} and consequently update the uncertain parameters.

\begin{figure}[ht]
\begin{tabular}{clc}
\includegraphics[width=0.475\textwidth]{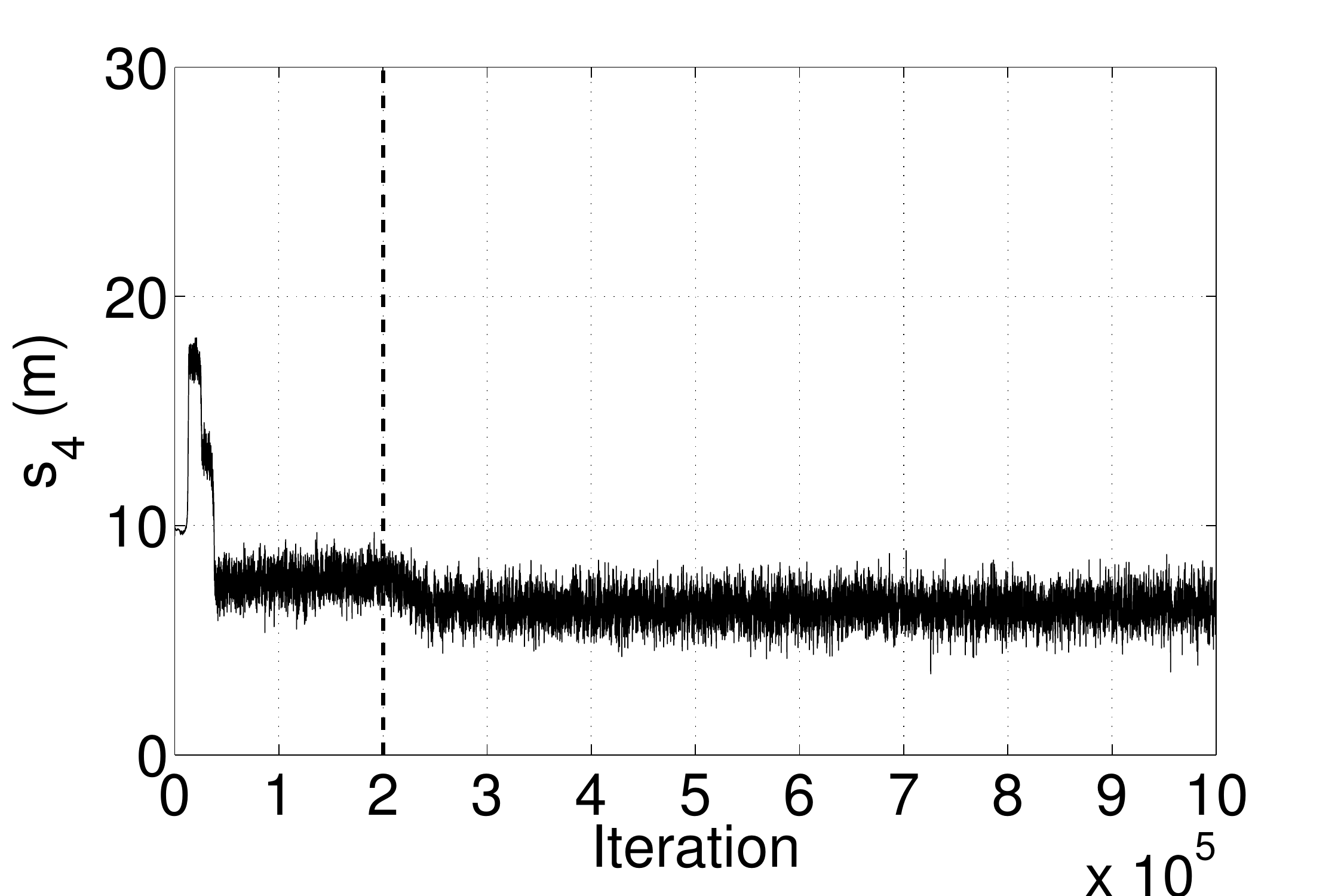} &
\includegraphics[width=0.475\textwidth]{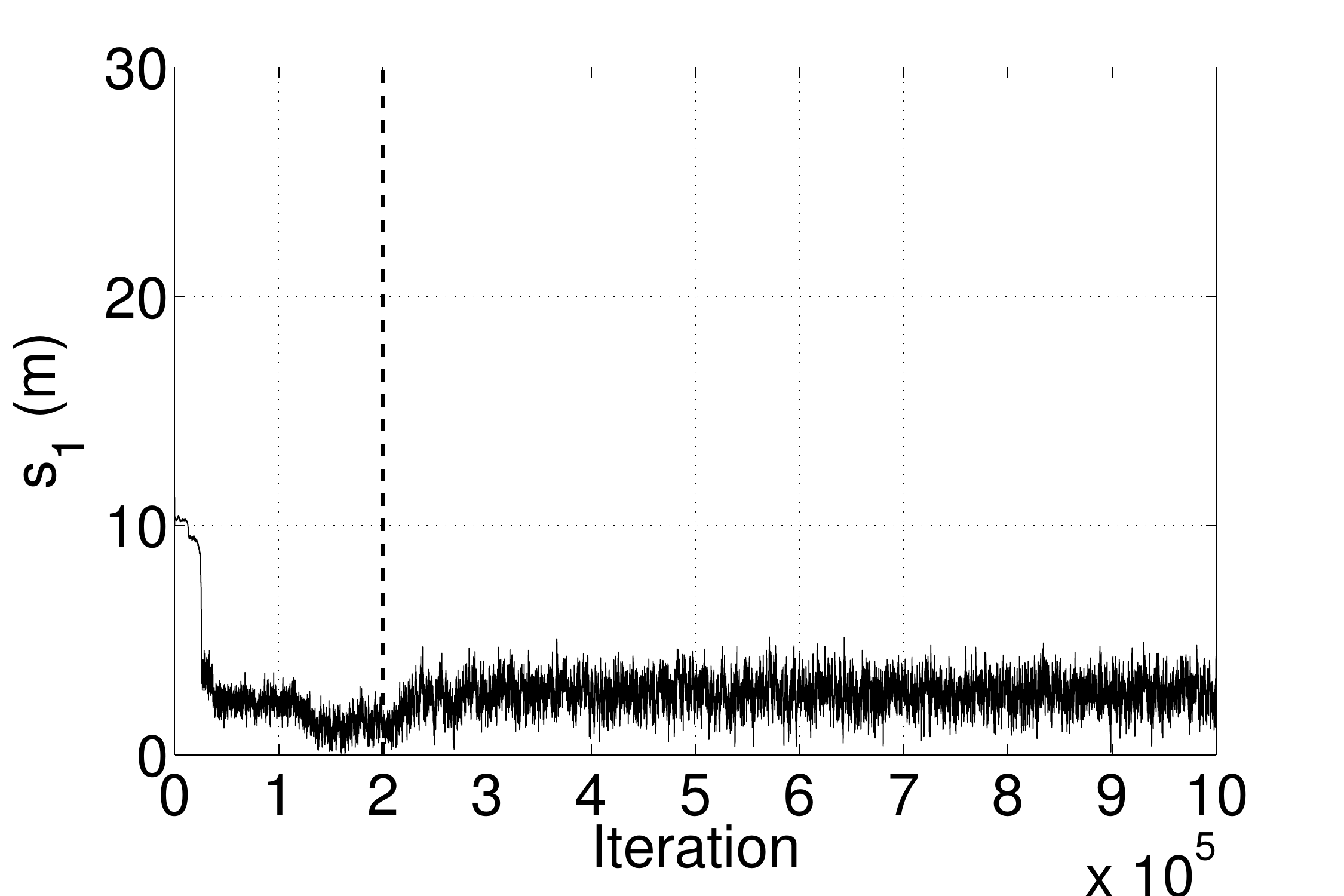} \\
\includegraphics[width=0.475\textwidth]{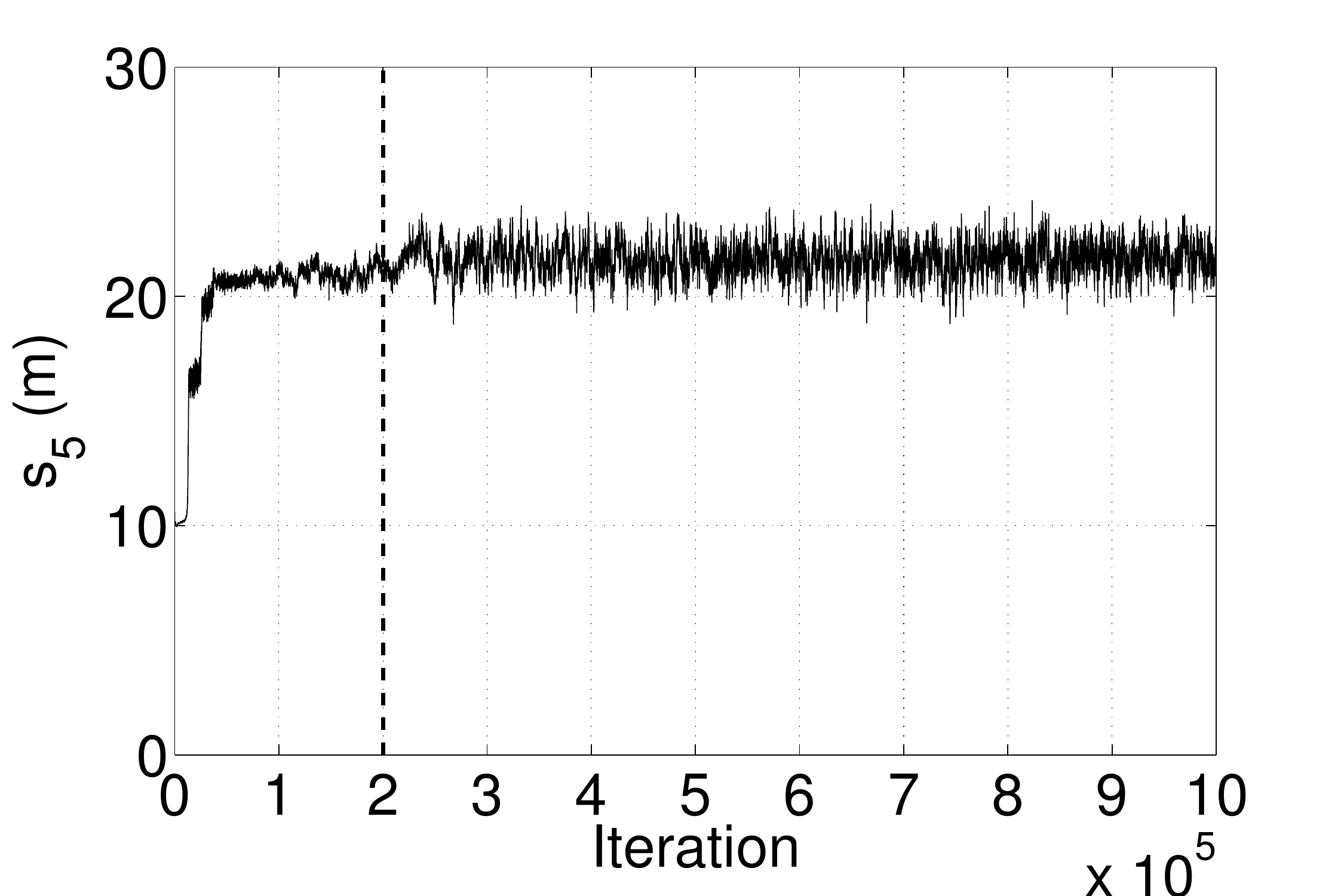} &
\includegraphics[width=0.475\textwidth]{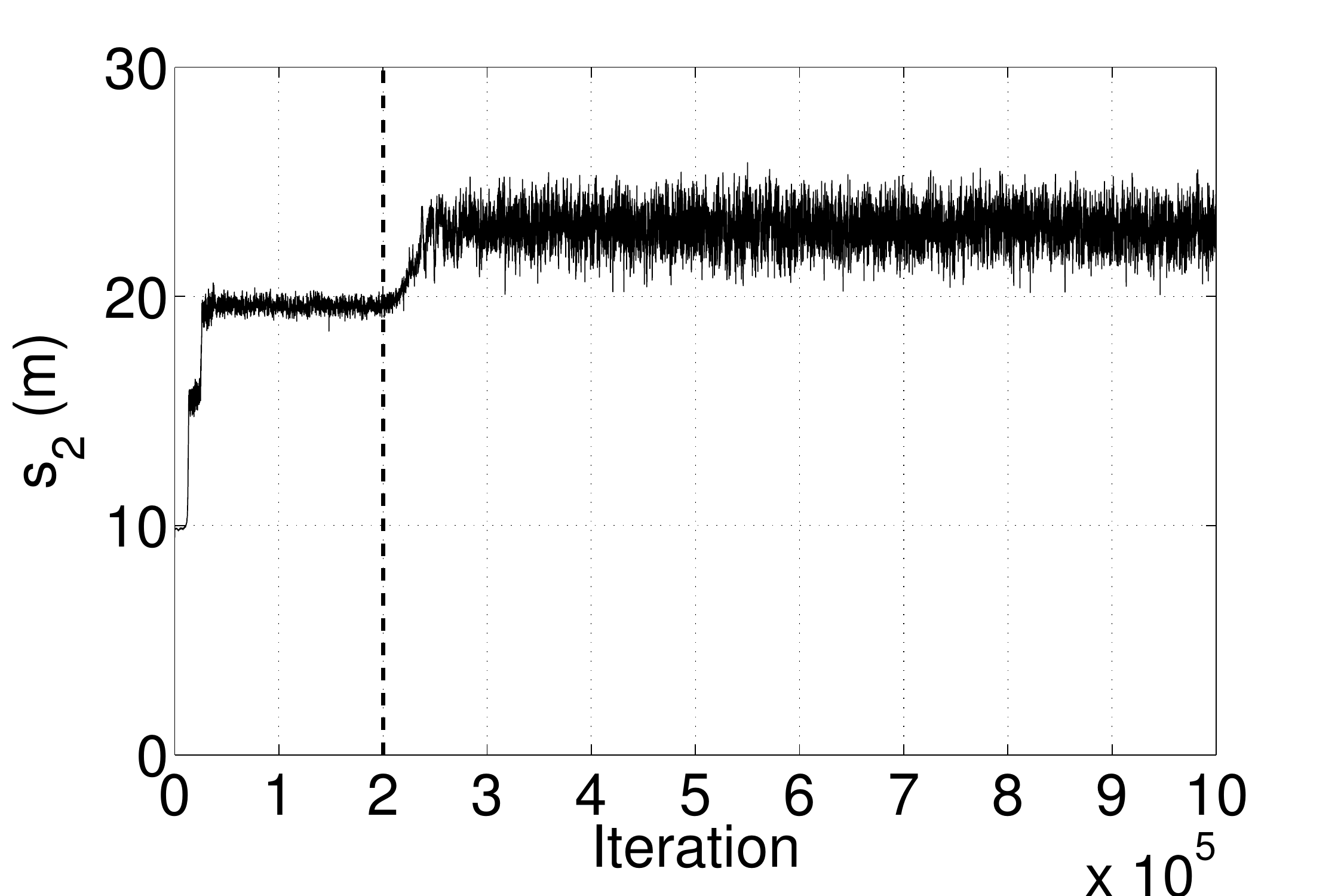} \\
\includegraphics[width=0.475\textwidth]{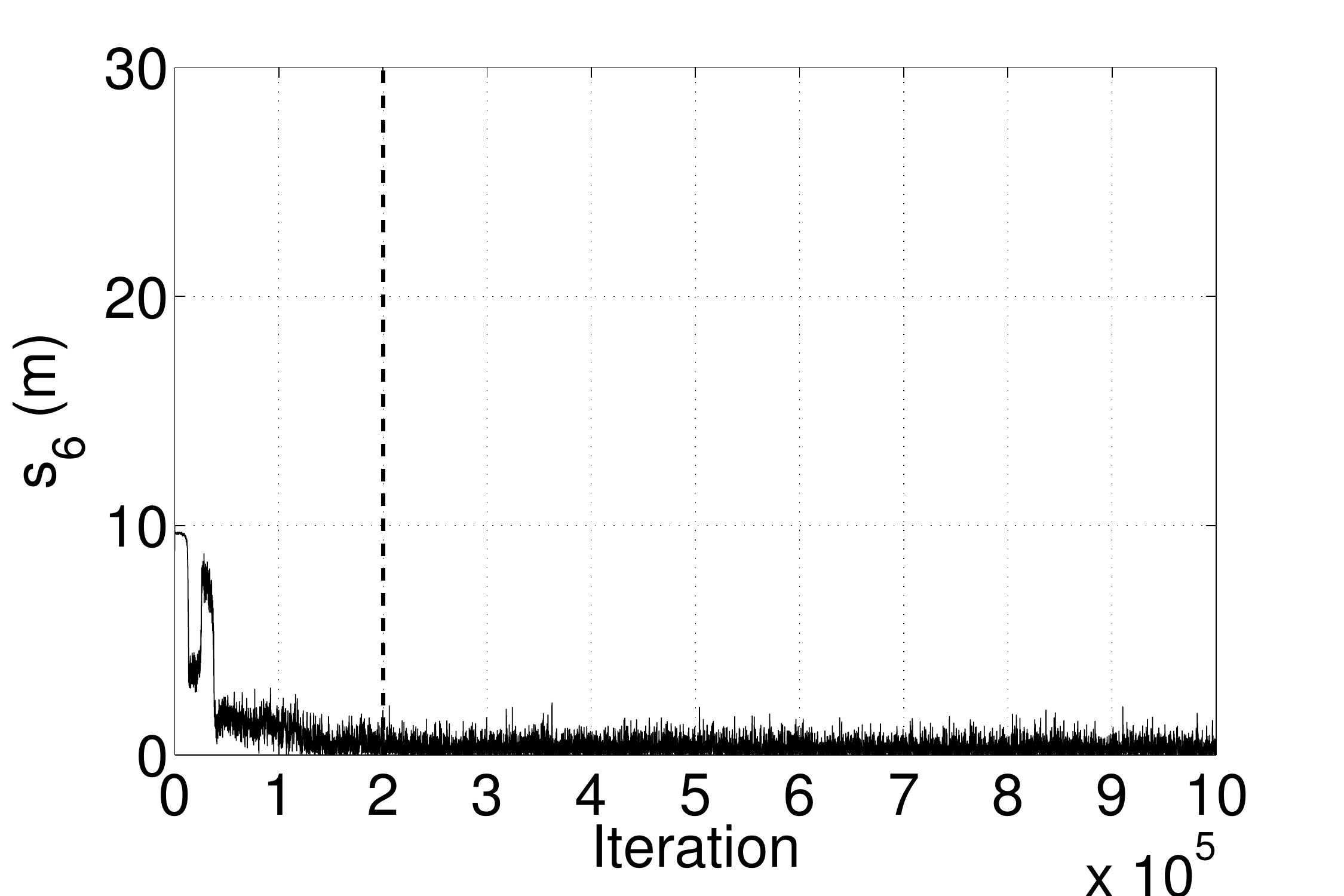} &
\includegraphics[width=0.475\textwidth]{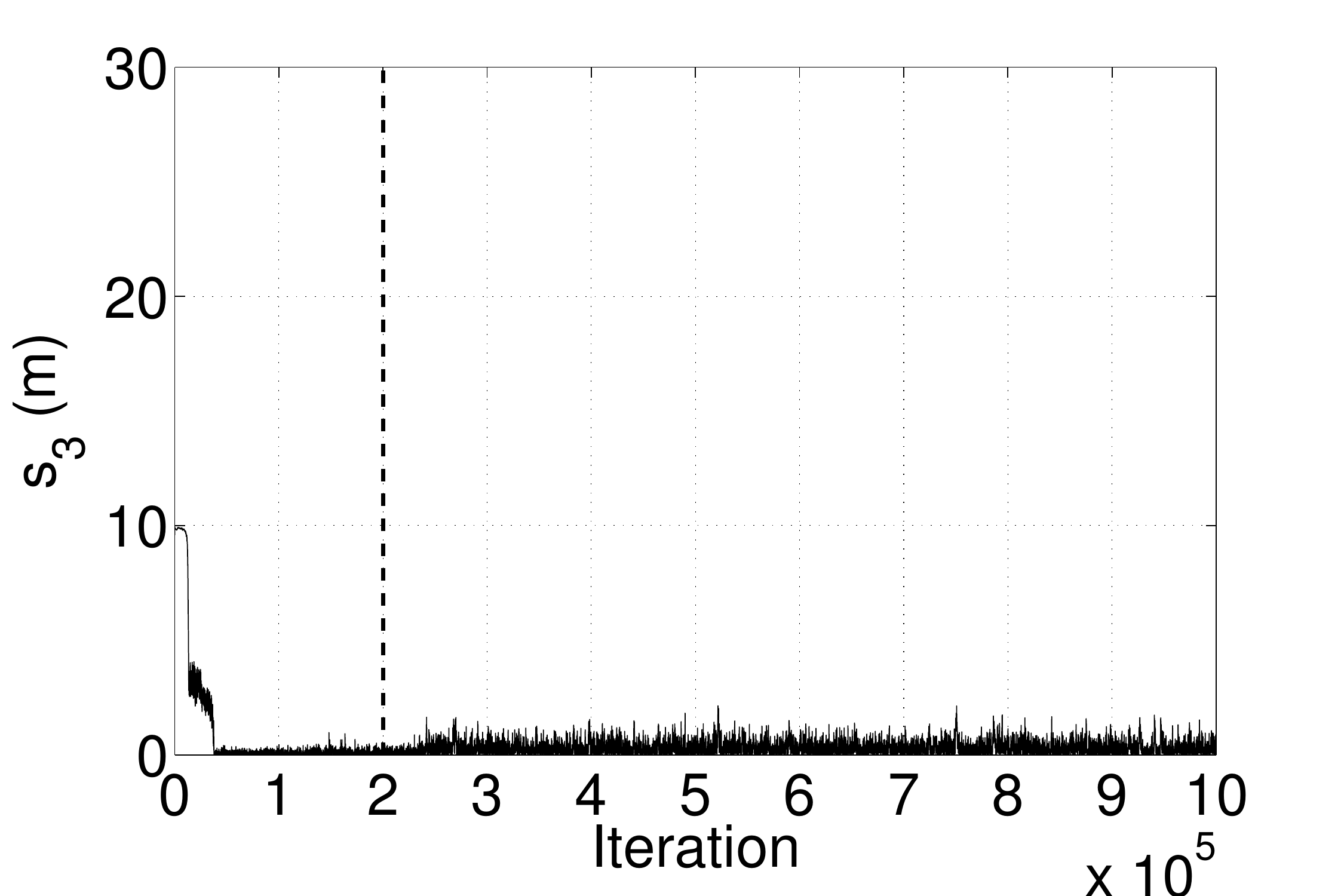} 
\end{tabular}
\caption{Chain samples for the six slip values $s_1,..,s_6$. The vertical dotted lines corresponds to the
burn-in iterations.}
\label{fig:chains_p} 
\end{figure}

\begin{figure}[ht]
\centering
\begin{tabular}{clc}        
\includegraphics[width=0.6\textwidth]{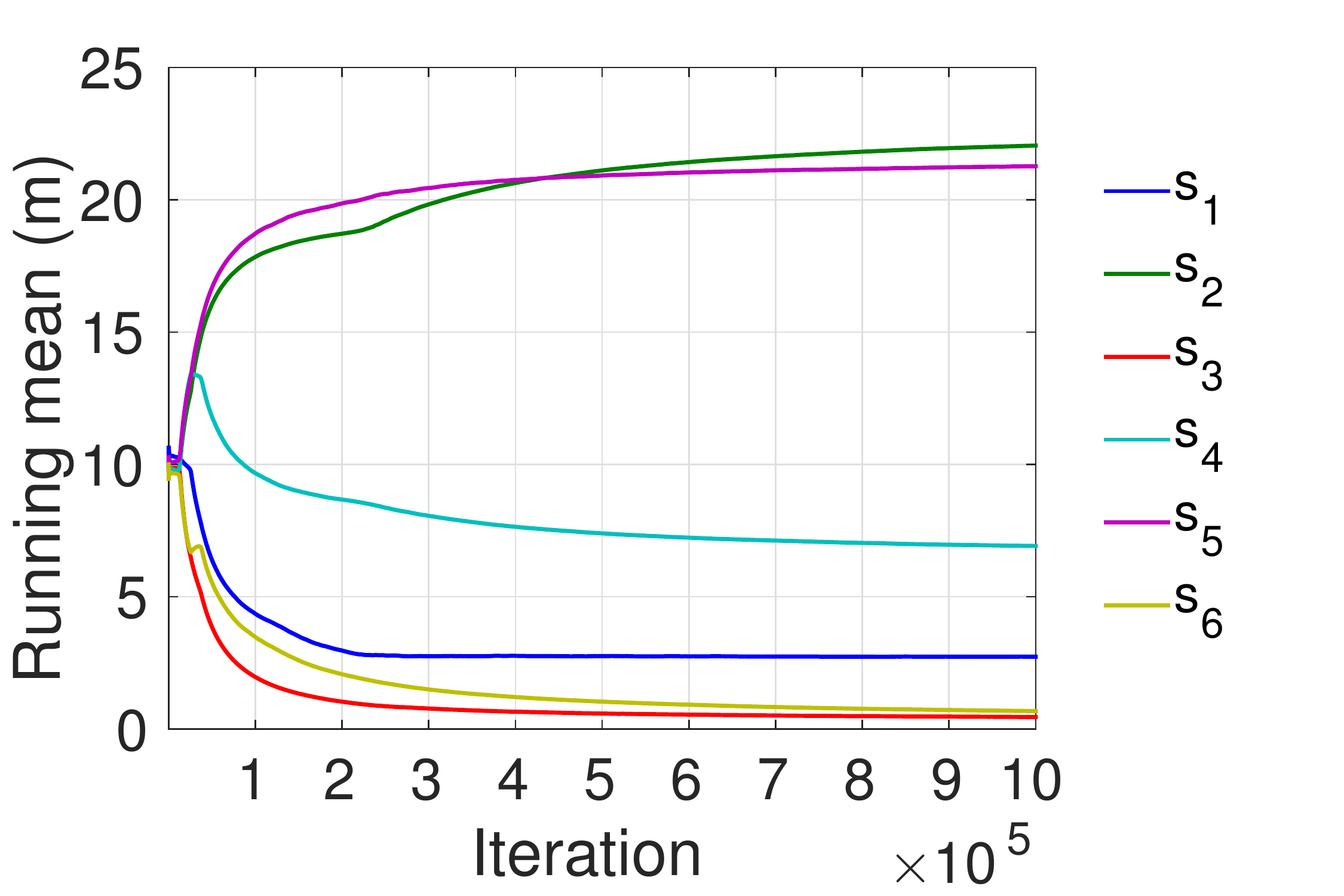}

\end{tabular}
\caption{Running mean of the slips values computed using their corresponding MCMC chains.}
\label{fig:running_mean}
\end{figure}

\begin{figure}[ht]
\begin{tabular}{clc}
\includegraphics[width=0.475\textwidth]{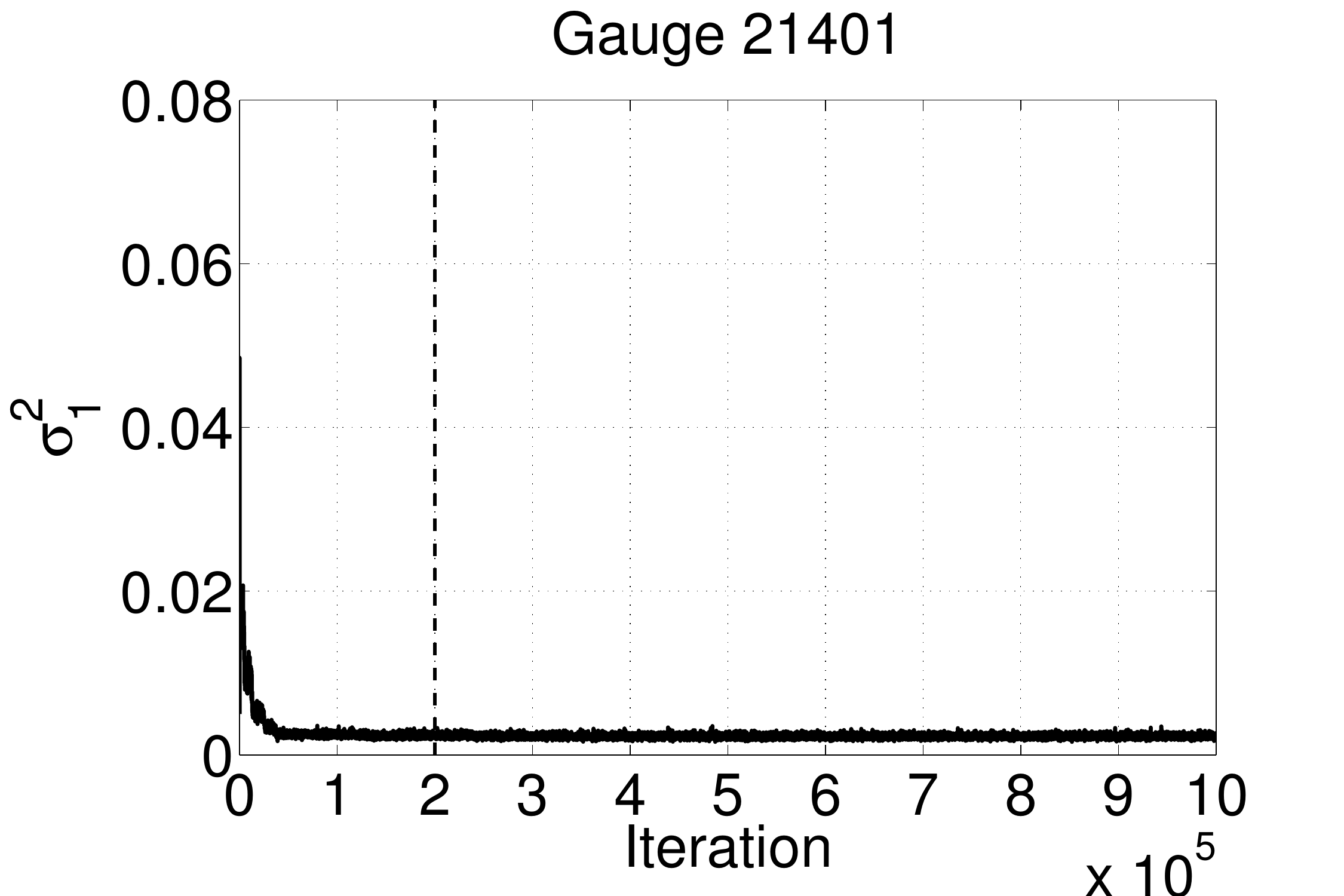} &
\includegraphics[width=0.475\textwidth]{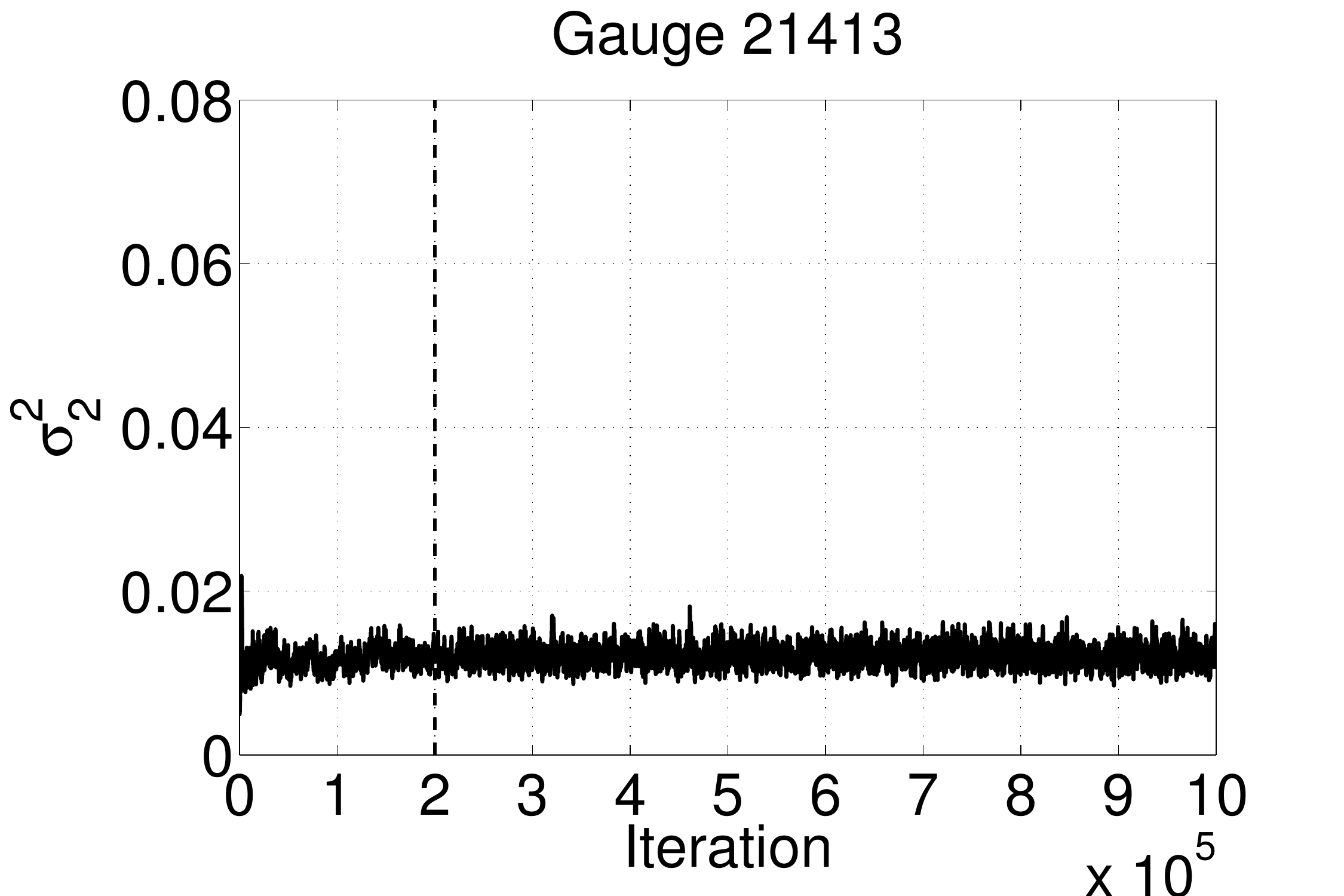} \\
\includegraphics[width=0.475\textwidth]{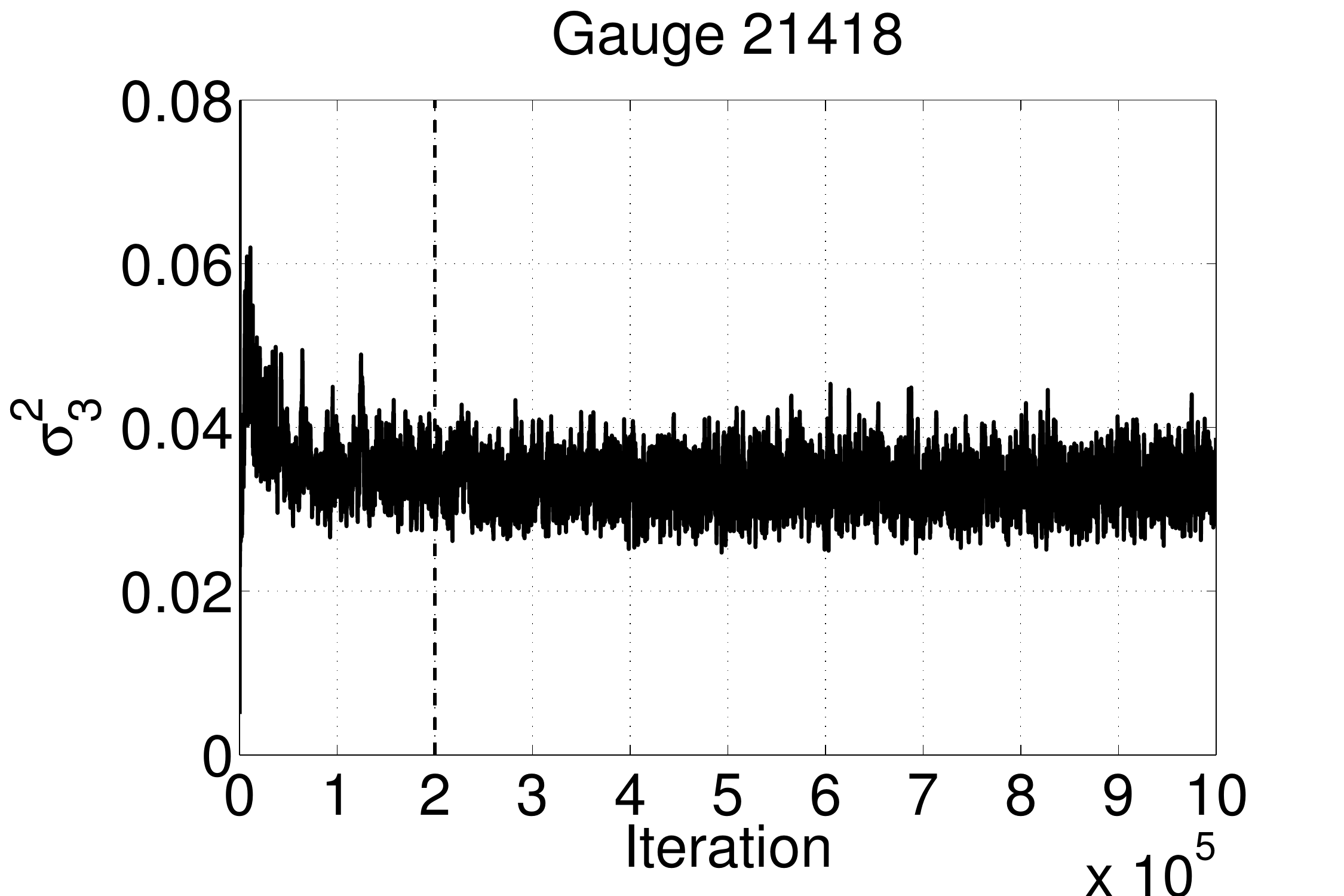} &
\includegraphics[width=0.475\textwidth]{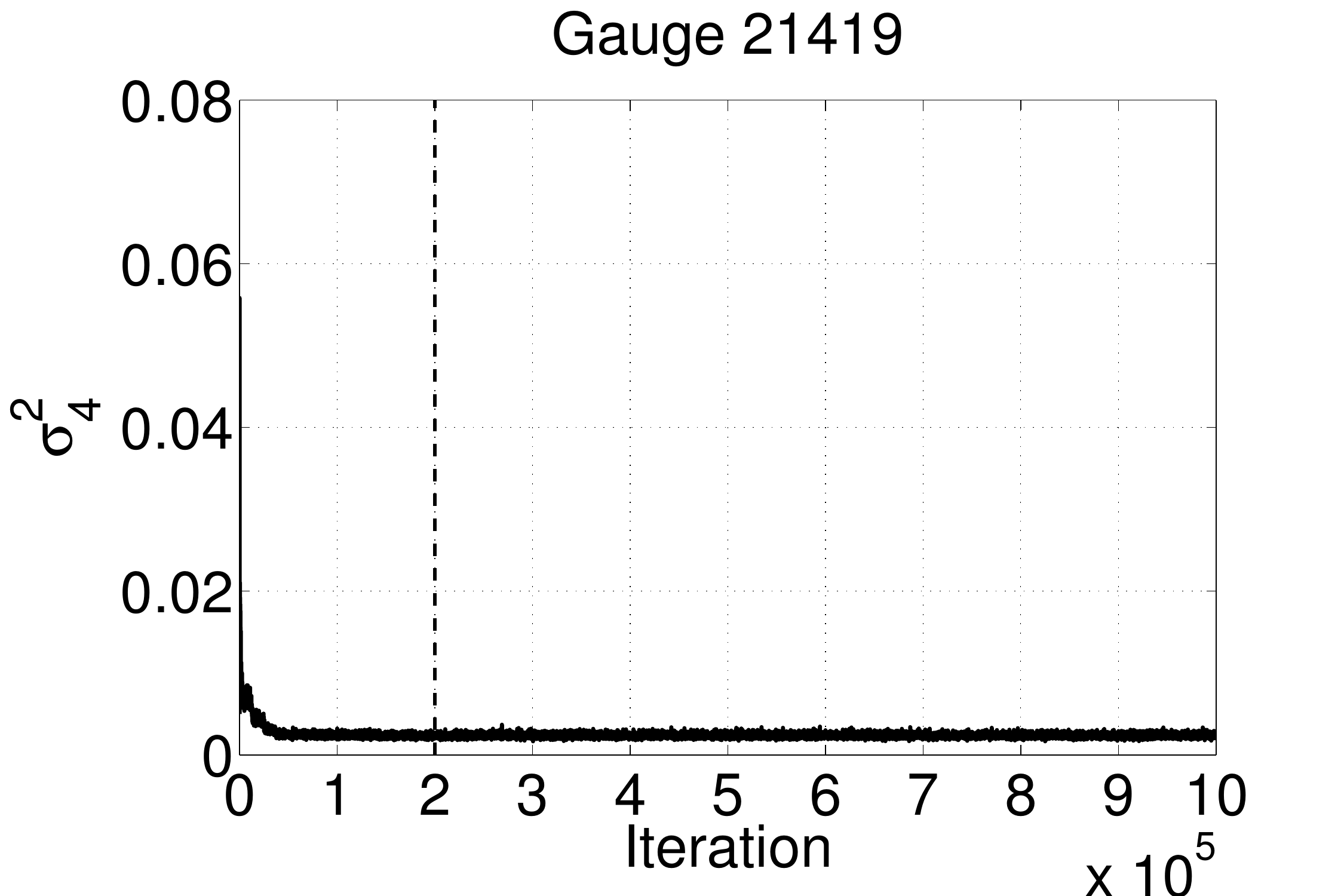} 
\end{tabular}
\caption{Chain samples for the four $\sigma^2$
the noise variance. The vertical dotted lines corresponds to the
burn-in iterations.}
\label{fig:chains_s} 
\end{figure}

The posterior was sampled $10^6$ times after which we find negligible change in the estimated posteriors of the fault slip values:  $s_{1} \ldots  s_6$ as well as for the noise variance $\sigma^2_1 \ldots \sigma^2_4$ with further iterations.
Figure \ref{fig:chains_p} plots the sample chains for the input parameters for different 
iterations of the MCMC algorithm. The different panels show well-mixed chains for all input parameters
where the chains of  $s_{1},s_{2},s_{4}$ and $s_{5}$ appear to be concentrated in an area of
the parameter prior range. In contrast, the $s_{3}$ and $s_6$ chains appear to be concentrated in the lower end of
the parameter range. The running mean plotted in Figure~\ref{fig:running_mean} is an indication of the convergence of the MCMC.
The chains for the noise variances ($\sigma^2_1 \ldots \sigma^2_4$) 
are shown in Figure~\ref{fig:chains_s} at the different gauges and appear to be well mixed with a well defined
posterior range. The maximum variance appears to be at gauge 21418 and its range lies between 0.025 and 0.045.

\begin{figure}[ht]
\begin{tabular}{clc}
\includegraphics[width=0.475\textwidth]{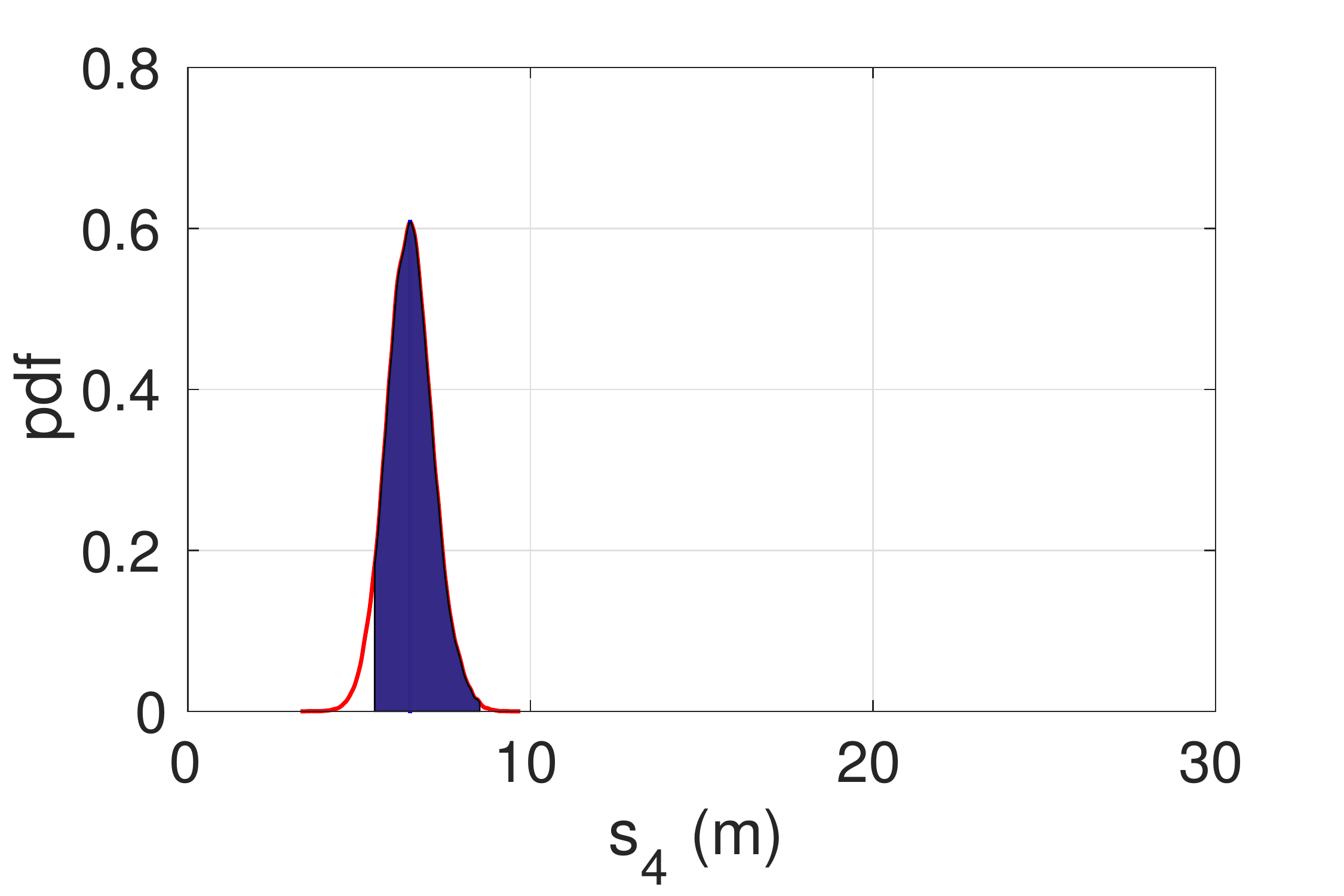} &
\includegraphics[width=0.475\textwidth]{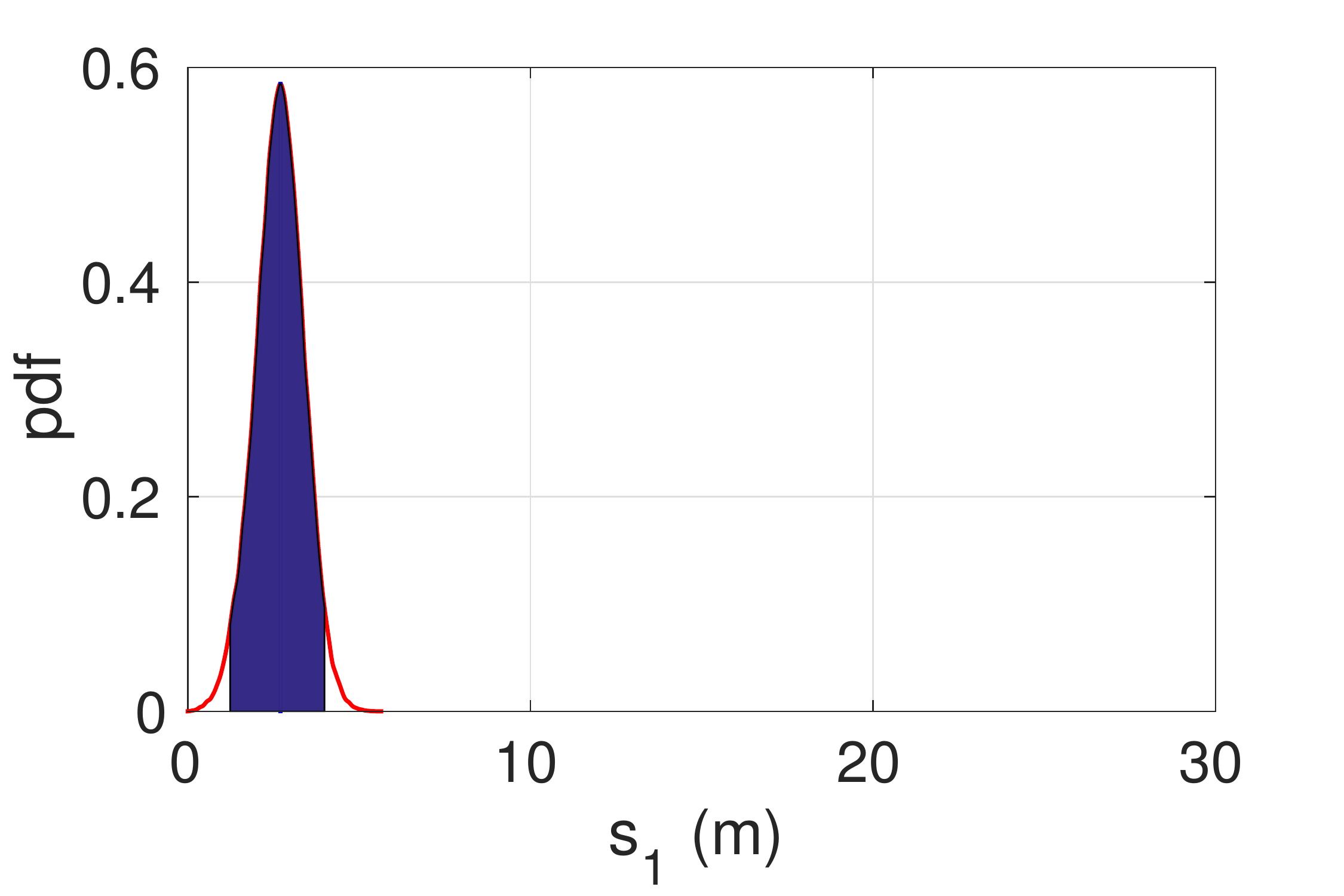} \\
\includegraphics[width=0.475\textwidth]{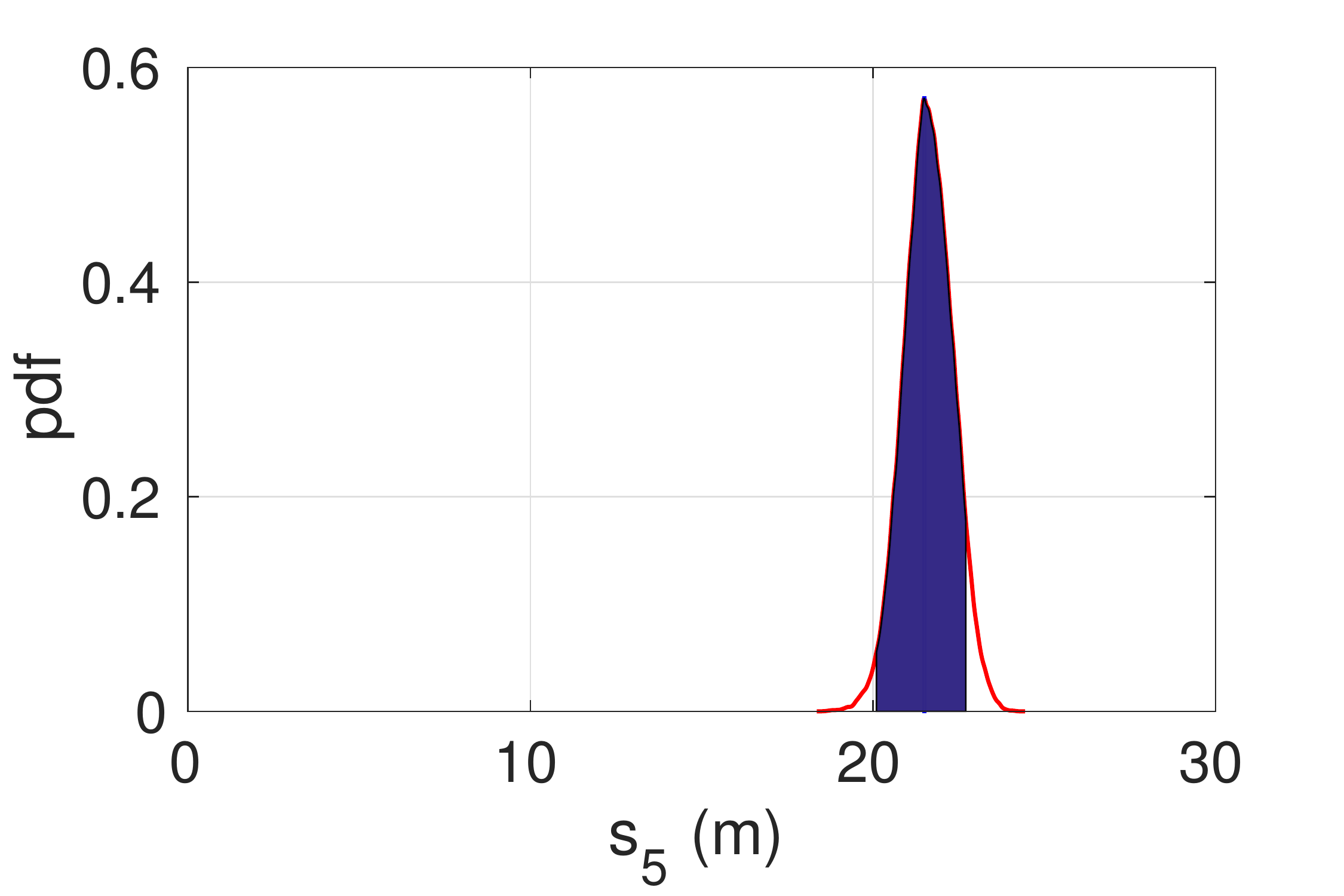} &
\includegraphics[width=0.475\textwidth]{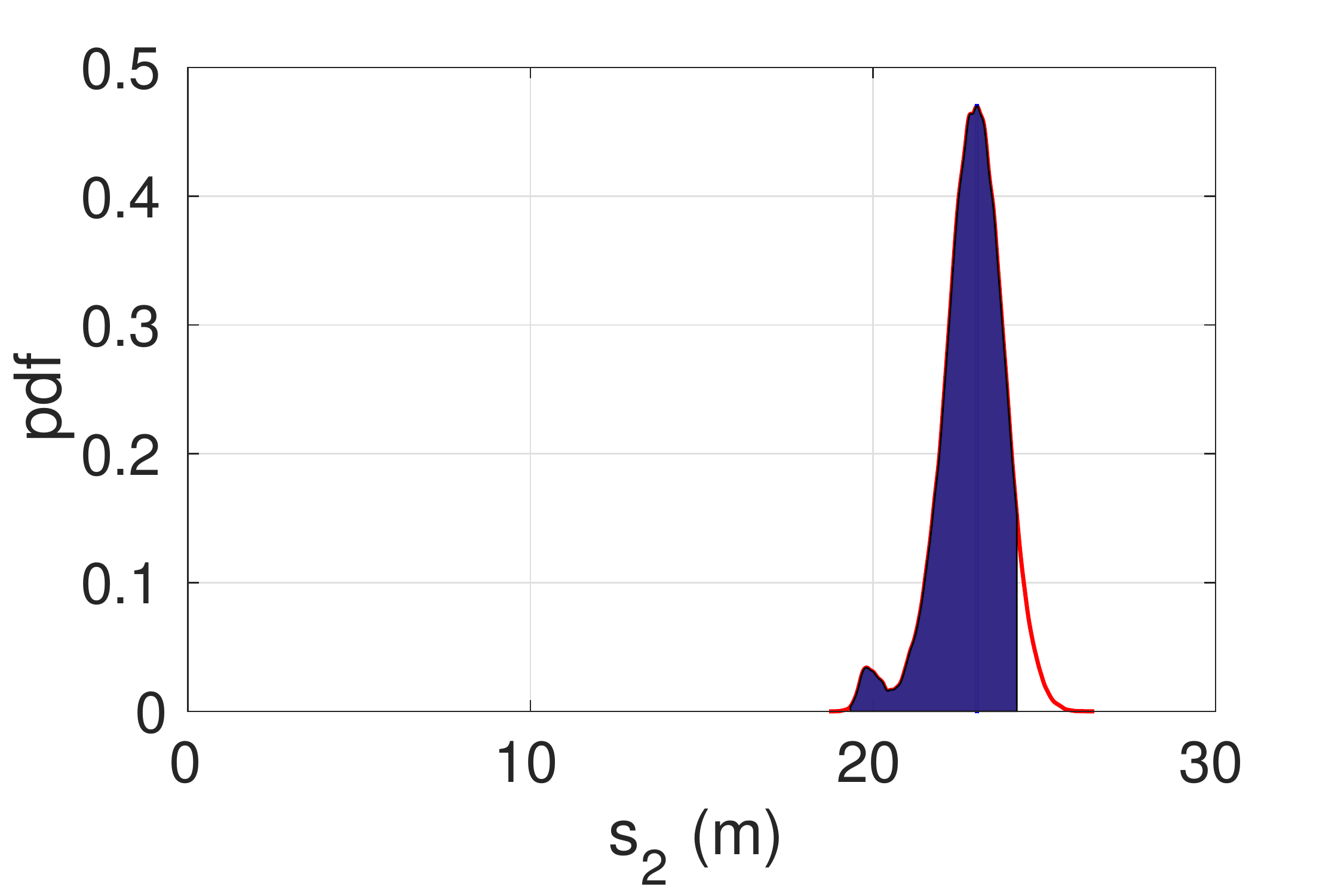} \\
\includegraphics[width=0.475\textwidth]{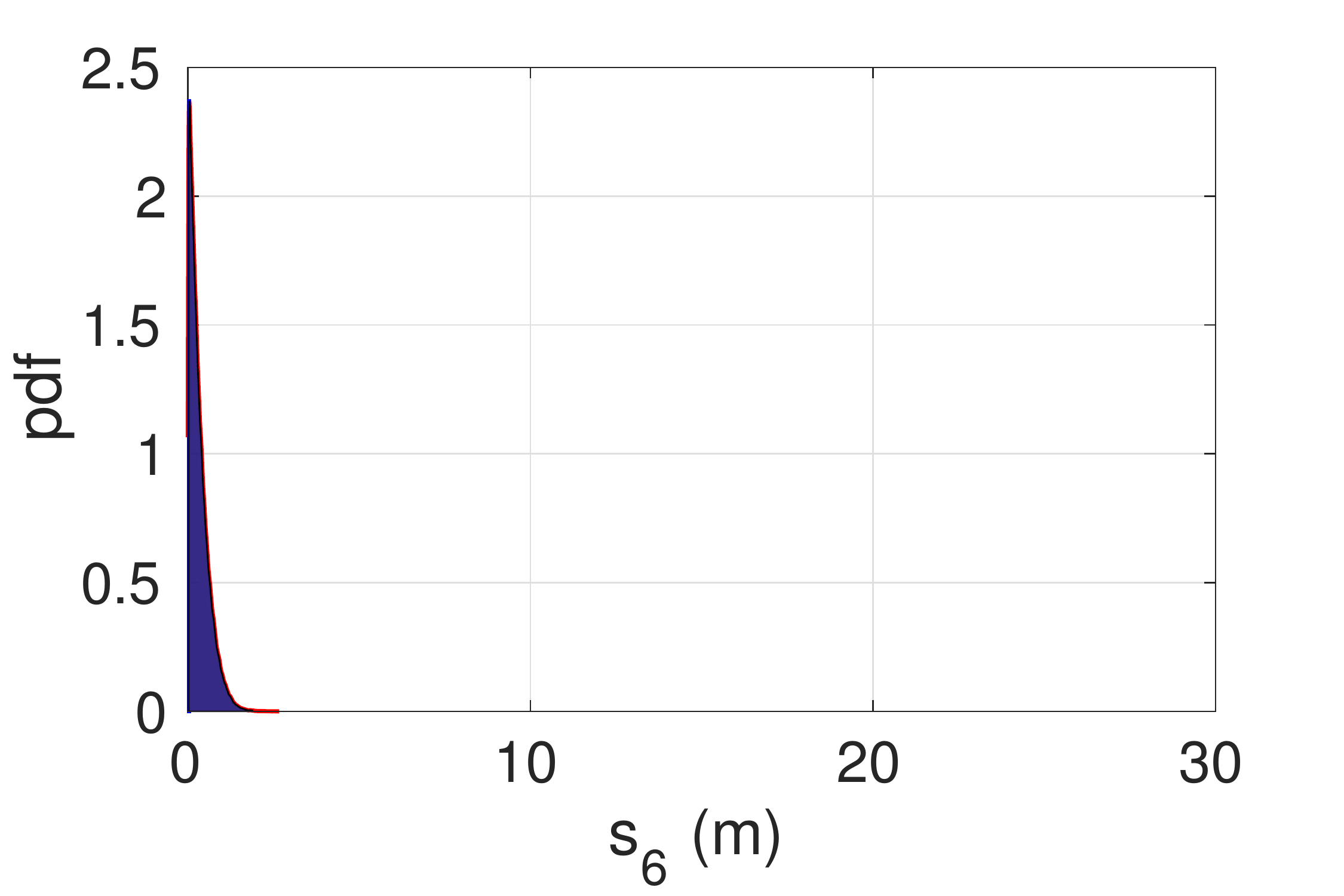} &
\includegraphics[width=0.475\textwidth]{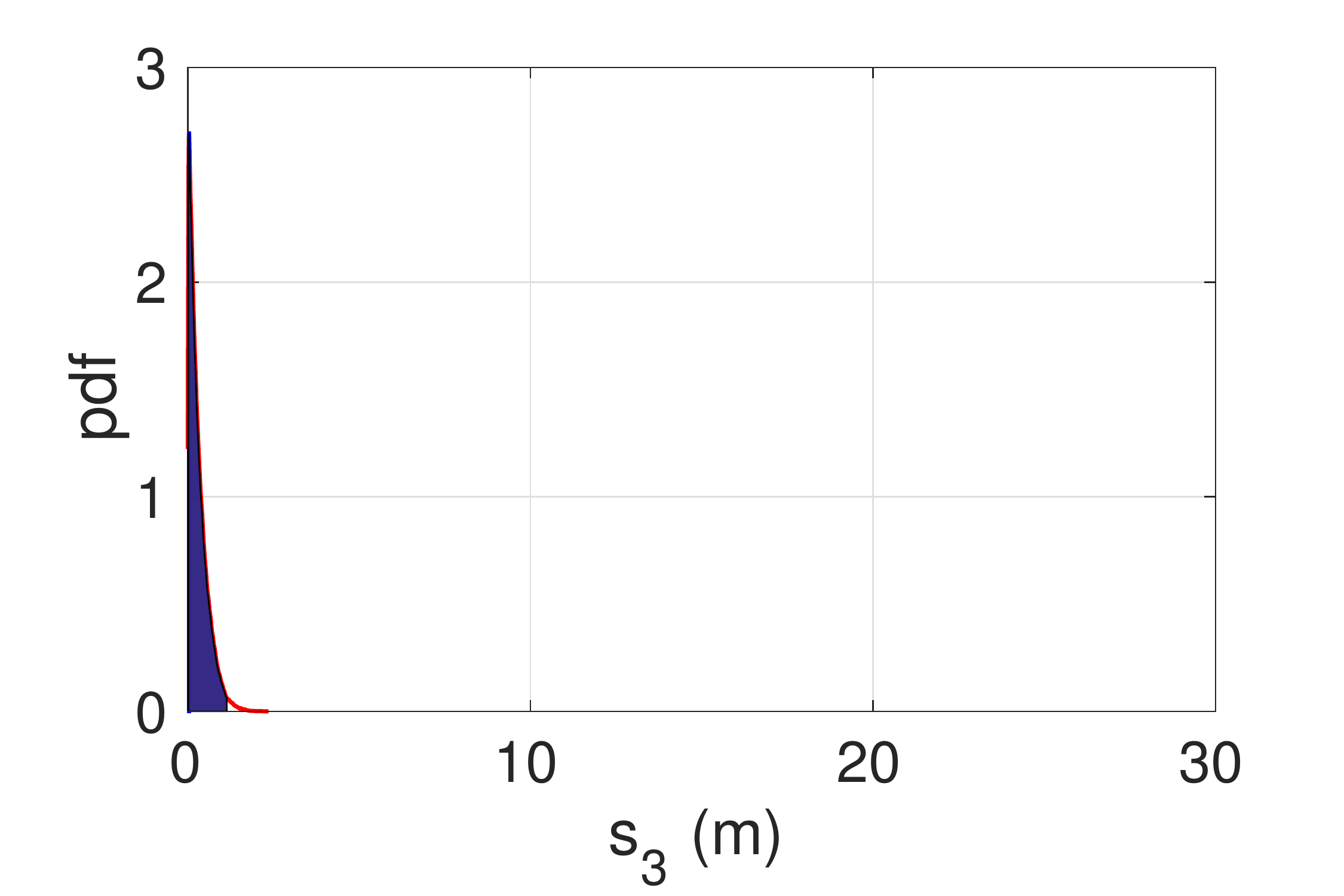}
\end{tabular}
\caption{KDE of the marginalized posterior distributions for the six slip values $\Pi(s_1)$ ... $\Pi(s_6)$.
The shaded regions corresponds to the 95\% intervals of high posterior
probability.
}
\label{fig:pdfs_p} 
        \end{figure}

We used Kernel Density Estimation (KDE)~\citep{Parzen1962,Silverman1986} to determine the marginalized posterior 
probability distribution functions (\emph{pdfs}) using the computed MCMC chains and plot them in Figure~\ref{fig:pdfs_p} for the different parameters. 
The first $2\times 10^5$ MCMC iterations were considered as the burn-in period and thus discarded.  
The shapes of the marginalized posterior \emph{pdfs} are consistent with the chains shown in Figure~\ref{fig:chains_p}
where the \emph{pdfs} of $s_{1},s_{2},s_{4}$ and $s_{5}$ appear to have a Gaussian-like shape with a well-defined peak;
the Maximum A Posteriori (MAP) values are estimated to be $2.7,~23,~6.5$ and $21.5$ respectively. On the other hand, for $s_3$ and $s_6$ and the \emph{pdfs} exhibit also a well-defined peak, 
but with an extended tail towards the smaller slip values; the mean values are
estimated to be $0.3$ for both. The 95\% intervals of high posterior probability
are shown as shaded regions for the inferred parameters.

\begin{figure}[ht]
        \begin{tabular}{clc}
\includegraphics[width=0.475\textwidth]{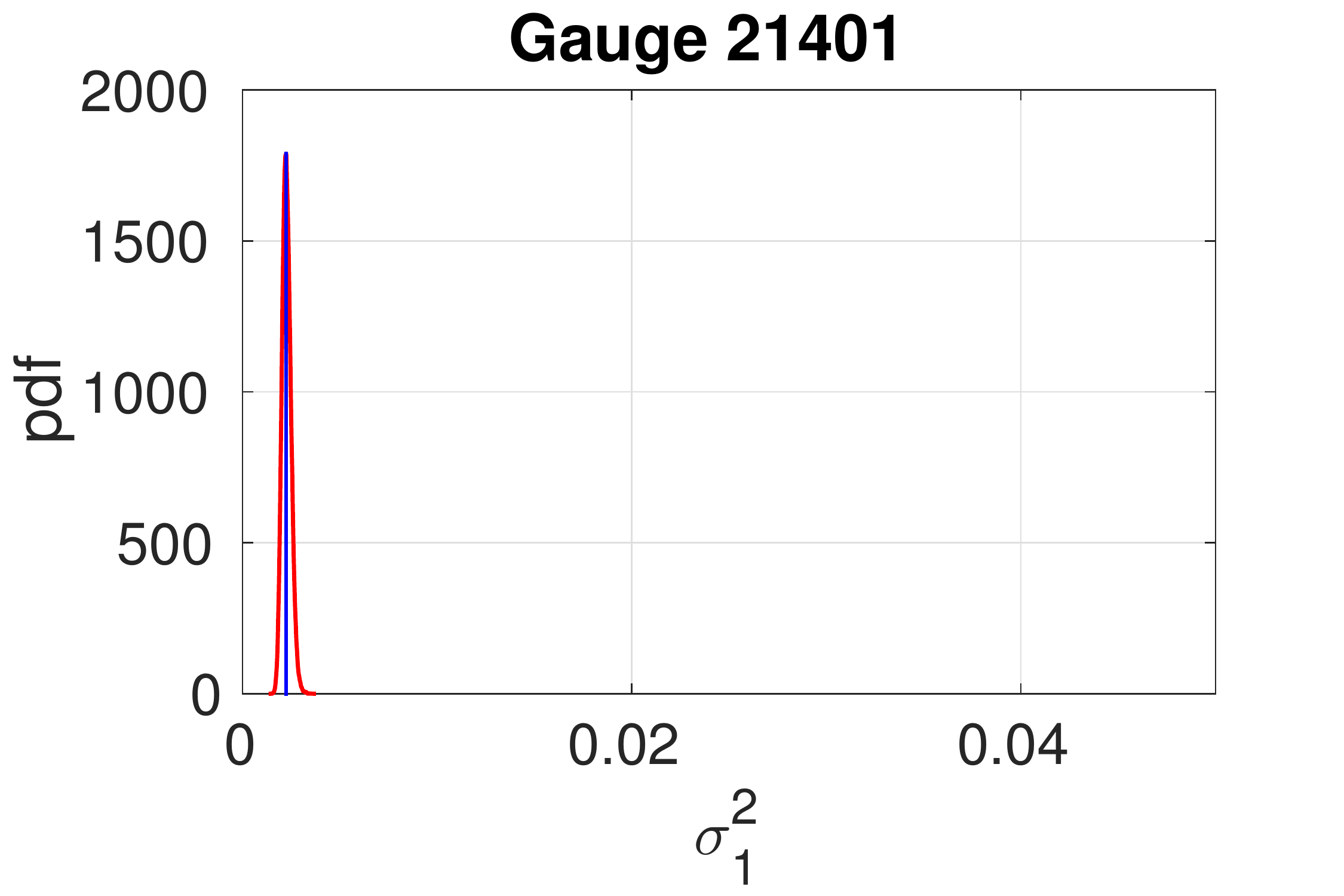} &
\includegraphics[width=0.475\textwidth]{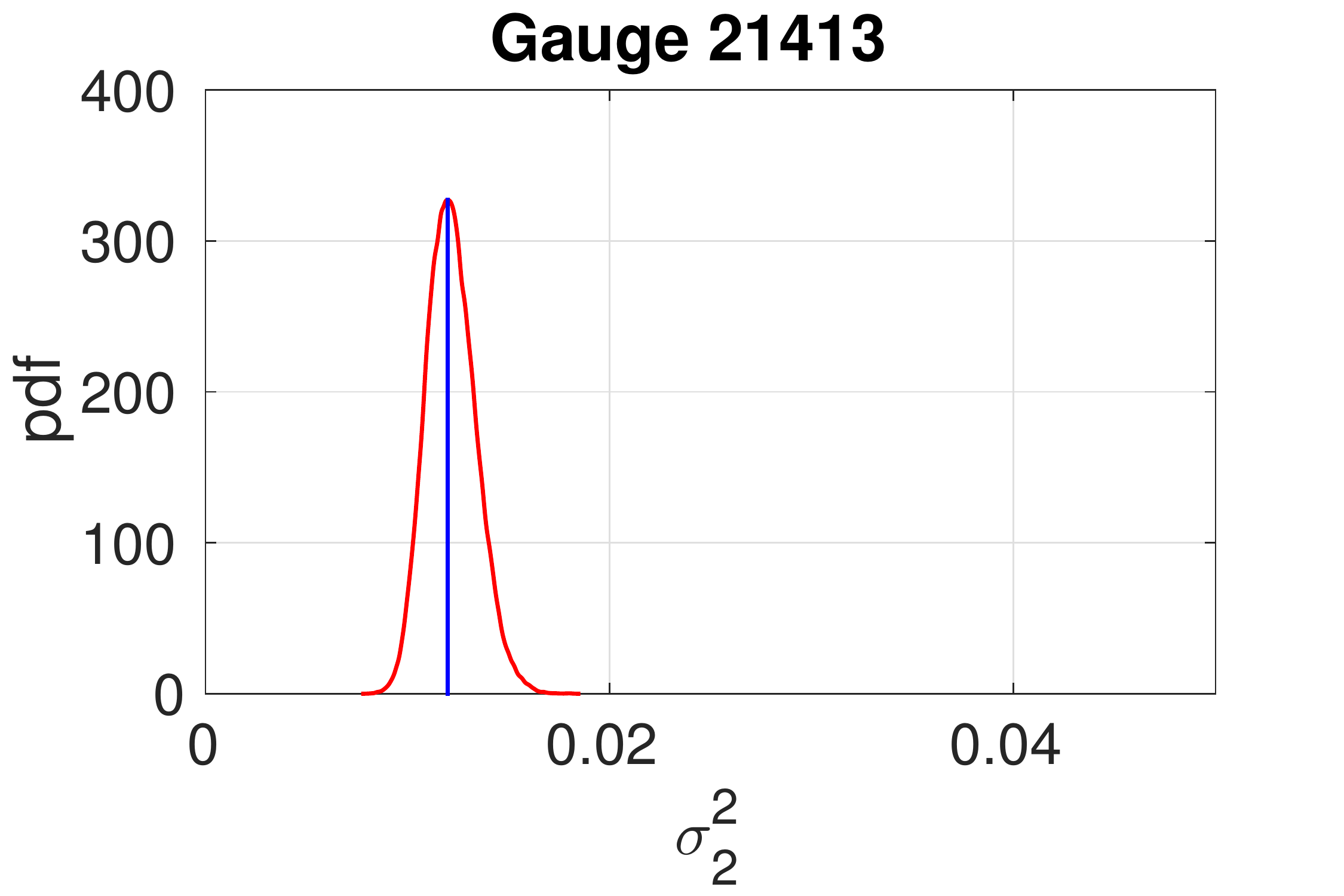} \\
\includegraphics[width=0.475\textwidth]{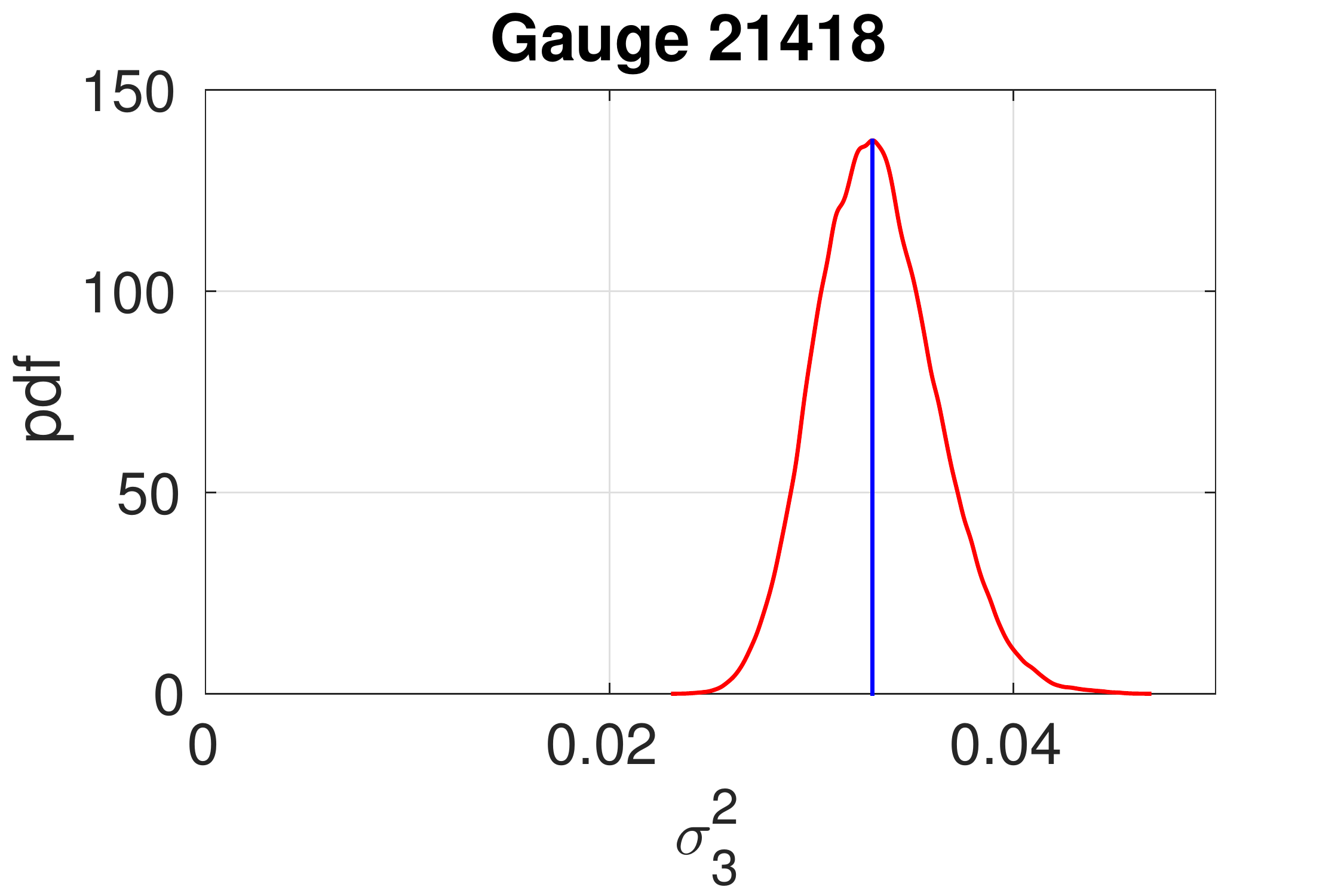} &
\includegraphics[width=0.475\textwidth]{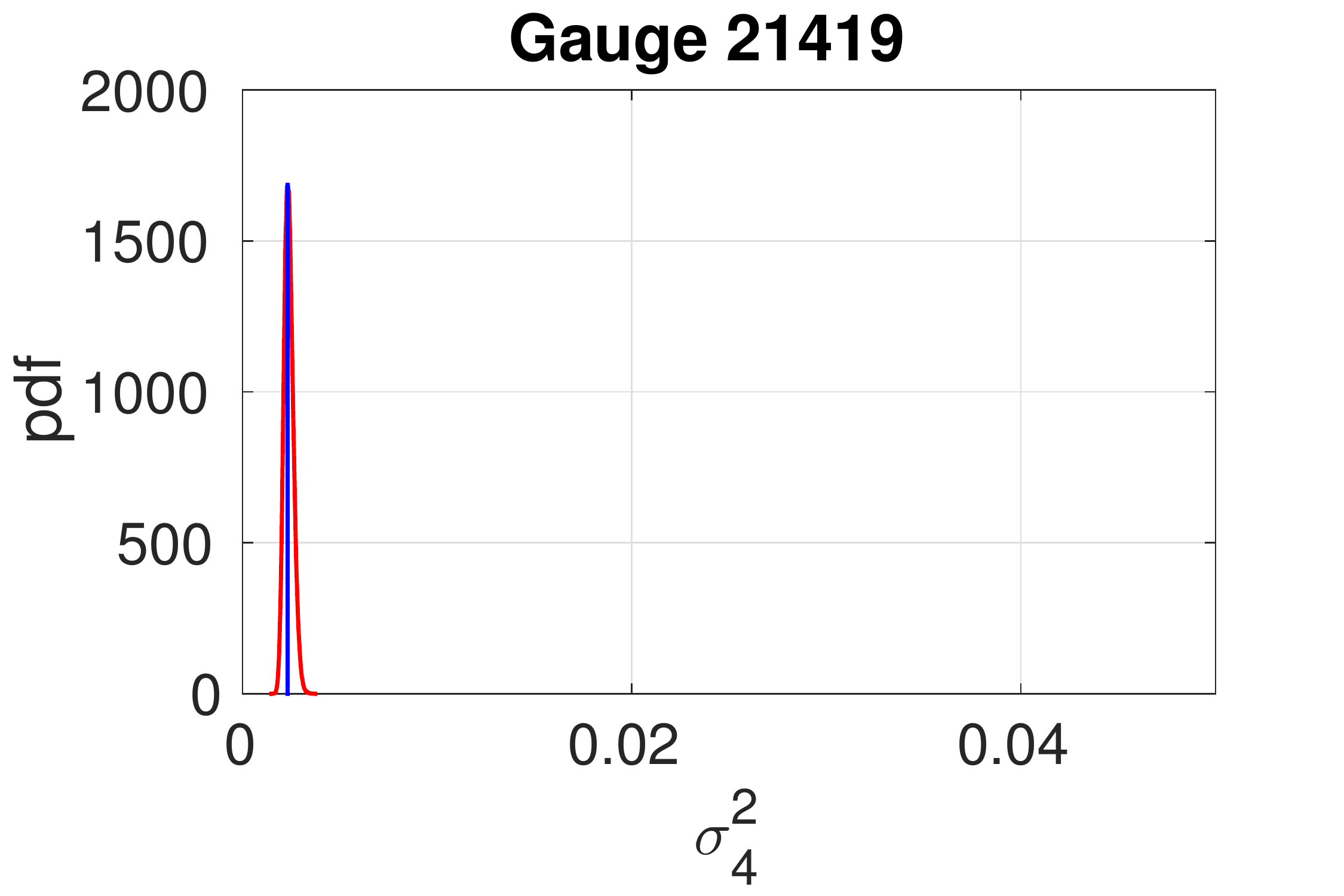} \\
\end{tabular}
\caption{KDE of the marginalized posterior distribution of the noise variance 
$\Pi(\sigma^2_i)$ at each gauge.}
\label{fig:pdfs_s} 
        \end{figure}

Regarding the noise variances, their \emph{pdfs} are shown in Figure~\ref{fig:pdfs_s} at the different gauges.
The \emph{pdfs} appear to be well-defined and Gaussian shaped with a clear MAP values.
These MAP values can be used to estimate the maximum water surface elevation standard deviation that was found to 
be $\sigma_3=0.182~m$ at gauge 21418. This value is a reflection of the mismatch between the model and 
observed data.  The $\sigma^2_i$ estimates are noticeably lower than those obtained with \geoclaw default slip distribution (shown in Figure~\ref{fig:observations} (Right)):
$2.27\times 10^{-3}~m^2$ versus $7.99\times 10^{-3}~m^2$, $1.22\times 10^{-2}~m^2$ versus $9.65\times 10^{-3}~m^2$, $3.32 \times 10^{-2}~m^2$ versus $4.62 \times 10^{-2}~m^2$, $2.38\times 10^{-3}~m^2$ versus $5.86\times 10^{-3}~m^2$ at gauges 20401, 21413, 21418 and 21419, respectively.  

\begin{figure}[ht]
    \centering
    \includegraphics[width=0.6\textwidth]{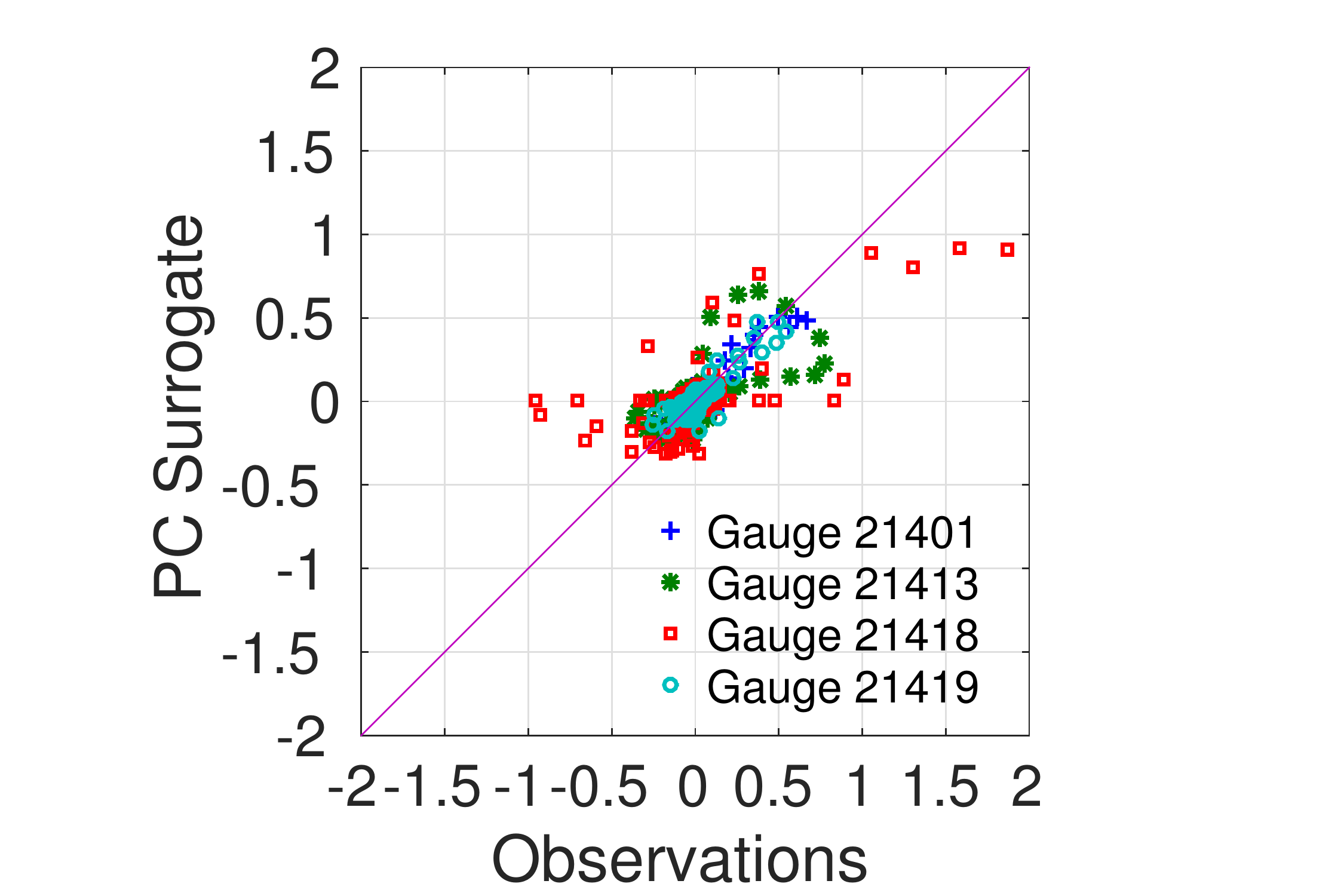}
    \label{fig:scatter_inf}
    \caption{Scatter plot of the measured water surface elevation against their
    PC model counterparts at the four gauges shown in different colors. The PC model
    was evaluated using the mean of the fault slip posteriors. The variance of the
    error between the two sets of values is: $2.27\times 10^{-3}~m^2, 1.22\times
    10^{-2}~m^2, 3.32 \times 10^{-2}~m^2, 2.38\times 10^{-3}~m^2$ at each gauge.}
\end{figure}

\begin{figure}[ht]
\centering
\begin{tabular}{clc}
\includegraphics[width=0.5\textwidth]{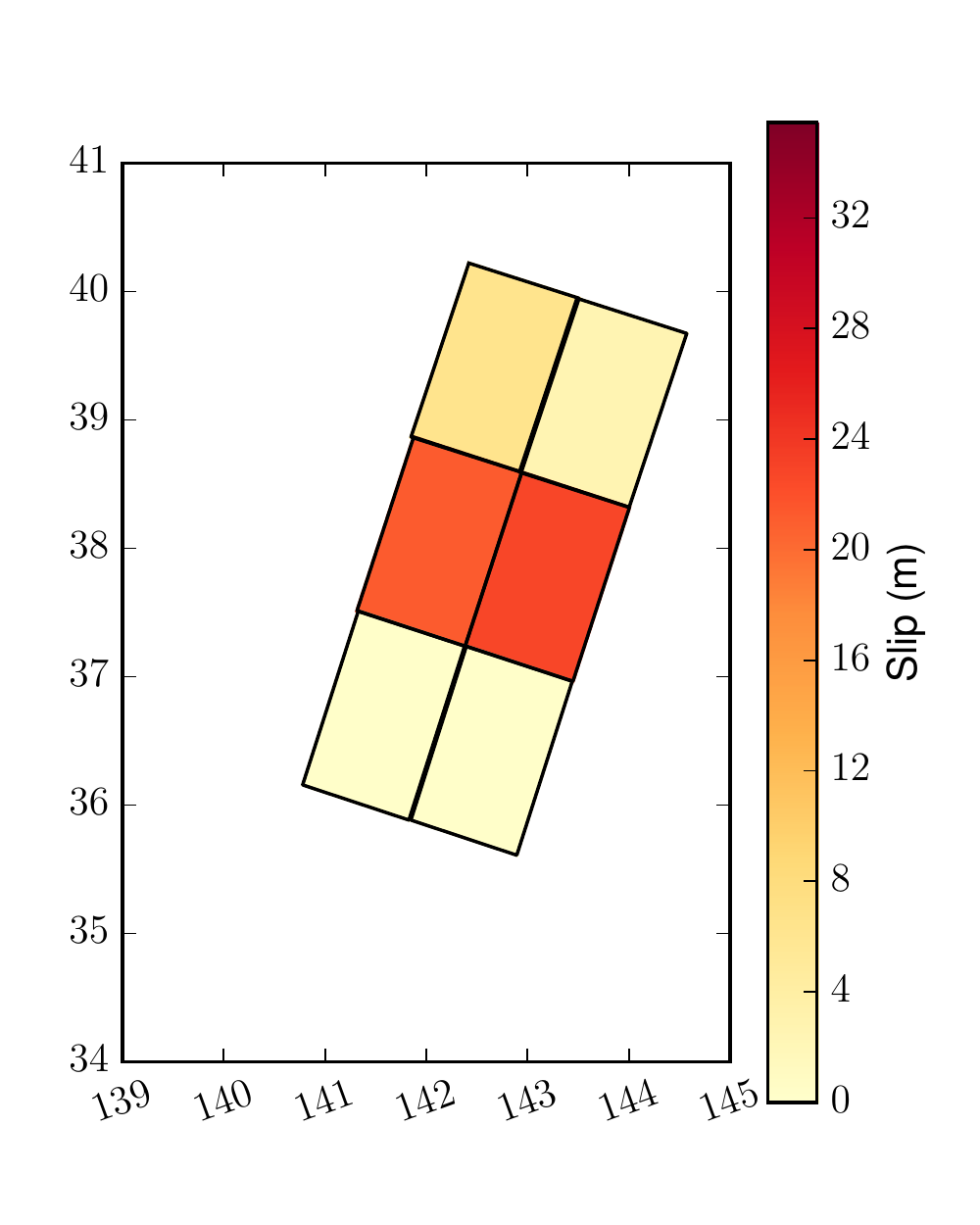} &
\includegraphics[width=0.5\textwidth]{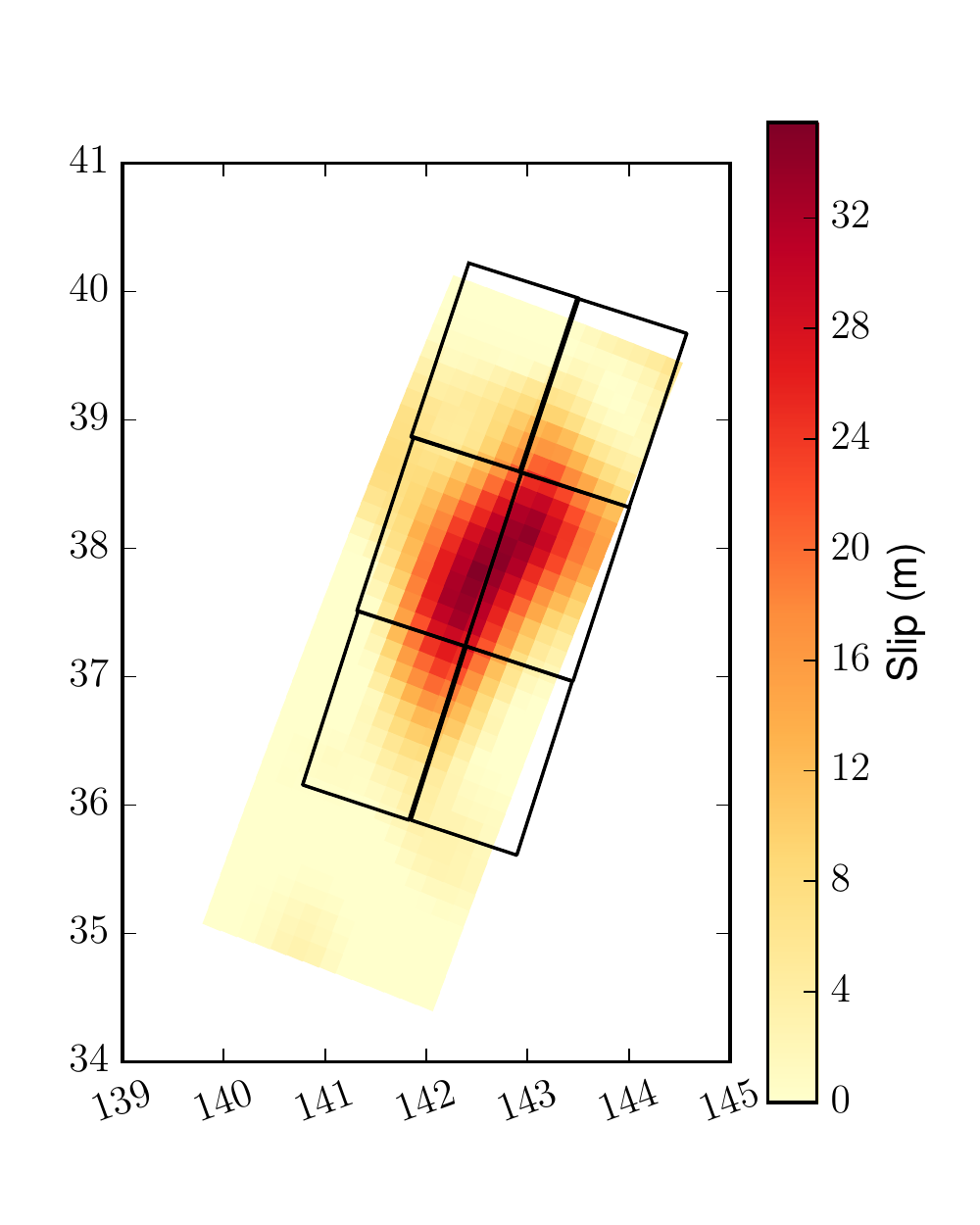} &
\end{tabular}
\caption{(Left) MAP of the inferred fault slip values. (Right) Fault slip distribution from Ammon \emph{et al.} }
\label{fig:map_inversion} 
\end{figure}

The scatter plot shown in Figure~\ref{fig:scatter_inf} uses inferred MAP
values. Thus, the parameters MAP values have reduced the discrepancies between
simulated water surface elevation and DART buoy data. This comparison can be
seen as an evaluation of the \emph{a posteriori} goodness-of-fit.  Additionally
Figure~\ref{fig:map_inversion} shows the comparison between the MAP values and
the Ammon \emph{et al.} model~\cite{Ammon:2011dm}.  Note that the inverted fault leads to a moment
and magnitude of $M_o = 3.43900\times 10^{22}$, $M_w = 8.99095$, respectively, whereas the Ammon \emph{et al.} model yields a moment and magnitude of $M_o = 3.63595\times 10^{22}$ and 
$M_w = 9.00708$.

\section{Discussion and Conclusions}
\label{sec:conc}

In this study, we sought to estimate the fault slip distribution that 
plays a critical role in earthquake and tsunami modeling, mainly in the 
prediction of water surface elevations. To this end, we proposed 
a low-dimensional parameterization of the fault slip distribution
in which we assumed the fault consists of six sub-faults that
have different slip magnitudes. The estimation of the fault slip distribution thus boiled down into a six-parameter inverse problem.
A Bayesian inference approach was employed
that sharpens the initial estimates of the six uncertain parameters based on measured observations.
In our test case, the \tohoku tsunami, we used water surface elevations information collected at four DART buoy gauges. 
Discrepancies with measurements were accounted for using a Gaussian noise model, whose variance was treated as a hyper parameter that was inferred along with the uncertain fault parameters.

Bayesian inference was accelerated using a surrogate
model constructed based on the Polynomial Chaos approach where
the output of the forward model \geoclaw was approximated using PC expansions.
The PC expansions were constructed based on a compressed sensing approach that uses basis-pursuit denoising technique,
to produce a faithful surrogate. This PC surrogate model was additionally 
used to quantify the uncertainties in the predicted water surface
elevations due to the uncertainties in slip values. This included the mean and standard deviation 
of water surface elevations.

The present study focused on formulating and estimating a low-dimensional
representation of the fault slip distribution using UQ techniques, namely
Bayesian inference and PC expansions. A high-dimensional representation of the fault slip distribution
would, however, require a large number of forward runs that is computational prohibitive.
Instead, one could exploit order-reduction techniques to
reduce the dimensionality such as Karhunen-Lo\`eve expansions~\cite{Sraj2016}. This will be the objective of a future study.

\section*{Acknowledgment}
Research reported in this publication was supported by the King Abdullah
University of Science and Technology (KAUST) in Thuwal, Saudi Arabia grant number CRG3-2156.
The authors would like to thank Dr. Olivier Le Maitre for the helpful discussions of the results.

\bibliographystyle{elsarticle-num}

\clearpage


\clearpage
\clearpage
\clearpage
\clearpage

\clearpage

\clearpage

\clearpage


\end{document}